\documentclass[aps,a4paper,showkeys,longbibliography,notitlepage,twocolumn]{revtex4-1}
\usepackage[utf8]{inputenc}
\usepackage{filecontents}
\usepackage{natbib}
\usepackage{amsmath,amssymb,bm,amsthm}
\usepackage{subfigure}
\usepackage{graphicx}
\usepackage{dcolumn}
\usepackage{bm}
\usepackage[mathlines]{lineno}
\usepackage{booktabs}
\usepackage{color}
\newcounter{one}
\usepackage{url}

\usepackage{textcase}

\usepackage{colortbl}
\usepackage{tabularx}
\usepackage{verbatim}
\usepackage{multirow}

\usepackage[T1]{fontenc} 
\usepackage{lmodern}
\usepackage{bbm}
\usepackage[utf8]{inputenc}
\usepackage{amsfonts}
\usepackage{array}

\usepackage{bbm}
\usepackage{enumitem}
\usepackage{umoline}
\usepackage[usenames,svgnames]{xcolor}
\usepackage{natbib}
\usepackage[hyperindex,breaklinks]{hyperref}
\hypersetup{
     colorlinks=true,       		
     linkcolor=Navy,          	
     citecolor=Navy,            
     filecolor=Navy,      		
     urlcolor=Navy,           	
    runcolor=cyan,
 }
\usepackage{graphicx}
\usepackage{amsfonts}
\usepackage{amssymb}
\usepackage{amsmath}
\usepackage{bbm}
\usepackage{enumitem}

\newcommand{\bra}[1]{\langle #1 |}
\newcommand{\ket}[1]{| #1 \rangle}

\newcommand{\tr}[0]{ {\rm tr}}

\newcommand{\half}[1]{{ \rm h}}
\newcommand{\Oorderof}{\mathcal{O}}
\newcommand{\orderof}[1]{\Oorderof(#1)} 

\newcommand{\for}[0]{\quad \textrm{for} \quad}

\newcommand{\dist}{d}

\newcommand{\co}{{\rm c}}

\newcommand{\poly}{{\rm poly}}

\newcommand{\ad}{{\rm ad}}

\def\beq{\begin{equation}}
\def\eeq{\end{equation}}
\def\nbeq{\begin{equation*}}
\def\neeq{\end{equation*}}
\def\<{\langle}
\def\>{\rangle}

\def\tr{{\rm tr}}

\newcommand{\PPT}{{\rm PPT}}
\newcommand{\SEP}{{\rm SEP}}

\newcommand{\mi}{{\mathcal{I}}}

\newtheorem{theorem}{Theorem}

\newtheorem{lemma}[theorem]{Lemma}
\newtheorem{corol}[theorem]{Corollary}

\newtheorem{prop}[theorem]{Proposition} 
\newtheorem{claim}[theorem]{Claim} 
\newtheorem{conj}[theorem]{Conjecture} 

\newcommand{\bal}[2]{#1[#2]}

\newcommand{\br}[1]{\left( #1 \right)}
\newcommand{\brr}[1]{\left[ #1 \right]}
\newcommand{\abs}[1]{\left| #1 \right|}
\newcommand{\norm}[1]{\left \|  #1 \right \|}

\newcommand{\QC}{{\rm QC}}
\newcommand{\C}{{\rm C}}

\usepackage{mathtools}
\def\multiset#1#2{\ensuremath{\left(\kern-.3em\left(\genfrac{}{}{0pt}{}{#1}{#2}\right)\kern-.3em\right)}}

\begin{document}



\title{Exponential clustering of bipartite quantum entanglement at arbitrary temperatures}





\author{Tomotaka Kuwahara$^{1,2}$ and Keiji Saito$^{3}$}
\email{tomotaka.kuwahara@riken.jp}
\affiliation{$^{1}$
Mathematical Science Team, RIKEN Center for Advanced Intelligence Project (AIP),1-4-1 Nihonbashi, Chuo-ku, Tokyo 103-0027, Japan
}
\affiliation{$^{2}$
JST PRESTO, 4-1-8 Honcho, Kawaguchi, Saitama 332-0012, Japan}

\affiliation{$^{3}$Department of Physics, Keio University, Yokohama 223-8522, Japan}

\begin{abstract}

Many inexplicable phenomena in low-temperature many-body physics are a result of macroscopic quantum effects. Such macroscopic quantumness is often evaluated via long-range entanglement, that is, entanglement in the macroscopic length scale. Long-range entanglement is employed to characterize novel quantum phases and serves as a critical resource for quantum computation. However, the conditions under which long-range entanglement is stable, even at room temperatures, remain unclear. In this regard, this study demonstrated the unstable nature of bi-partite long-range entanglement at arbitrary temperatures, which exponentially decays with distance. 
The proposed theorem is a no-go theorem pertaining to the existence of long-range entanglement.
The obtained results were consistent with existing observations, indicating that long-range entanglement at non-zero temperatures can exist in topologically ordered phases, where tripartite correlations are dominant. The derivation in this study introduced a quantum correlation defined by the convex roof of the standard correlation function. Further, an exponential clustering theorem for generic quantum many-body systems under such a quantum correlation at arbitrary temperatures was established, which yielded the primary result by relating quantum correlation with quantum entanglement. Moreover, a simple application of analytical techniques was demonstrated by deriving a general limit on the Wigner-Yanase-Dyson skew and quantum Fisher information; this is expected to attract significant attention in the field of quantum metrology. Notably, this study reveals the novel general aspects of low-temperature quantum physics and clarifies the characterization of long-range entanglement.

\end{abstract}

\maketitle





\maketitle


\section{Introduction}

\subsection{Background}

In quantum many-body physics, macroscopic quantum effects such as superconductivity, Bose-Einstein condensation, quantum spin liquid, and quantum topological order are critical features of exotic quantum phenomena. 
In these phenomena, the length scale of the quantum effect is comparable to that in the real world. 
However, the clarification of such macroscopic quantum effects remains a crucial problem in modern physics, and various methods for characterizing quantumness in the macroscopic length scale have been proposed~\cite{RevModPhys.34.694,10.1143/PTP.69.80,Leggett_2002,PhysRevLett.110.160403}. 
In particular, over the last two decades, quantum entanglement has become a representative measure for the quantumness~\cite{RevModPhys.81.865,RevModPhys.90.025004}. 
Several studies have investigated the entanglement
behaviors in quantum many-body systems from various perspectives~\cite{PhysRevA.66.032110,PhysRevLett.87.017901,PhysRevLett.90.227902,PhysRevLett.92.027901,RevModPhys.80.517,Latorre_2009,PhysRevLett.101.010504,Calabrese_2009,RevModPhys.82.277,LAFLORENCIE20161}.
These advances in quantum entanglement have significantly contributed toward improving our understanding and establishing efficient classical and quantum algorithms to simulate quantum many-body systems~\cite{PhysRevLett.99.220405,doi:10.1080/14789940801912366,SCHOLLWOCK201196,0034-4885-75-2-022001,gharibian2015quantum}.


\begin{figure}[tt]
\centering
{\includegraphics[clip, scale=0.7]{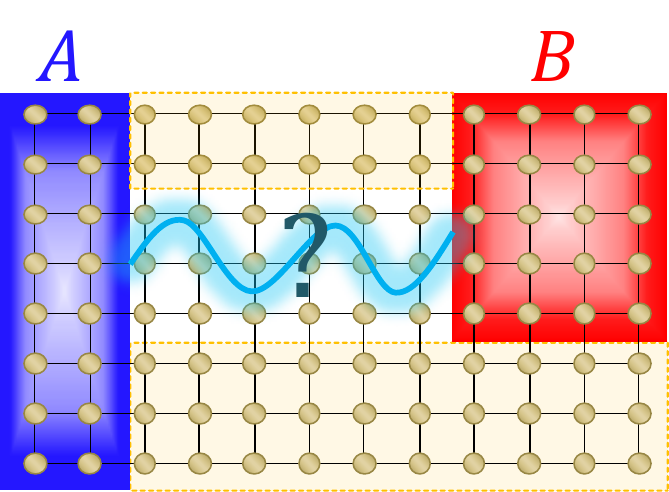}}
\caption{(Entanglement between two separated subsystems $A$ and $B$) 
Considering the tripartite entanglement between subsystems $A$, $B$, and $C$, long-range entanglement at non-zero temperatures can be detected. It can also be observed in topologically ordered quantum many-body systems. 
This study aimed to elaborate the observation and prove that (more than) tripartite entanglement is required for long-range entanglement.
The quantum entanglement between two systems is shown to decay exponentially with distance at arbitrary non-zero temperatures, in any quantum Gibbs state with short-range interacting Hamiltonians.
Here, the entanglement-length scale, at most, grows in a polynomial manner with the inverse temperature $\beta$, as stated in inequalities~\eqref{quantum_corr_clustering}, \eqref{Ineq:main_conj_partial_proof}, and \eqref{Ineq:main_conj_partial_proof_1D}. 
}
\label{fig_entanglement_clustering}
\end{figure}

A critical question regarding many-body quantum entanglement is whether entanglement can exist in the macroscopic length scale. Such entanglement is often referred to as long-range entanglement, which plays a crucial role not only in characterizing quantum phases~\cite{PhysRevB.82.155138,RevModPhys.89.041004} but also in realizing quantum computing~\cite{PhysRevA.71.062313,RevModPhys.80.1083,PhysRevLett.111.080503}.
It can be inferred that temperature plays an essential role in this context. 
Moreover, owing to the fragility of quantumness, thermal noise destroys the entanglement, making the length scale of the entanglement short range. 
Indeed, at a sufficiently high temperature where the possibility of thermal phase transition is eliminated, the quantum thermal state can be classified as the trivial phase~\cite{CMI_clustering} (i.e., generated by the finite-depth quantum circuit~\cite{PhysRevB.82.155138}). 
By contrast, at zero temperature, various quantum systems are known to exhibit long-range entanglement~\cite{PhysRevLett.96.247206,Gottesman_2010,Vitagliano_2010,PhysRevLett.106.050501,Sahling2015}.
However, at non-zero but low temperatures, where thermal phase transition can occur, the effect of temperature on the entanglement remains highly unclear. 
In this case, the effect of thermal noise is sufficiently suppressed, and it is possible to observe long-range entanglement in this temperature regime. 
Consequently, in such quantum systems, quantum phases protected by the topological order can exhibit long-range entanglement. 
It has been shown that, in the 4D toric code model~\cite{KITAEV20032}, long-range entanglement can occur even at room temperatures~\cite{doi:10.1142/S1230161210000023,PhysRevLett.107.210501} (see also~\cite{eldar2019robust,anshu2020circuit}).


The purpose of this study was to identify the limitations associated with the structure of long-range entanglement at arbitrary non-zero temperatures. 
In the known example involving long-range entanglement, the protection afforded by the topological order plays an essential role.
Moreover, the topological order is inherently a tripartite correlation~\cite{PhysRevLett.96.110404,PhysRevLett.96.110405,PhysRevA.93.022317}.  
However, these findings pose the following fundamental question: \textit{``Can the long-range entanglement at non-zero temperatures only exist as (more than) tripartite correlations, or equivalently, does bi-partite entanglement necessarily decay to zero at long distances under arbitrary temperatures?''}  
We conjecture that the answer to this question is yes (Fig.~\ref{fig_entanglement_clustering}). 
The possibility of this conjecture being true can provide crucial information related to identifying the essence of long-range entanglement in the quantum phases at non-zero temperatures, which can further serve a guideline in the search for candidate systems suitable for quantum devices.
The conjecture is trivially true for arbitrary commuting Hamiltonians~\footnote{In general, it can be easily shown~\cite[Theorem~1]{Kuwahara_2012} that any quantum Gibbs state has no bipartite entanglement between arbitrary subsystems $A$ and $B$ if the Hamiltonian $H$ is decomposed to $H=H_1 + H_2$ with $[H_1,H_2]=0$, provided the Hamiltonian $H_1$ ($H_2$) includes interactions on subset $A$ ($B$) and not on $B$ ($A$)}, where all the local interaction terms commute with each other.
Hence, in the toric code model with the commuting Hamiltonian, bipartite long-range entanglement is strictly prohibited, regardless of the existence of the tripartite long-range entanglement.
Thus, as long as the commuting Hamiltonian is considered, the conjecture does not contradict the observations. 

Thus far, rigorous and general studies on low-temperature phases remain scarce.
At low temperatures, in contrast to high-temperature phases, the structures of quantum many-body systems are considerably influenced by the system details. 
Therefore, analyses of the low-temperature properties are often considered as computationally hard problems~\cite{Barahona_1982,ref:KempeKitaevRegev}.
In such situations, all the long-range quantum effects are not strictly prohibited (e.g., off-diagonal long-range order~\cite{RevModPhys.34.694}), 
with only a fraction of them being forbidden at low temperatures. 
In the latter example, the thermal area law is known as a representative characterization of the low-temperature phases of many-body systems, which is universally true at arbitrary temperatures~\cite{RevModPhys.80.517,PhysRevLett.100.070502,PhysRevX.11.011047}.  
It states that the entanglement between two adjacent subsystems can reach the maximum of the size of their boundaries.
In other words, the area law implies that entanglement should be localized around the boundary and thus indirectly supports the argument presented.

\subsection{Brief description of main results}



Here, we provide an overview of the contributions of this study. 
The quantum Gibbs state is denoted as $\rho_\beta$ at inverse temperature $\beta$, where a short-range interacting Hamiltonian is considered (further details are provided in Sec.~\ref{sec:Hamiltonian and quantum Gibbs state}).  
Let $\rho_{\beta, AB}$ be a reduced density matrix on the subsystems $A$ and $B$, which are separated by distance $R$. 
For an arbitrary choice of $A$ and $B$, we focus on the entanglement between $A$ and $B$ (Fig.~\ref{fig_entanglement_clustering}).

First, the primary challenge faced when addressing the main problem is that the entanglement for a mixed state cannot be described in an analytically tractable form (e.g., Eqs.~\eqref{def:relative_entanglement_ent_general} and \eqref{def_Entanglement_form_corr}). 
Moreover, owing to the computational hardness~\cite{10.5555/2011725.2011730,10.5555/2011350.2011361}, the entanglement cannot be computed even at numerical levels, except for specific cases~\cite{PhysRevLett.80.2245}.
However, in free fermion and harmonic chains, analytical forms of entanglement negativity~\cite{PhysRevA.65.032314} [see Eq.~\eqref{def_Entanglement_Negativity_corr}] have been obtained~\cite{Calabrese_2014,Eisler_2014,Shapourian_2019} at finite temperatures.
These studies considered the entanglement negativity between adjacent subsystems $A$ and $B$ (i.e., $R=0$) on one-dimensional chains and consequently analyzed the manner in which the negativity is saturated with an increase in the sizes of $A$ and $B$ (e.g., setting $|A|=|B|=\ell$ and tuning length $\ell$).  
In these systems, the saturation rate is approximately expressed as $e^{-\ell/\orderof{\beta}}$, and Ref.~\cite{Shapourian_2019} concluded that 
quantum coherence can only be maintained for length scales of $\orderof{\beta}$. 
Similar observations have been numerically obtained for a more general class of many-body systems~\cite{PhysRevResearch.2.043345,PhysRevLett.125.140603}.  
Thus, these results strongly support the clustering of bi-partite entanglement in specific models.

To overcome the difficulties in the analysis of the entanglement, first, a quantum correlation~$\QC_\rho (O_A,O_B)$ was introduced, which is defined based on the analogy of the entanglement measure and obtained from the convex roof of the standard correlation function $\C_\rho(O_A,O_B)=\tr(\rho O_AO_B) - \tr(\rho O_A) \tr(\rho O_B)$, as in Eq.~\eqref{def_Quantum_corr}.  
The quantum correlation $\QC_\rho(O_A,O_B)$ is strongly associated with entanglement (see Sec.~\ref{sec:Quantum correlation}).
In particular, the upper bound of the quantum correlation yields an upper bound for the entanglement measure of the positive-partial-transpose (PPT) relative entanglement (Proposition~\ref{prop:quantum_correlation_negativity}).  
In general, the exponential clustering of the quantum correlation at arbitrary temperatures of arbitrary dimensions can be proven (see Theorem~\ref{thm:quantum_correlation}): 
\begin{align}
\label{quantum_corr_clustering}
&\QC_{\rho_\beta} (O_A,O_B)\lesssim (|\partial A|+| \partial B|) e^{-R/\xi_\beta},
\end{align}
with $\xi_\beta=\orderof{\beta}$, whose explicit form is expressed as Eq.~\eqref{parameters_definitions}, where $O_A$ and $O_B$ are supported on subsets $A$ and $B$, respectively.  
The inequality~\eqref{quantum_corr_clustering} provides a quantum version of the clustering theorem that generally holds \textit{at arbitrary temperatures}.

\begin{figure}[tt]
\centering
{\includegraphics[clip, scale=0.35]{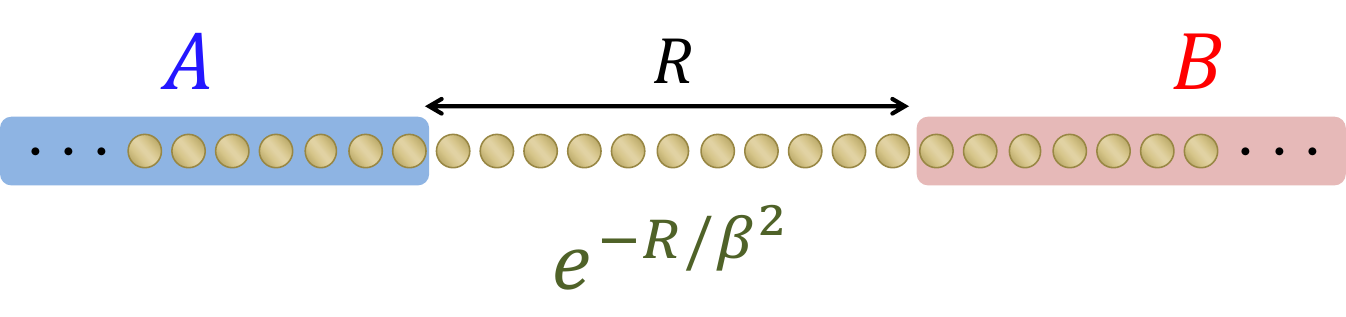}}
\caption{(Schematic of 1D entanglement clustering) 
In the obtained bound~\eqref{Ineq:main_conj_partial_proof}, the subset dependence $e^{\orderof{|A| + |B|}}$ prohibits its application to the upper bound of the entanglement between two large blocks. 
In one-dimensional systems, this problem can be resolved to obtain better subset dependence, as in \eqref{Ineq:main_conj_partial_proof_1D}. 
Here, the characteristic length of bipartite entanglement becomes $\orderof{\beta^2}$ instead of $\orderof{\beta}$.}
\label{fig_entanglement_clustering_1D}
\end{figure}

Based on the upper bound~\eqref{quantum_corr_clustering}, it may be possible to avoid the intractability of the quantum entanglement.
Further, using the association between the quantum correlation and the entanglement,   
the following statement on entanglement clustering is proven (see Corollary~\ref{corol:quantum_correlation_negativity}): 
\begin{align}
\label{Ineq:main_conj_partial_proof}
&E_R^{\PPT}(\rho_{\beta,AB}) \lesssim e^{-R/\xi_\beta + \orderof{|A| + |B|}},
\end{align}
where $E_R^{\PPT}(\rho_{AB})$ is the PPT relative entanglement~\eqref{def_Entanglement_PPT_relative}. 
Herein, two points can be improved: i) a bound is obtained for $E_R^{\PPT}$ instead of the standard relative entanglement $E_R$, and
ii) the subset dependence is exponential (i.e., $e^{\orderof{|A| + |B|}}$) instead of polynomial [i.e., $\poly(|A|,|B|)$]. 
To address the first point, the zero-quantum correlation must be related to the separable condition 
instead of the PPT condition (Lemma~\ref{lem:QC_Entanglement_condition_2}).
However, this point remains to be addressed (Conjecture~\ref{conj_entanglement_exist}).
Regarding the second point, the inequality~\eqref{Ineq:main_conj_partial_proof} in one-dimensional systems (Theorem~\ref{thm:quantum_correlation_negativity_1D}, Fig.~\ref{fig_entanglement_clustering_1D}) can be improved by refining the analyses based on the belief propagation~\cite{PhysRevB.76.201102,PhysRevB.86.245116}:
\begin{align}
\label{Ineq:main_conj_partial_proof_1D}
&E_R^{\PPT}(\rho_{\beta,AB}) \lesssim (|A|+|B|) e^{-\orderof{R/\xi^2_\beta}}. 
\end{align}
Thus, a significantly improved clustering theorem for the bipartite entanglement measure in one-dimensional systems can be obtained.

Finally, as a related quantity, another type of quantum correlation that is based on the Wigner-Yanase-Dyson (WYD) skew information~\cite{PhysRevLett.117.130401,Frerot2019} is considered: 
$\bar{Q}_{\rho}(O_A,O_B) :=\int_0^1Q_{\rho}^{(\alpha)}(O_A,O_B) d\alpha $ with
$Q_{\rho}^{(\alpha)}(O_A,O_B) := \tr (\rho O_AO_B) - \tr \br{ \rho^{1-\alpha} O_A \rho^{\alpha} O_B }$. 
In a previous study~\cite{PhysRevLett.117.130401}, it was numerically verified that the quantity $\bar{Q}_{\rho}(O_A,O_B)$ exhibits an exponential decay with distance, even at the critical point. Because the WYD skew information is considered as a measure of quantum coherence~\cite{RevModPhys.89.041003},
the decay rate of $\bar{Q}_{\rho}(O_A,O_B)$ has been dubbed as the ``quantum coherence length''~\cite{PhysRevLett.117.130401}.
Consequently, using a similar analysis for the proof of Ineq.~\eqref{quantum_corr_clustering}, it is proven that the numerical observations in Ref.~\cite{PhysRevLett.117.130401,Frerot2019} are universally true (Theorem~\ref{quantum_correlation_skey_type}):
\begin{align}
\label{quantum_corr_clustering_intri}
&Q_{\rho_\beta}^{(\alpha)}(O_A,O_B)\lesssim (|\partial A|+| \partial B|) e^{-R/\xi'_\beta}
\end{align}
for arbitrary $\alpha$, 
where $\xi'_\beta=\orderof{\beta}$ is explicitly expressed as Eq.~\eqref{def_C'_beta_si_beta}.
The above inequality also yields the general limits on the WYD skew information as well as the quantum Fisher information:
\begin{align}
\label{WYD_Fisher}
\mathcal{I}_{\rho_\beta}^{(\alpha)}(K) \lesssim \beta^D n \quad {\rm and} \quad \mathcal{F}_{\rho_\beta} (K) \lesssim \beta^D n,
\end{align}
with $K$ being an arbitrary operator in the form of $K=\sum_{i\in \Lambda} O_i$ ($\Lambda$: total set of the sites), 
where $\mathcal{I}_{\rho_\beta}^{(\alpha)}(K)$ and $\mathcal{F}_{\rho_\beta} (K)$ are the WYD skew ~\eqref{def:quantum_information_skey_type} and quantum Fisher ~\eqref{def_quantum_Fisher} information, respectively. 
These general limits provide useful information related to the application of quantum many-body systems to quantum metrology~\cite{PhysRevLett.121.020402,PhysRevLett.124.120504,PhysRevResearch.2.043329,PhysRevX.8.021022,PhysRevLett.126.010502}.

The remainder of this paper is organized as follows.
In Sec.~\ref{sec:Setup and Preliminaries}, 
the precise setting and notations used throughout the paper are formulated, coupled with the introduction to certain preliminaries such as the Lieb-Robinson bound and entanglement measure. 
In Sec.~\ref{sec:Quantum correlation}, 
the quantum correlation $\QC_{\rho} (O_A,O_B)$ is introduced as the convex roof of the standard correlation function.
In addition, several rigorous results on the relationships between the quantum correlation and quantum entanglement are provided. 
Further, in Sec.~\ref{sec:Exponential clustering for quantum correlations}, the main results on the clustering theorem for the quantum correlation [Ineq.~\eqref{quantum_corr_clustering}] and the PPT relative entanglement [Ineqs.~\eqref{Ineq:main_conj_partial_proof} and \eqref{Ineq:main_conj_partial_proof_1D}] are provided. 
Thereafter, in Sec.~\ref{Quantum correlations based on the skew information}, the obtained results are demonstrated on the WYD skew and quantum Fisher information [Ineqs.~\eqref{quantum_corr_clustering_intri} and \eqref{WYD_Fisher}]. 
In Sec.~\ref{sec:Further discussions}, the following topics relevant to the obtained results are discussed: 
i) relationship between the macroscopic quantum effect and quantum entanglement (Sec.~\ref{sec:Macroscopic quantum effect v.s. quantum entanglement}), 
ii) relationship between entanglement clustering and quantum Markov property (Sec.~\ref{sec:Macroscopic quantum effect v.s. quantum entanglement}), 
iii) relationship between the quantum correlation and entanglement of formation (Sec.~\ref{sec:General upper bound on the quantum correlation}), 
iv) optimality of the proposed main theorems (Sec.~\ref{sec:Optimality of the obtained bounds}), and 
v) extension of the results obtained to more general quantum states based on the Bernstein-Widder theorem (Sec.~\ref{sec:Beyond quantum Gibbs states}). 
Finally, in Sec. \ref{sec:Summary and Future works}, the study is summarized, along with a discussion regarding the scope for future work.




%



\section{Set up and Preliminaries} \label{sec:Setup and Preliminaries}
Consider a quantum system on a $D$-dimensional lattice with $n$ sites. 
On each of the sites, the Hilbert space with dimension $d_0$ is assigned. 
Let $\Lambda$ be the set of total sites. Further, for an arbitrary subset $X\subseteq \Lambda$, the cardinality (the number of sites contained in $X$) is denoted as $|X|$.
In addition, a complementary subset of $X$ is denoted as $X^\co := \Lambda\setminus X$.
For an arbitrary subset $X\subseteq \Lambda$, $\mathcal{D}_X$ is defined as the dimension of the Hilbert space on $X$, that is, 
$\mathcal{D}_X=d_0^{|X|}$. 
Finally, $X\cup Y$ is denoted as $XY$.

For arbitrary subsets $X, Y \subseteq \Lambda$, $\dist_{X,Y}$ is defined as the shortest path length on the graph connecting $X$ and $Y$; in other words, if $X\cap Y \neq \emptyset$, $\dist_{X,Y}=0$. 
However, when $X$ comprises only one element (e.g., $X=\{i\}$), the distance $\dist_{\{i\},Y}$ is denoted as $\dist_{i,Y}$ for simplicity.
In addition, the surface subset of $X$ is denoted as $\partial X:=\{ i\in X| \dist_{i,X^\co}=1\}$.

For a subset $X\subseteq \Lambda$, the extended subset $\bal{X}{r}$ is defined as
\begin{align}
\bal{X}{r}:= \{i\in \Lambda| \dist_{X,i} \le r \}, \label{def:bal_X_r}
\end{align}
where $\bal{X}{0}=X$, and $r$ is an arbitrary positive number (i.e., $r\in \mathbb{R}^+$).
Based on the notation, for $i\in \Lambda$, the subset $i[r]$ is concluded to be a ball region with radius $r$ centered at the site $i$. 
A geometric parameter $\gamma$ is introduced, which is determined based on the lattice structure alone.
Further, $\gamma \ge 1$ is defined as a constant of $\orderof{1}$ that satisfies the following inequalities:
\begin{align}
\max_{i\in \Lambda} \br{ |\partial i[r] | } \le \gamma r^{D-1}, \quad \max_{i\in \Lambda} \br{ | i[r] |} \le \gamma r^{D}, \label{def:parameter_gamma} 
\end{align}
where $r\ge 1$.

\subsection{Hamiltonian and quantum Gibbs state} \label{sec:Hamiltonian and quantum Gibbs state}
Throughout the study, generic Hamiltonians with few-body interactions are considered.
Here, the Hamiltonian is expressed in the following $k$-local form:
\begin{align}
H= \sum_{|Z| \le k} h_Z, \quad \max_{i\in \Lambda}\sum_{Z:Z\ni i} \|h_Z\| \le g, \label{supp_def:Ham}
\end{align}
where each of the interaction terms $\{h_Z\}_{|Z|\le k}$ acts on the spins on $Z \subset \Lambda$.
For an arbitrary subset $L\subset \Lambda$, the subset Hamiltonian, which includes interactions in a subset $L$, is denoted as $H_L$:
\begin{align}
H_L= \sum_{Z: Z\subset L} h_Z. \label{supp_def:Ham_L}
\end{align}

To characterize the interaction strength of the Hamiltonian, the following assumption is imposed:
\begin{align}
\max_{\{i,j\}\subset \Lambda}\sum_{Z \supset \{i,j\}} \|h_Z\| \le J(\dist_{i,j}), 
\label{supp_def:interaction length}
\end{align}
where $J(x)$ is a function that monotonically decreases with $x\ge 0$.
Here, the short-range interaction is primarily considered, where the decay of the function $J(x)$ is faster than exponential decay; in other words, 
\begin{align}
\label{int_short_range}
J(x) \le g_0 e^{-\mu_0 x} \quad (\textrm{short-range interaction})
\end{align}
with $g_0=\orderof{1}$ and $\mu_0=\orderof{1}$.
The results can be generalized to a broader class of interactions, as discussed in Appendix~\ref{sec:Beyond short/range interacting spin systems}.

Using the Hamiltonian, the quantum Gibbs state can be defined as follows:
\begin{align}
\rho_\beta = \frac{e^{-\beta H}}{Z_\beta}, \quad Z_\beta= \tr(e^{-\beta H}), \label{def:Ham_Gibbs}
\end{align}
where $\beta$ is the inverse temperature.
Throughout the paper, by appropriately choosing the energy origin, $Z_\beta=1$ is enforced, that is, 
\begin{align}
\rho_\beta = e^{-\beta H}. \label{def:Ham_Gibbs_2}
\end{align}
However, when considering a reduced density matrix on a region $L$ ($L\subset \Lambda$), it is denoted as $\rho_{\beta,L}$:
\begin{align}
\label{reduced_rho_def}
\rho_{\beta,L} := \tr_{L^\co} (\rho_\beta),
\end{align}
where $\tr_{L^\co}$ implies the partial trace for the Hilbert space on the subset $L^\co$.

\subsection{Lieb-Robinson bound}

\begin{table*}[tt]
 \caption{Fundamental parameters in our statements}
  \label{tab:fund_para} 
\begin{ruledtabular}
\begin{tabular}{lr}
\textrm{\textbf{Definition}}&\textrm{\textbf{Parameters}} 
\\
\colrule
Spatial dimension    
&  $D$ \\
Local Hilbert space dimension    
&  $d_0$ \\
Structure parameter of the lattice (see Eq.~\eqref{def:parameter_gamma})      
& $\gamma$  \\
 Maximum number of sites involved in interactions (see Eq.~\eqref{supp_def:Ham})    
&    $k$ \\
Upper bound on the one-site energy (see Eq.~\eqref{supp_def:Ham})     
&  $g$   \\
 parameters in the Lieb-Robinson bound (see Ineq.~\eqref{Lieb-Robinson_main_short})   
& $C$, $v$, $\mu$
\end{tabular}
\end{ruledtabular}
\end{table*}

Herein, we present the Lieb-Robinson bound that characterizes the quasi-locality via time evolution~\cite{ref:LR-bound72,ref:Hastings2006-ExpDec,Nachtergaele2006,ref:Nachtergaele2006-LR}. 
The Lieb-Robinson bound is central to most of the derived results in this study,
and it is formulated as follows:
\begin{lemma}[Lieb-Robinson bound~\cite{nachtergaele2010lieb}] \label{Lieb-Robinson_lemma_corol}
For arbitrary operators $O_X$ and $O_Y$ with unit norm and $\dist_{X,Y}=R$, the norm of the commutator $[O_X(t), O_Y]$ satisfies the following inequality:
\begin{align}
\label{Lieb-Robinson_main_short}
\norm{ [O_X(t), O_Y]} \le C \min(|\partial X|, |\partial Y|) \br{e^{v|t|}-1} e^{-\mu R },
\end{align}
where $C,v,\mu$ are constants of $\orderof{1}$, which depend on the system parameters, that is, $k$, $g$, $g_0$, $\mu_0$, $D$, and $\gamma$.   
\end{lemma}

Using the Lieb-Robinson bound~\eqref{Lieb-Robinson_main_short}, the approximation of $O_X(t)$ 
onto a local region $Y \supset X$ can be obtained. 
We define $O_X(t,Y)$ as
\begin{align}
O_X(t,Y) := \frac{1}{\tr_{Y^\co}(\hat{1})} \tr_{Y^\co} \left[O_X(t)\right] \otimes \hat{1}_{Y^\co},
\label{supp_def:W_X_local_approx}
\end{align}
where $\tr_{Y^\co}(\cdots)$ is the partial trace for subset $Y^\co$; hence,
the operator $O_X(t,Y)$ is supported on the subset $Y \subseteq \Lambda$.
Note that $O_X(t,\Lambda)=O(t)$. 
As shown in Ref.~\cite{PhysRevLett.97.050401}, for arbitrary subsets $Y \supseteq X$, the following can be derived
\begin{align}
\norm{ O_X(t)- O_X(t,Y)} \le \inf_{U_{Y^\co}} \norm{ [O_X(t), U_{Y^\co} ] },
\end{align}
where $\inf_{U_{Y^\co}}$ accepts all unitary operators $U_{Y^\co}$ that are supported on $Y^\co$. 
On selecting $Y=X[R]$ with $R\in \mathbb{N}$, the following inequality can be obtained using the Lieb-Robinson bound~\eqref{Lieb-Robinson_main_short}: 
\begin{align}
\label{Lieb-Robinson_main_short_region}
\norm{ O_X(t)- O_X(t,X[R])} \le C |\partial X| \br{e^{v|t|}-1} e^{-\mu R },
\end{align}
where the inequality~\eqref{Lieb-Robinson_main_short} is applied to $[O_X(t),U_{X[R]^\co}]$ with $U_{X[R]^\co}$, an arbitrary unitary operator.  
Based on the above inequality, it can be ensured that $O_X(t) \approx O_X(t,X[R])$ for $R \gtrsim (v/\mu) t$. 
Often, $(v/\mu)$ is referred to as the ``Lieb-Robinson velocity:'' $v_{\rm LR} = v/\mu$.
In Table~\ref{tab:fund_para}, the fundamental parameters used are summarized.  

Provided the Lieb-Robinson bound holds, the primary results of this study can be extended to more general quantum systems such as long-range interacting systems with power-law decaying interactions (see also Appendix~\ref{sec:Beyond short/range interacting spin systems}).

\subsection{Quantum entanglement} \label{sec:Quantum entanglement}

Here, the basic definition of quantum entanglement~\cite{RevModPhys.81.865,Plenio2014} is presented. 
First, ${\rm SEP}(A:B)$ is defined as a set of separable quantum states on the subset $AB$. 
For an arbitrary quantum state $\rho$, 
the reduced density matrix $\rho_{AB}$ satisfies $\rho_{AB} \in {\rm SEP}(A:B)$ if and only if the following decomposition exists: 
 \begin{align}
\label{Sep_def_ent}
\rho_{AB} = \sum_s p_s \rho_{s,A} \otimes \rho_{s,B}.
\end{align}
When $\rho_{AB}$ is a pure state, $\rho_{AB} \in {\rm SEP}(A:B)$ implies that $\rho_{AB}$ is given by the product state. 
Further, a quantum state $\rho_{AB}$ is defined to be entangled if and only if $\rho_{AB} \notin {\rm SEP}(A:B)$. 

In quantifying the entanglement, the relative entanglement~\cite{PhysRevLett.78.2275,PhysRevA.57.1619,RevModPhys.74.197} can be adopted as follows: 
\begin{align}
\label{def:relative_entanglement_ent_general}
E_R^{\mathcal{X}}(\rho_{AB}) := \inf_{\sigma_{AB} \in \mathcal{X}} S(\rho_{AB} || \sigma_{AB}),
\end{align}
where $\mathcal{X}$ is the arbitrary class of quantum states (focus of this study) and 
 $S(\rho_{AB} || \sigma_{AB})$ is the relative entropy: 
\begin{align}
\label{def:relative_ent}
&S(\rho_{AB} || \sigma_{AB}) \notag \\
& := \tr\brr{\rho_{AB} \log (\rho_{AB})} - \tr\brr{\rho_{AB} \log (\sigma_{AB})}. 
\end{align}
In particular, on choosing $\mathcal{X}= {\rm SEP}(A:B)$, the following is denoted  
\begin{align}
\label{def:relative_entanglement_ent}
E_R(\rho_{AB}) := \inf_{\sigma_{AB} \in {\rm SEP}(A:B)} S(\rho_{AB} || \sigma_{AB})
\end{align}
for simplicity. 
 
The relative entanglement $E_R(\rho_{AB})$ is also related to the closeness of the target state to the zero-entangled state.
Pinsker's inequality entails
\begin{align}
\label{def:Pinsker's inequality}
& \|\rho_{AB} - \sigma_{AB} \|_1 \le \sqrt{2 S(\rho_{AB}||\sigma_{AB})} 
\end{align}
for an arbitrary $\sigma_{AB}$. Hence, definition~\eqref{def:relative_entanglement_ent} immediately yields
\begin{align}
\label{sep_closeness_ineq}
\delta_{\rho_{AB}} &:= \inf_{\sigma_{AB} \in {\rm SEP}(A:B) } \norm{ \rho_{AB} - \sigma_{AB}}_1 \notag \\
&\le \sqrt{2E_R(\rho_{AB})}. 
\end{align}
The quantity $\delta_{\rho_{AB}}$ yields meaningful upper bounds for various entanglement measures. 
Using the continuity of the information measures~\cite{Alicki_2004,doi:10.1063/1.1498001}, 
most of the entanglement measures are upper-bounded by $\orderof{\delta_{\rho_{AB}}} \times \log(\mathcal{D}_{AB})$, such as the entanglement of formation~\cite{PhysRevA.61.064301}, 
the entanglement of purification~\cite{doi:10.1063/1.1498001}, the relative entanglement~\cite{DONALD1999257}, and 
the squashed entanglement~\cite{Alicki_2004,doi:10.1063/1.1643788}.

\subsection{Clustering theorem at high-temperatures: Known results} \label{Sec:Clustering theorem at high-temperatures}

This section reviews an established clustering theorem that holds above a threshold temperature, which is usually determined by the convergence of the cluster expansion.
In high-temperature regimes, clustering of the entanglement can be immediately derived by combining Pinsker's inequality and the exponential decay of the mutual information (Corollary~\ref{corol_no_entanglement_high_T} below). 

For an arbitrary quantum state $\rho$, the standard correlation function $\C_{\rho}(O_A,O_B)$ between observables $O_A$ and $O_B$ can be defined as 
\begin{align}
\label{def:Correlation_classical}
\C_{\rho}(O_A,O_B) := \tr (\rho O_A O_B) - \tr (\rho O_A) \cdot \tr (\rho O_B). 
\end{align}
As a stronger concept of the bi-partite correlation, the mutual information~$\mi_{\rho} (A:B)$ between two subsystems $A$ and $B$ can be defined as follows:
\begin{align}
\label{def_mutual_information_AB}
\mathcal{I}_{\rho}(A:B) := S_{\rho} (A) + S_{\rho} (B) - S_{\rho} (AB),
\end{align}
where $S_{\rho} (A) $ is the von Neumann entropy for the reduced density matrix on subset $A$, that is, 
$S_{\rho} (A) := \tr\brr{-\rho_A \log(\rho_A)}$, with $\rho_A$ being the reduced density matrix on $A$ [see Eq.~\eqref{reduced_rho_def}] 

Previous studies~\cite{Araki1969,bluhm2021exponential} have provided the following clustering theorem, which holds at arbitrary temperatures as $\beta \lesssim \log(n)$ (see also~\footnote{In general, the results in Refs.~\cite{Araki1969,bluhm2021exponential} are restricted to Hamiltonians with finite-range translation-invariant interactions and $n=\infty$ (i.e., the thermodynamic limit). 
Therefore, Lemma~\ref{clustering_1D_previous} cannot be applied to our setup in a strict sense, although the above restrictions are inessential and expected to be removed.
Moreover, the correlation length should be at least as large as $e^{c\beta}$ ($c>0$), that is, $e^{\Omega(\beta)}$.
This estimation results from the 1D classical Ising chain, where the correlation length is known to increase exponentially with $\beta$ as $e^{c\beta}$ ($c>0$).}): 
\begin{lemma}[1D clustering theorem] \label{clustering_1D_previous}
Let $O_A$ and $O_B$ be arbitrary operators supported on subsets $A$ and $B$, respectively.
When a quantum Gibbs state~$\rho_\beta$ as in Eq.~\eqref{def:Ham_Gibbs_2} with $D=1$ is considered, 
the following inequality holds at arbitrary temperatures $\beta \lesssim \log(n)$ ($n$: system size)~\cite{Araki1969}: 
\begin{align}
\C_{\rho_\beta}(O_A,O_B) \le \poly(|A|,|B|) \exp\br{ -\frac{R}{e^{\Omega(\beta)}} },
\end{align}
where $\dist_{A,B}=R$, and the notation $\Omega(\beta)$ denotes $\Omega(\beta) \propto \beta^{1+z}$ ($z\ge0$). 
In addition, the mutual information $\mathcal{I}_{\rho}(A:B)$ decays exponentially with distance~\cite{bluhm2021exponential}:
\begin{align}
\label{bound_mutual_information_1D}
\mathcal{I}_{\rho}(A:B) \le \poly(|A|,|B|) \exp\br{ -\frac{R}{e^{\Omega(\beta)}} }.
\end{align}
\end{lemma}
A similar result holds in arbitrary dimensional systems:
\begin{lemma}[2D-- clustering theorem]\label{clustering_2D_previous}
Under the same setup as in statement~\ref{clustering_1D_previous}, 
the following inequality holds at arbitrary temperatures, such that $\beta < \beta_c$ in arbitrary dimensional systems~\cite{Gross1979,Park1995,ueltschi2004cluster,PhysRevX.4.031019,frohlich2015some}: 
\begin{align}
\C_{\rho_\beta}(O_A,O_B) \le \poly(|A|,|B|) \exp\br{ -\frac{R}{\orderof{1}} },
\end{align}
where $\beta_c$ is a constant that does not depend on the system size. 
Furthermore, the mutual information $\mathcal{I}_{\rho}(A:B)$ decays exponentially with distance~\cite{CMI_clustering}:
\begin{align}
\label{bound_mutual_information_high_T}
\mathcal{I}_{\rho}(A:B) \le \poly(|A|,|B|) \exp\br{ -\frac{R}{\orderof{1}} }.
\end{align}
\end{lemma}

Lemmas~\ref{clustering_1D_previous} and \ref{clustering_2D_previous} immediately imply the exponential decay of the bi-partite quantum entanglement. Consequently, using Pinsker's inequality~\eqref{def:Pinsker's inequality} 
and the equation 
\begin{align}
\mathcal{I}_{\rho}(A:B) 
= S(\rho_{AB}||\rho_{A} \otimes \rho_{B}), 
\end{align}
the following corollary is obtained:
\begin{corol} \label{corol_no_entanglement_high_T}
In the temperature regimes $\beta \lesssim \log(n)$ (1D) and $\beta<\beta_c$ (2D--), the trace distance of 
$
\norm{\rho_{AB}- \rho_{A} \otimes \rho_{B}}_1 
$
exponentially decays with the distance between regions $A$ and $B$:
 \begin{align}
 \label{Product_high_temp}
\norm{\rho_{AB}- \rho_{A} \otimes \rho_{B}}_1 \le \poly(|A|,|B|) e^{-\orderof{R}}.
\end{align}
\end{corol}

\noindent 

Owing to $\rho_{A} \otimes \rho_{B} \in \SEP(A:B)$, the above corollary implies $\delta_{\rho_{AB}}\lesssim e^{-\orderof{R}}$. 
For the relative entanglement~\eqref{def:relative_entanglement_ent}, $E_R(\rho_{AB}) \le \poly(|A|,|B|)e^{-\orderof{R}}$ is obtained from the continuity bound~\cite{DONALD1999257}. 
Therefore, in high-temperature regimes, 
the problem of bipartite entanglement clustering can be easily proved using the established results~\footnote{
Choosing $R\gtrsim \log(n)$ in Ineq.~\eqref{Product_high_temp} yields $\delta_{\rho_{AB}} =1/{\rm poly}(n)$ with $\delta_{\rho_{AB}} $ defined in \eqref{sep_closeness_ineq}; hence, the continuity bound yields $E(\rho_{AB})\le 1/{\rm poly}(n)$ as well. However, for specific entanglement measures, this discussion cannot be applied. 
For example, regarding the entanglement negativity, the distance $\delta_{\rho_{AB}}$ should be as small as $1/\sqrt{\mathcal{D}_A}$ to obtain a meaningful upper bound~\cite[Inequality~(16)]{PhysRevB.102.235110}), where $\mathcal{D}_A$ may be as large as $e^{\orderof{n}}$ if $|A|=\orderof{n}$. Therefore, even at high temperatures, the possibility of certain entanglement measures satisfying the clustering property remains unclear.}.
Consequently, this study focused on the low-temperature regimes, where thermal phase transitions can occur and the clustering of bi-partite correlations may no longer be satisfied.

\section{Quantum correlation} \label{sec:Quantum correlation}

Before discussing the entanglement clustering theorem, the quantum correlation function, defined as a convex roof of the standard correlation function $\C_{\rho}(O_A,O_B)$ in Eq.~\eqref{def:Correlation_classical}, must be considered. 
Quantum correlation is a natural quantum analog of the standard correlation function and has a significant relationship with quantum entanglement (Sec.~\ref{sec:Relation between the quantum correlation and the quantum entanglement}).
Quantum correlation is introduced for two primary reasons:
\begin{enumerate}
\item{} The clustering theorem for quantum correlation can be proved in a completely general manner (Theorem~\ref{thm:quantum_correlation}). 
\item{} The clustering of quantum correlation is also utilized to prove the entanglement clustering theorems 
(Corollary~\ref{corol:quantum_correlation_negativity} and Theorem~\ref{thm:quantum_correlation_negativity_1D})
\end{enumerate}

\subsection{Definition}

For an arbitrary many-body quantum state $\rho$, 
the quantum correlation for observables $O_A$ and $O_B$ can be defined by the convex roof of the standard correlation function~\eqref{def:Correlation_classical}, that is, $\C_{\rho}(O_A,O_B)$ [\ $= \tr (\rho O_A O_B) - \tr (\rho O_A) \cdot \tr (\rho O_B)$]: 
\begin{align}
\label{def_Quantum_corr}
\QC_\rho (O_A,O_B):= \inf_{\{p_s,\rho_s\}} 
\sum_{s}p_s |\C_{\rho_s}(O_A,O_B)|, 
\end{align}
where minimization is performed for all possible decompositions of $\rho$ such that $\rho= \sum_{s}p_s \rho_s$ with $p_s>0$, and $\rho_s$ is a quantum state.
Herein, the mixed convex roof was adopted instead of the pure convex roof, for which decomposed states $\{\rho_s\}$ are restricted to the pure state; in other words, $\rho_s=\ket{\phi_s}\bra{\phi_s}$ for $\forall s$.
This is because using it ensures inequality~\eqref{QC_prop_2_ineq} in Lemma~\ref{lema:basic_property_QC}.
For example, the mixed convex roof has been considered in Refs.~\cite{Synak_Radtke_2006,PhysRevLett.95.190501,Paz_Silva_2009,5075874}.

Subsequently, the definition immediately implies 
\begin{align}
\label{def_Quantum_corr_pure}
\QC_\rho (O_A,O_B)=\bigl |\C_{\rho}(O_A,O_B) \bigr | 
\end{align}
when $\rho$ is given by the pure state. 

The quantum correlations for a density matrix $\rho$ may be different from those for a reduced density matrix $\rho_L$ ($L\subset \Lambda$), that is, $\QC_{\rho_L} (O_A,O_B) \neq \QC_{\rho} (O_A,O_B)$ [Ineq.~\eqref{QC_prop_2_ineq}].
For example, consider the case wherein $\rho$ is given by the Greenberger-Horne-Zeilinger (GHZ) state, as follows:
\begin{align}
\label{GHZ_example_QC}
\frac{1}{2} ( \ket{0_\Lambda} + \ket{1_\Lambda}) ( \bra{0_\Lambda} + \bra{1_\Lambda}),
\end{align}
where $\ket{0_\Lambda}$ ($\ket{1_\Lambda}$) is the product state of $\ket{0_i}$ ($\ket{1_i}$) states ($i\in \Lambda$). 
Then, the quantum state $\rho$ has a non-zero quantum correlation, based on Eq.~\eqref{def_Quantum_corr_pure}, 
while the reduced density matrices in arbitrary subsystems $L\subset \Lambda$ are given by a mixed state of $\ket{0_L}$ and $\ket{1_L}$, each of which exhibits no correlations. 
Hence, no quantum correlations exist in the reduced density matrix of the GHZ state.

As the basic properties of $\QC_\rho (O_A,O_B)$, the following lemma is proven:
\begin{lemma} \label{lema:basic_property_QC}
Let $O_A$ and $O_B$ be arbitrary operators supported on $A$ and $B$, respectively.
Subsequently, the following inequalities are obtained:
\begin{align}
\label{QC_prop_1_ineq}
\QC_\rho (O_A,O_B) \le |\C_\rho (O_A,O_B)|,
\end{align}
and 
\begin{align}
\label{QC_prop_2_ineq}
\QC_{\rho_L} (O_A,O_B) \le \QC_{\rho} (O_A,O_B),
\end{align}
where $A \subseteq L$ and $B\subseteq L$. 
The second inequality is consistent with the example of the GHZ state~\eqref{GHZ_example_QC}.

In addition, the quantum correlation satisfies the following continuity bound.
For arbitrary two quantum states $\rho$ and $\sigma$, the difference between $\QC_{\rho} (O_A,O_B)$ and $\QC_{\sigma} (O_A,O_B)$ is upper-bounded as 
\begin{align}
\label{continuity_main_ineq}
&|\QC_\sigma(O_A,O_B) - \QC_\rho(O_A,O_B) | \le 7\sqrt{2} \epsilon^{1/2},
\end{align}
where $\|O_A\|=\|O_B\|=1$ and $\epsilon=\|\sigma- \rho\|_1$ are set.
\end{lemma}

{~}\\

\noindent
\textit{Proof.} The proof of inequality~\eqref{QC_prop_1_ineq} is obtained by choosing the decomposition as $\rho=p_1 \rho_1$ with $p_1=1$ and $\rho_1=\rho$ in definition~\eqref{def_Quantum_corr}.
Regarding the second inequality, the decomposition $\{p_s,\rho_s\}$ is considered such that 
\begin{align}
\sum_{s}p_s |\C_{\rho_s}(O_A,O_B)| = \QC_\rho(O_A,O_B). 
\end{align}
For the reduced density matrix $\rho_L$, the decomposition using $\{p_s,\rho_s\}$ is chosen as
\begin{align}
\rho_L = \sum_s p_s \rho_{s,L}, \quad \rho_{s,L} = \tr_{L^\co} (\rho_s).
\end{align}
Subsequently, $|\C_{\rho_s} (O_A,O_B)|= |\C_{\rho_{s,L}} (O_A,O_B)|$ is obtained, and hence, inequality~\eqref{QC_prop_2_ineq} is derived as 
\begin{align}
\QC_{\rho_L} (O_A,O_B) &\le \sum_s p_s |\C_{\rho_{s,L}} (O_A,O_B)| \notag \\
&= \QC_{\rho} (O_A,O_B).
\end{align}

Finally, the inequality~\eqref{continuity_main_ineq} is proven via the application of the method in Ref.~\cite[Proposition~5]{Synak_Radtke_2006}.
For the standard correlation $\C_\rho(O_A,O_B)$, straightforward calculations yield
\begin{align}
|\C_\rho(O_A,O_B)| \le 1,
\end{align}
and 
\begin{align}
\abs{ \C_\rho(O_A,O_B) - \C_\sigma(O_A,O_B)} 
\le 3 \| \rho - \sigma\|_1,
\end{align}
where $\|O_A\|=\|O_B\|=1$.
Hence, we can choose parameters $K$ and $M$ in Ref.~\cite[Ineqs. (29) and (30)]{Synak_Radtke_2006} as $K=3/\log(d_X)$ and $M=1/\log(d_X)$, where $d_X$ is the total Hilbert space dimension for $\rho$, that is, $d_X=\mathcal{D}_\Lambda$ according to this study’s notations. 
Thus, inequality~\eqref{continuity_main_ineq} can be obtained from Ref.~\cite[Ineqs.~(31), (51) and Proposition~5]{Synak_Radtke_2006}.
This completes the proof.
$\square$

\subsection{Condition for zero quantum correlation} \label{sec:Relation between the quantum correlation and the quantum entanglement}

As a trivial statement, we first prove the following lemma:
\begin{lemma} \label{lem:QC_Entanglement_condition}
For a quantum state $\rho_{AB}$ supported on $A\cup B$, the quantum correlation $\QC_{\rho_{AB}}(O_A,O_B)$ is equal to zero for arbitrary operators $O_A$ and $O_B$ if $\rho_{AB}$ is not entangled between the subsystems $A$ and $B$ (i.e., $\rho_{AB}\in \SEP(A:B)$):
\begin{align}
\label{QC_Entanglement_condition_statement}
&\textrm{$\rho_{AB} \in \SEP(A:B)$} \longrightarrow  \textrm{$\QC_{\rho_{AB}}(O_A,O_B) = 0$ }
\end{align}
for arbitrary pairs of $O_A,O_B$.
Considering the contraposition of statement~\eqref{QC_Entanglement_condition_statement}, it can be concluded that 
\begin{align}
&\textrm{$\QC_{\rho_{AB}}(O_A,O_B) \neq 0$ for a pair of $O_A,O_B$} \notag \\
&\longrightarrow \rho_{AB} \notin \SEP(A:B).
\end{align}
\end{lemma}

\noindent
\textit{Proof.}
Considering definition~\eqref{Sep_def_ent} for $\SEP(A:B)$, there exists a decomposition of
\begin{align}
\rho_{AB} = \sum_s p_s \rho_{s,A} \otimes \rho_{s,B}  
\end{align}
when the quantum state $\rho_{AB}$ is not entangled. 
For such a decomposition, the state $\rho_{AB}$ exhibits no quantum correlations for operators $O_A$ or $O_B$:
\begin{align}
\QC_{\rho_{AB}}(O_A,O_B)& \le\sum_s p_s |\C_{\rho_{s,A} \otimes \rho_{s,B}}(O_A,O_B) |\notag \\
&=0.
\end{align}
This completes the proof. $\square$

{~}\\

Thus, zero entanglement has been proven to be a sufficient condition for the zero-quantum correlation, as in Eq.~\eqref{QC_Entanglement_condition_statement}. 
However, this leads to the immediate question of whether the converse is also true, that is,
\begin{align}
\label{QC_Entanglement_condition_statement_converse}
&\textrm{$\QC_{\rho_{AB}}(O_A,O_B) = 0$ for arbitrary pairs of $O_A,O_B$} \notag \\
&\xrightarrow{\ ?\ } 
\textrm{$\rho_{AB}\in \SEP(A:B)$ }.
\end{align}
To address this question, the following conjecture is proposed:
\begin{conj}\label{conj_entanglement_exist}
Statement \eqref{QC_Entanglement_condition_statement_converse} is true. 
In other words, the zero-quantum correlation for arbitrary pairs of $O_A,O_B$ is necessary and sufficient for zero entanglement. 
\end{conj}

\noindent
The reason for considering the conjecture to be true is that the following relationship exists for the standard correlation function:
\begin{align}
&\textrm{$\C_{\rho_{AB}}(O_A,O_B) = 0$ for arbitrary pairs of $O_A,O_B$} \notag \\
&\longleftrightarrow \textrm{$\rho_{AB}$ is a product state}.
\end{align}
Hence, it is natural to expect that the quantum version of the above relationship is true as well. 
Regarding the above conjecture, at the very least, the following statement can be proven: 
\begin{lemma} \label{lem:QC_Entanglement_condition_2}
If $\QC_{\rho_{AB}}(O_A,O_B) = 0$ for arbitrary pairs of $O_A,O_B$, the Peres-Horodecki separability criterion~\cite{PhysRevLett.77.1413,HORODECKI19961}, i.e., the positive partial transpose (PPT) condition, is satisfied. 
Thus, the operator $\rho_{AB}^{T_A}$ has no negative eigenvalues, where $T_A$ is the partial transpose with respect to the Hilbert space on the subset $A$. 
\end{lemma}

\noindent
\textit{Proof.} 
The statement is immediately followed by Proposition~\ref{prop:quantum_correlation_negativity} below. 
The condition that $\QC_{\rho_{AB}}(O_A,O_B) = 0$ for arbitrary pairs of $O_A,O_B$ implies $\epsilon=0$ in~\eqref{cond_M_Quantum_corr}. 
Hence, by applying $\epsilon=0$ to inequality~\eqref{ineq:prop:quantum_correlation_negativity}, $\rho_{AB} \in \PPT$ is obtained, where $\PPT$ is a set of states such that the PPT condition is satisfied [Eq.~\eqref{def_Entanglement_PPT_relative} below]. 
This completes the proof. $\square$\\

\noindent
The above lemma shows that conjecture~\ref{conj_entanglement_exist} rigorously holds for a certain class of quantum systems, such as $2\times 2$, $2\times 3$ quantum systems~\cite{HORODECKI19961,HORODECKI2003589}.
Thus, any attempt to prove/disprove the conjecture in general cases must consider the existence of the bound entanglement~\cite{PhysRevLett.80.5239,PhysRevLett.82.1056}. 
A possible route to proving conjecture~\ref{conj_entanglement_exist} relies on the entanglement witness~\cite{PhysRevLett.80.2261,PhysRevA.63.044304,PhysRevA.69.022308,PhysRevA.72.022310}.
However, appropriately reducing the calculations of the witnesses to those of quantum correlations is a challenging task.
As shown in the proofs of Proposition~\ref{prop:quantum_correlation_negativity} and Lemma~\ref{two_qubits_cases_qc} below, the calculation of the partial transpose can be related to the quantum correlations.

\subsection{Positive partial transpose (PPT) relative entanglement}
Finally, in this section, quantum correlation is related to the PPT relative entanglement. 
As shown in Lemma~\ref{lem:QC_Entanglement_condition_2}, quantum correlation is proven to be strongly related with the PPT condition. 
Consequently, using this property, quantum correlations can be related to the following PPT relative entanglement~\cite{PhysRevLett.87.217902,PhysRevA.66.032310,PhysRevA.78.032310,Girard_2014}: 
\begin{align}
\label{def_Entanglement_PPT_relative}
E_{R}^{\PPT}(\rho_{AB}) :=\inf_{\sigma_{AB} \in \PPT} S(\rho_{AB}||\sigma_{AB}),
\end{align}
where $\mathcal{X}=\PPT$ is used in Eq.~\eqref{def:relative_entanglement_ent_general} with $\PPT$ as a set of the quantum states $\sigma_{AB}$ that satisfy the PPT condition, that is, $\sigma_{AB}^{T_A}\succeq 0$ for $\sigma_{AB}\in \PPT$.
Because the $\PPT$ set includes the separable set $\SEP$ ($\PPT\supseteq \SEP$), 
$E_{R}^{\PPT}(\rho_{AB})$ is smaller than or equal to $E_R(\rho_{AB})$, except for special cases.
As shown in Ref.~\cite{PhysRevA.57.1619}, the PPT relative entanglement satisfies all basic conditions for the entanglement measure (i.e., the four conditions in Ref.~\cite{Plenio2014}). 
In addition, it provides an upper bound for Rains’ bound~\cite{959270,PhysRevA.95.062322}, which is strongly related to the distillable entanglement~\cite{959270,PhysRevLett.87.217902}.

As shown in the following proposition, the quantum correlation~\eqref{def_Quantum_corr} provides an upper bound for the PPT relative entanglement (see Appendix~\ref{App_Proof of Proposition_prop:quantum_correlation_negativity} for the proof): 
\begin{prop} \label{prop:quantum_correlation_negativity}
Let $\rho_{AB}$ be an arbitrary quantum state such that 
\begin{align}
\label{cond_M_Quantum_corr}
\QC_{\rho_{AB}}(O_A,O_B)\le\epsilon\|O_A\|\cdot\|O_B\|
\end{align}
for two arbitrary operators $O_A$ and $O_B$. 
Thus,
\begin{align}
 \label{ineq:prop:quantum_correlation_negativity}
&E_{R}^{\PPT}(\rho_{AB})\le 4\mathcal{D}_{AB} \bar{\delta} \log(1/\bar{\delta}) 
\le 4\mathcal{D}_{AB} \bar{\delta}^{1/2}, \notag \\
&\bar{\delta} := 4\epsilon \min(\mathcal{D}_{A},\mathcal{D}_{B}), 
\end{align}
where the second inequality is trivially derived from $x\log(1/x) \le x^{1-1/e} \le x^{1/2}$ for $0\le x\le 1$. 
Recall that $\mathcal{D}_{AB}$ is the Hilbert space dimension in the region $AB$. 
\end{prop} 

\noindent
Based on the proposition, if there are no quantum correlations, that is, if $\epsilon=0$ in \eqref{cond_M_Quantum_corr}, it can be ensured that $E_{R}^{\PPT}(\rho_{AB})=0$, which also yields Lemma~\ref{lem:QC_Entanglement_condition_2}. 
Consequently, the clustering theorem for the quantum correlation can be associated with that for quantum entanglement.
In the following section, the generic quantum Gibbs states are presented to satisfy the exponential clustering for quantum correlations at arbitrary temperatures, thereby indicating that the entanglement clustering theorem also holds.

\section{Exponential clustering for quantum correlations} \label{sec:Exponential clustering for quantum correlations}

In this section, the main theorems of this study on the exponential clustering of the quantum correlations as well as quantum entanglement are presented.
The theorems capture the universal structures of generic quantum Gibbs states at arbitrary temperatures.

First, consider the following theorem on quantum correlation (Appendix~\ref{app_Proof of Theorem_thm:quantum_correlation} for the proof): 
\begin{theorem} \label{thm:quantum_correlation}
Let $O_A$ and $O_B$ be arbitrary operators with the unit norm that are supported on the subsets $A\subset \Lambda$ and $B \subset \Lambda$, respectively ($\dist_{A,B}=R$).
Then, when a quantum state $\rho$ is given by a quantum Gibbs state with the short-range Hamiltonian~\eqref{int_short_range} ($\rho=\rho_\beta$), 
the quantum correlation $\QC_{\rho_\beta} (O_A,O_B)$ is upper-bounded as follows:
\begin{align}
\label{ineq:thm:quantum_correlation}
&\QC_{\rho_\beta} (O_A,O_B) \notag \\
&\le C_\beta (|\partial A| +|\partial B| )\br { 1 + \log|AB| } e^{-R/\xi_\beta},
\end{align}
where $C_\beta= c_{\beta, 1}+c_{\beta, 2}$, and the parameters $c_{\beta, 1}$, $c_{\beta, 2}$, and $\xi_\beta$ can be defined as follows:
 \begin{align}
 \label{parameters_definitions}
&\xi_\beta:=\frac{4}{\mu} \br{ 1 + \frac{v\beta}{\pi}}, \quad c_{\beta, 1}:= e^{2/\xi_\beta}\br{ \frac{24}{\pi} + \frac{12C}{v\beta}}, \notag\\
&c_{\beta,2}:= e^{2/\xi_\beta} \br{\frac{12+3C}{\pi}
+
\frac{3C}{v\beta}} [ 2+ \log (1+2g\beta)],
 \end{align}
The basic parameters are summarized in Table~\ref{tab:fund_para}. 
\end{theorem} 

{\bf Remark.}
The constant $C_\beta$ depends on the inverse temperature $\beta$; however, it increases, at most, logarithmically with $\beta$, that is, $C_\beta=\orderof{\log(\beta)}$. 
By contrast, in the limit of $\beta\to +0$, the upper bound for $\QC_{\rho_\beta} (O_A,O_B)$ apparently breaks down.
However, the temperatures of $\beta\ll 1$ correspond to the high-temperature regime; hence,
a significantly stronger statement (e.g., exponential decay of the mutual information, see Sec.~\ref{Sec:Clustering theorem at high-temperatures}) can be proven using the cluster expansion technique~\cite{CMI_clustering}. 
Therefore, the important temperature regime is $\beta \gg 1$, which cannot be captured by cluster expansion.
Finally, it must be considered that the inequality~\eqref{ineq:thm:quantum_correlation} yields non-trivial upper bounds even for $\beta = \orderof{n^z}$ ($z>0$).


\subsection{Exponential entanglement clustering}

%
%
%

%


The combination of Proposition~\ref{prop:quantum_correlation_negativity} with Theorem~\ref{thm:quantum_correlation} yields the following corollary:
\begin{corol} \label{corol:quantum_correlation_negativity}
Let $\rho_\beta$ be a quantum state given by a quantum Gibbs state with the short-range Hamiltonian~\eqref{int_short_range}.
Then, for arbitrary subsystems $A$ and $B$ separated by a distance $R$ (i.e., $\dist_{A,B}=R$), the PPT relative entanglement is upper-bounded by 
\begin{align}
\label{ineq_quantum_correlation_negativity}
E_{R}^{\PPT} ( \rho_{\beta,AB})
\le 8 C_\beta^{1/2} e^{-R/(2\xi_\beta) + 3\log(\mathcal{D}_{AB} )}
\end{align}
with $\{C_\beta, \xi_\beta\}$ defined in Eq.~\eqref{parameters_definitions}, 
where we use $|\partial A| +|\partial B| \le\mathcal{D}_{AB}$, $1 + \log |AB| \le \mathcal{D}_{AB}$, and $\min( \mathcal{D}_A,\mathcal{D}_B)\le \mathcal{D}_{AB}$ in applying inequality~\eqref{ineq:thm:quantum_correlation} to \eqref{ineq:prop:quantum_correlation_negativity}.
\end{corol}

In the above upper bound, the bi-partite entanglement decays exponentially beyond a distance $R\gtrsim \orderof{|A|+|B|}$. 
Hence, the inequality is meaningless when $A$ and $B$ depend on the system size (i.e., $\mathcal{D}_{AB}=e^{\orderof{n}}$). 
However, it cannot be improved using the decay of quantum correlations alone. 
To highlight this, consider a random state $\ket{\psi_{\rm rand}}$ that has the same property as the infinite temperature states, provided the local regions are considered.
As shown in Ref.~\cite{brandao2013area,Brandao2015}, the state $\ket{\psi_{\rm rand}}$ satisfies exponential clustering for the standard correlation functions~\eqref{def:Correlation_classical}, which clearly implies the exponential decay of quantum correlations from inequality~\eqref{QC_prop_1_ineq}. 
However, the state $\ket{\psi_{\rm rand}}$ exhibits a large quantum entanglement between $A$ and $B$, implying that the characteristics of the quantum Gibbs state must be exploited. 

Further, using the quantum belief propagation technique~\cite{PhysRevB.76.201102,PhysRevB.86.245116}, 
inequality~\eqref{ineq_quantum_correlation_negativity} can be significantly improved for one-dimensional cases (Appendix~\ref{Theorem_thm_quantum_correlation_PPT_rela_1D} for the proof):
\begin{theorem} \label{thm:quantum_correlation_negativity_1D}
Let $H$ be a 1D quantum Hamiltonian with a finite interaction length of $k$, at most. 
Thus, the PPT relative entanglement is upper-bounded by 
\begin{align}
\label{ineq_quantum_correlation_negativity_1D_main}
E_{R}^{\PPT} ( \rho_{\beta,AB})
&\le \bar{C}_\beta \log(\mathcal{D}_{AB}) e^{-R/ [6 \log(d_0) \xi^2_\beta] + 7gk\beta},
%
\end{align}
where $d_0$ is defined as the one-site Hilbert space dimension and $\bar{C}_\beta:= 24\br{\tilde{C}_\beta + 16d_0^4C_\beta }^{1/2}$, with $C_\beta$ defined in Eq.~\eqref{parameters_definitions} and $\tilde{C}_\beta$ defined in Eq.~\eqref{ineq:lemma_quantum_belief_para_tilde_C} as
\begin{align}
\label{thm_ineq:lemma_quantum_belief_para_tilde_C}
&\tilde{C}_\beta := 1280 \br{\frac{5+2C e^{\mu k}}{\pi^2} + \frac{2C e^{\mu k}}{\pi v\beta}}^2.
\end{align}
\end{theorem} 

{\bf Remark.} The assumption of the finite interaction length in the statement is not essential.
However, without this assumption, inequality~\eqref{Lieb-Robinson_1D_finite} in the proof becomes slightly more complicated. 

Here, the PPT relative entanglement has been considered. 
In addition, the definition of $E_{R}^{\PPT} ( \rho_{\beta,AB})$ is significantly associated with that of entanglement negativity~\cite{PhysRevA.65.032314}, which is another popular entanglement measure, particularly in the context of numerical calculations. Furthermore, a part of the above results pertaining to PPT relative entanglement can be applied to entanglement negativity (~Appendix~\ref{sec:Remark on the entanglement negativity})

\section{Quantum correlations based on the skew information} \label{Quantum correlations based on the skew information}

 Herein, another type of quantum correlation  
 based on the WYD skew information~\cite{Wigner910,LIEB1973267,Hansen2006} is considered: 
\begin{align}
\label{def:quantum_information_skey_type}
&\mathcal{I}_{\rho}^{(\alpha)}(K):= \tr (\rho K^2) - \tr \br{ \rho^{1-\alpha} K \rho^{\alpha} K}  
\end{align} 
for $0 < \alpha < 1$, where $K$ is an arbitrary operator. 
The WYD skew information is considered as a measure of the non-commutability between $\rho$ and $K$. 
However, as a representative application, it is utilized in formulating the Heisenberg uncertainty relation for mixed states~\cite{PhysRevLett.91.180403,PhysRevA.72.042110,YANAGI201012,e20020132}.
More recently, the WYD skew information has garnered attention in the context of the quantum coherence theory~\cite{Adesso_2016,RevModPhys.89.041003,PhysRevA.95.042337,Takagi2019,PhysRevA.100.032116}. 

In Refs.~\cite{PhysRevLett.117.130401, De_Chiara_2018,frerot2017quantum,Frerot2019}, the following quantity has been defined to characterize quantum correlations:
\begin{align}
\label{QC_Wigner_Yanase_alpha_average}
&\bar{Q}_{\rho}(O_A,O_B) 
:=\int_0^1Q_{\rho}^{(\alpha)}(O_A,O_B) d\alpha \notag\\
&= \tr (\rho O_AO_B) -\int_0^1 \tr \br{ \rho^{1-\alpha} O_A \rho^{\alpha} O_B } d\alpha 
\end{align}
with
\begin{align}
\label{QC_Wigner_Yanase_alpha}
Q_{\rho}^{(\alpha)}(O_A,O_B) := \tr (\rho O_AO_B) - \tr \br{ \rho^{1-\alpha} O_A \rho^{\alpha} O_B }. 
\end{align}
The quantity $Q_{\rho}^{(\alpha)}(O_A,O_B)$ is reduced to the standard correlation function $\C_\rho(O_A,O_B)$ when $\rho$ is a pure state.

The authors in Refs.~\cite{PhysRevLett.117.130401,Frerot2019} numerically verified that 
the quantum correlation defined by $\bar{Q}_{\rho}(O_A,O_B)$ decays exponentially with a finite correlation length, even at critical points, in hard-core bosons and quantum rotors on a 2D square lattice.
However, whether these observations hold universally at arbitrary temperatures remains unclear. 
This problem can be resolved through the following theorem (see Appendix~\ref{proof_WYD_app} for the proof): 

\begin{theorem} \label{quantum_correlation_skey_type}
The quantum correlation~\eqref{QC_Wigner_Yanase_alpha} is upper-bounded for $0\le \alpha\le 1$ as follows:
\begin{align}
Q_{\rho_\beta}^{(\alpha)}(O_A,O_B) \le C'_\beta\min(|\partial A |,|\partial B |) e^{- R/ \xi'_\beta},
\end{align}
where $C'_\beta $ and $\xi'_\beta$ are characterized solely by the parameters in the Lieb-Robinson bound~\eqref{Lieb-Robinson_main_short} as follows:
\begin{align}
\label{def_C'_beta_si_beta}
C'_\beta= \frac{12+2C}{\pi} 
+
\frac{4C}{v\beta},\quad 
\xi_\beta^{'-1}=\frac{\mu}{2+(v\beta)/\pi}.
\end{align}
It is evident that the same upper bound trivially holds for $\bar{Q}_{\rho}(O_A,O_B)$ in Eq.~\eqref{QC_Wigner_Yanase_alpha_average}. 
\end{theorem}
\noindent
As shown in Appendix~\ref{proof_WYD_app}, the proof technique employed here is similar to that in Refs.~\cite{PhysRevLett.93.126402,PhysRevLett.119.110601}, where the clustering theorem for specific operators in fermion systems at arbitrary temperatures has been proven.

Thus, using Theorem~\ref{quantum_correlation_skey_type}, 
a general upper bound for the WYD skew information (see Appendix~\ref{proof_WYD_app_corol} for the proof) can be obtained:
\begin{corol} \label{corol:quantum_information_skey_type}
Let $K$ be an operator expressed as
\begin{align}
\label{op_K_O_i_i}
K= \sum_{i\in \Lambda} O_i \quad (\|O_i\|\le 1).
\end{align}
Then, the WYD skew information $\mathcal{I}_{\rho_\beta}^{(\alpha)}(K)$ ($0\le \alpha\le 1$) is upper-bounded by
\begin{align}
\label{Ineq:corol:quantum_information_skey_type}
\mathcal{I}_{\rho_\beta}^{(\alpha)}(K)
&:= \tr (\rho_\beta K^2) - \tr \br{ \rho_\beta^{1-\alpha} K \rho_\beta^{\alpha} K} \notag \\
&\le \tilde{C}_\beta' \xi_\beta'^{D} n = \orderof{\beta^D} n,
\end{align}
where $\tilde{C}_\beta':= C'_\beta \brr{(\mu/2)^D + \gamma e^{\mu/2} D! }$.
\end{corol}

\subsection{Quantum Fisher information}

As a relevant quantity, the quantum Fisher information  $\mathcal{F}_{\rho}(K)$, which is defined as follows~\cite{frowis2012}, is considered:
\begin{align}
\label{def_quantum_Fisher}
\mathcal{F}_{\rho}(K) = \sum_{s,s'} \frac{2(\lambda_s - \lambda_{s'})^2}{\lambda_s + \lambda_{s'}} |\bra{\lambda_s} K \ket{\lambda_{s'}} |^2,
\end{align}
where $K=\sum_{i\in \Lambda} O_i$ and $\rho=\sum_{s}\lambda_s \ket{\lambda_s} \bra{\lambda_s}$ ($\lambda_s>0$).
Here, $\lambda_s$ and $\ket{\lambda_s}$ are defined by the spectral decomposition $\rho=\sum_s \lambda_s \ket{\lambda_s}\bra{\lambda_s}$. 
When considering the quantum Gibbs states (i.e., $\rho_\beta$), $\lambda_s = e^{-\beta E_s}$ and $\ket{\lambda_s}=\ket{E_s}$ are obtained, 
where $\ket{E_s}$ is the eigenstate of the Hamiltonian with the corresponding eigenenergy $E_s$. Subsequently, the quantum Fisher information is expressed as
\begin{align}
\mathcal{F}_{\rho_\beta}(K) = \sum_{s,s'=1}^{\mathcal{D}_\Lambda} \frac{2(e^{-\beta E_s} - e^{-\beta E_{s'}})^2}{e^{-\beta E_s} + e^{-\beta E_{s'}}} |\bra{E_s} K \ket{E_{s'}} |^2, \notag
\end{align}
where $\mathcal{D}_\Lambda$ is the dimension of the total Hilbert space.


The quantum Fisher information was introduced in the field of quantum metrology~\cite{PhysRevLett.72.3439,BRAUNSTEIN1996135,T_th_2014,Yadin2021}.
As per the definition~\eqref{def_quantum_Fisher}, the quantum Fisher information $\mathcal{F}_{\rho}(K)$ characterizes the sensitivity of the quantum state $\rho$ to the unitary transformation $e^{-iK\theta}$. Specifically, the uncertainty in estimating the parameter $\theta$ is lower-bounded by the quantum Cram\'er–Rao bound~\cite{PhysRevLett.72.3439,BRAUNSTEIN1996135}:
\begin{align}
\Delta \theta \ge \frac{1}{\sqrt{m \mathcal{F}_{\rho}(K)}},
\end{align}
where $m$ is the number of independent measurements on $e^{-iK\theta}\rho e^{iK\theta}$.
Thus, with an increase in the quantum Fisher information, the required number of measurements decreases.  
In the context of the entanglement theory, this is also regarded among the representative measures for macroscopic quantum entanglement~\cite{ref:Shimizu05-macroE, frowis2012,PhysRevA.92.022356,Hauke2016,PhysRevA.85.022321, RevModPhys.90.025004}.
In recent studies, the quantum Fisher information has garnered attention in the development of quantum technologies 
(see Refs.~\cite{RevModPhys.90.025004,Liu_2019,meyer2021fisher} for recent reviews).

The quantum Fisher information is associated with the WYD skew information through the inequality $(\mathcal{F}_{\rho_\beta} (K)/4) \le 2 \mathcal{I}_{\rho_\beta}^{(\alpha=1/2)}(K)$, which was proven in Ref.~\cite[Theorem~2]{10.2307/1194711} (~\footnote{In Ref.~\cite{10.2307/1194711}, the quantum Fisher information is defined in a manner different to that in this study; it is defined as $\mathcal{F}_{\rho_\beta} (K) /4$ according to this study’s notation.}). 
Hence, based on inequality~\eqref{Ineq:corol:quantum_information_skey_type}, the upper bound can be obtained as
\begin{align}
\label{Ineq2:corol:quantum_information_skey_type}
\mathcal{F}_{\rho_\beta} (K) \le 8\tilde{C}_\beta' \xi_\beta'^{D} n,
\end{align} 
where $\tilde{C}_\beta'$ and $\xi_\beta'$ are defined in Corollary~\ref{corol:quantum_information_skey_type}. 
By contrast, a general lower bound for the quantum Fisher information is provided in Ref.~\cite{Gabbrielli2018}. 
Further, in Appendix~\ref{sec:Relation to quantum Fisher information matrix}, several discussions related to the fundamental properties of the quantum Fisher information and quantum Fisher information matrix, which plays an important role in quantum correlation, are presented.

To discuss macroscopic entanglement using the quantum Fisher information, the scaling exponent, 
$\mathcal{F}_{\rho_\beta} (K) \propto n^p$ ($p \le 2$) is considered.
When $p=2$, the state is composed of the superposition of macroscopically different quantum states; for example, the GHZ state has $p=2$~\cite{ref:Shimizu05-macroE,frowis2012}.
By contrast, when $p=1$, scaling is the same as the product states, and macroscopic superposition does not exist. 
Based on inequality~\eqref{Ineq2:corol:quantum_information_skey_type}, the scaling of the Fisher information is always given by $\orderof{n}$ (i.e., $p=1$), provided $\beta=\textrm{poly-log}(n)$.
Thus, the results obtained offer rigorous proof for the absence of macroscopic superposition at finite temperatures.  

At the quantum critical point (i.e., $\beta=\infty$), scaling of the quantum Fisher information typically behaves as $p>1$~\cite[Eq.~(22)]{Gabbrielli2018}; for example, $p=7/4$ for the critical transverse Ising model~\cite{PhysRevB.93.195127,PhysRevA.94.052105}. 
The obtained upper bound~\eqref{Ineq2:corol:quantum_information_skey_type} characterizes the necessary temperature required when applying the many-body macroscopic entanglement to quantum metrology~\cite{PhysRevLett.121.020402,PhysRevLett.124.120504,PhysRevResearch.2.043329,PhysRevX.8.021022,PhysRevLett.126.010502}; this has attracted considerable attention in recent studies.


\section{Further discussion} \label{sec:Further discussions}

\subsection{Macroscopic quantum effect v.s. quantum entanglement} \label{sec:Macroscopic quantum effect v.s. quantum entanglement}
The entanglement properties have been discussed in the finite-temperature Gibbs state. 
This section shows that, in general, the observations on the entanglement properties for the finite-temperature mixed state are considerably different from those for pure states. Nevertheless, the typical unusual wave function at low temperatures in condensed matter physics is worth discussing, such as Bardeen-Cooper-Schrieffer states in a superconductor, which exhibit off-diagonal long-range orders (ODLRO~\cite{RevModPhys.34.694}). In Refs.~\cite{Vedral_2004,Vedral2008}, Vedral discussed $\eta$-pairing states, which are eigenstates in the Hubbard, and similar models to explain high-temperature superconductivity. It was argued that such states have a vanishing entanglement between two sites as the distance diverges, whereas the classical correlations remain finite even in the thermodynamic limit. In addition, maximally mixed states with $\eta$-paring states also exhibit this property. Consequently, this observation suggests that ODLRO is not directly associated with the quantum entanglement discussed in this study. The quantum entanglement properties in the finite-temperature Gibbs state have not been analytically scrutinized under a general framework thus far. However, recent large-scale numerical computations involving two-dimensional transverse field Ising models revealed that entanglement measured via the R\'enyi negativity is short-ranged, even at finite critical temperatures~\cite{PhysRevResearch.2.043345,PhysRevLett.125.140603}.
This observation is consistent with the general statement declared.

\subsection{Relation to the quantum Markov property} \label{sec:Relation to quantum Markov property}

In this subsection, a brief derivation of the relation between the clustering of quantum entanglement and the approximate quantum Markov property is presented. 

For this purpose, the squashed entanglement~\cite{doi:10.1063/1.1643788,Brandao2011,Li2014}, defined using the conditional mutual information $\mi_{\rho_{ABE}}(A:B|E)$ 
for tripartite quantum systems, is considered:  
\begin{align}
\mi_{\rho_{ABE}}(A:B|E) := &S_{\rho_{ABE}} (AE) +S_{\rho_{ABE}}(BE) \notag\\
&-S_{\rho_{ABE}}(ABE) -S_{\rho_{ABE}}(E). \label{cond_info}
\end{align}
Recall that $S_{\rho_{ABE}} (L) $ is the von Neumann entropy for the reduced density matrix on the subset $L\subseteq ABE$.  
Thus, the squashed entanglement is defined as follows:
\begin{align}
\label{Squashed_entanglement_def}
&E_{\rm sq}(\rho_{AB}) \notag \\
& := \inf_E \left\{\frac{1}{2}\mi_{\rho_{ABE}}(A:B|E) \biggl | \tr_E (\rho_{ABE}) = \rho_{AB} \right\},
\end{align}
where $\inf_E$ is considered over all extensions of $\rho_{AB}$, such that $\tr_E (\rho_{ABE}) = \rho_{AB}$. 
In contrast to the PPT relative entanglement~\eqref{def_Entanglement_PPT_relative}, 
squashed entanglement is equal to zero if and only if the quantum state is not entangled~\cite{Brandao2011}. 

In addition, squashed entanglement is strongly related with the quantum Markov property, which implies the following equation for the arbitrary tripartition of total systems ($\Lambda=A\sqcup C \sqcup B$):
 \begin{align}
&\mi_\rho(A:B|C) = 0 \for \dist_{A,B} \ge r_0,
\end{align}
where $r_0$ is a constant of $\orderof{1}$. 
When the Hamiltonian is short-range and commuting, the above Markov property strictly holds for quantum Gibbs states at arbitrary temperatures~\cite{brown2012quantum,Jouneghani2014}. 
Further, the quantum Markov property has a useful operational meaning~\cite{Fawzi2015}, and it is crucial to preparing the quantum Gibbs states on a quantum computer~\cite{Kastoryano2016,Brandao2019,kato2016quantum,CMI_clustering}.
Thus, for non-commuting Hamiltonians with short-range interactions, it is conjectured that, in general, the quantum Markov property holds in an approximate sense:
\begin{conj} [Quantum Markov conjecture] \label{conj:Quantum_Markove}
For arbitrary quantum Gibbs states, the conditional mutual information $\mi_{\rho_\beta}(A:B|E)$ ($\Lambda=A\sqcup E \sqcup B$) exponentially decays with the distance between $A$ and $B$:
 \begin{align}
 \label{quantum_Markov_Conj_ineq}
&\mi_{\rho_\beta}(A:B|E) \le {\rm poly}(|A|,|B|) e^{-\dist_{A,B}/\xi_\beta}  
\end{align}
with $\xi_\beta = {\rm poly}(\beta)$.
\end{conj}
\noindent
If the inequality~\eqref{quantum_Markov_Conj_ineq} holds, the exponential clustering for the squashed entanglement is obtained as 
\begin{align}
\label{Sq_exp_decay_coj}
E_{\rm sq}(\rho_{\beta, AB}) &\le \frac{1}{2} \mi_{\rho_\beta}(A:B|E) \notag \\
&\le {\rm poly}(|A|,|B|) e^{-\dist_{A,B}/\xi_\beta}, 
\end{align}
where $E= \Lambda\setminus (AB)$ and $\rho_{ABE}=\rho_{\beta}$ are considered in Eq.~\eqref{Squashed_entanglement_def}.
Thus far, the above conjecture has been proven only in high-temperature regimes, where thermal phase transition cannot occur, that is, 
$\beta \lesssim \log(n)$ in 1D cases~\cite{kato2016quantum}, and $\beta <\beta_c$ ($\beta_c=\orderof{1}$) in high-dimensional cases~\cite{CMI_clustering}.  
Moreover, in these temperature regimes, regarding entanglement, considerably stronger statements than~\eqref{Sq_exp_decay_coj} (i.e., Corollary~\ref{corol_no_entanglement_high_T}) have already been derived.

Finally, it is shown that inequality~\eqref{Sq_exp_decay_coj} cannot be used to prove the exponential clustering of other quantum entanglement measures [e.g., the relative entanglement~\eqref{def:relative_entanglement_ent} or the entanglement of formation~\eqref{def_Entanglement_form_corr}] in general.

To upper-bound the other entanglement measures, it is necessary to upper-bound the quantity $\delta_{\rho_{AB}}$, which is defined in Eq.~\eqref{sep_closeness_ineq} as $
\delta_{\rho_{AB}} := \inf_{\sigma_{AB} \in {\rm SEP}(A:B) } \norm{ \rho_{AB} - \sigma_{AB}}_1 .
$
This characterizes the distance between the quantum $\rho_{AB} $ and non-entangled states.  
The squashed entanglement yields the following upper bound for $\delta_{\rho_{AB}}$~\cite{Brandao2011,Li2014}: 
\begin{align}
\delta_{\rho_{AB}} \le 42 \sqrt{\mathcal{D}_{AB} E_{\rm sq}(\rho_{AB})},
\end{align}
where $\mathcal{D}_{AB}$ is the dimension of the Hilbert space of $AB$.
If $E_{\rm sq}(\rho_{AB}) \ll 1/\mathcal{D}_{AB}$, it can be ensured that $\delta_{\rho_{AB}}$ is sufficiently small.
However, $\mathcal{D}_{AB}$ is exponentially large, with a size of $|AB|$.  
Hence, regardless of the quantum Markov conjecture~\ref{conj:Quantum_Markove} being proven, 
the distance $\delta_{\rho_{\beta,AB}}$ for the quantum Gibbs state may still be considerably large when subsets $A$ or $B$ are as large as the system size $n$.   
Thus, a problem similar to that in inequality~\eqref{ineq:prop:quantum_correlation_negativity} of Proposition~\ref{prop:quantum_correlation_negativity} is encountered. 
Therefore, the clustering problem of bi-partite entanglement cannot be generalized to other entanglement measures by simply clarifying the quantum Markov property.

\subsection{General upper bound on the quantum correlation} \label{sec:General upper bound on the quantum correlation}

Here, it is shown that the entanglement formation~\cite{PhysRevA.54.3824,PhysRevLett.80.2245} is a simple upper bound for the quantum correlation $\QC_{\rho_{AB}}(O_A,O_B)$. 
The relation between the entanglement of formation and the quantum correlation $\QC_{\rho_{AB}}(O_A,O_B)$ is derived from that between mutual information $\mi_{\rho_{AB}}(A:B)$ and standard correlation function $\C_{\rho_{AB}}(O_A,O_B)$.
The entanglement of formation is defined as follows:
\begin{align}
\label{def_Entanglement_form_corr}
E_F(\rho_{AB}) &:= \inf_{\{p_s, \ket{\psi_{s,AB}}\}} 
\sum_{s}\frac{p_s}{2} \mi_{\ket{\psi_{s,AB}}} (A:B) \notag \\
&=\inf_{\{p_s, \ket{\psi_{s,AB}}\}} 
\sum_{s}p_s S_{\ket{\psi_{s,AB}}} (A),
\end{align}
where $\mi_{\ket{\psi_{s,AB}}} (A:B) $ and $S_{\ket{\psi_{s,AB}}}(A)$ are the mutual information and the von Neumann entropy for the reduced density matrix on the subset $A$, respectively. Furthermore, $\inf_{\{p_s, \ket{\psi_{s,AB}}\}}$ is considered for arbitrary decomposition $\rho=\sum_s p_s \ket{\psi_{s,AB}}\bra{\psi_{s,AB}}$ with $p_s>0$. 
In addition, $\mi_{\rho_s} (A:B)= 2S_{\rho_{s,AB}} (A)$ when $\rho_s$ is a pure state.

The mutual information $\mi_{\ket{\psi_{s,AB}}} (A:B)$ captures the entire correlations between two subsystems~\cite{PhysRevLett.100.070502}. 
Hence, it is quite plausible that the entanglement of formation provides an upper bound for quantum correlations. 
Indeed, the following lemma connects the quantum correlation $\QC_{\rho_{AB}}(O_A,O_B)$ and the entanglement of formation: 
\begin{lemma}
For arbitrary operators $O_A$ and $O_B$, the quantum correlation $\QC_{\rho_{AB}}(O_A,O_B)$ is upper-bounded using the entanglement of formation $E_F(\rho_{AB})$, as follows: 
\begin{align}
\label{def_Entanglement_form_corr_upperbound}
\QC_\rho (O_A,O_B) \le 2\|O_A\| \cdot \|O_B\| \sqrt{E_F(\rho_{AB})}.
\end{align}
\end{lemma}

\noindent
\textit{Proof.} 
First, we note that 
\begin{align}
&\sum_s p_s | \C_{\rho_{s,AB}}(O_A,O_B) |^2 \notag\\
&\ge \br{ \sum_s p_s | \C_{\rho_{s,AB}}(O_A,O_B) | }^2,
\end{align}
which yields 
\begin{align}
\label{ave_square_corre_ineq}
\inf_{\{p_s,\rho_{s,AB}\}} \sum_s p_s | \C_{\rho_{s,AB}}(O_A,O_B) |^2 \ge  [\QC_{\rho_{AB}} (O_A,O_B)]^2.
\end{align}
Hence, the aim is to provide an upper-bound for the LHS in the above inequality. 

Second, the classical squashed (c-squashed) entanglement~\cite{5075874}, which is obtained from the mixed convex roof of mutual information, is considered~\footnote{
The c-squashed entanglement $E_{\rm sq}^c(\rho_{AB})$ is obtained from the infimum of $I_{\rho_{AB}}(A:B|E)$ under the restriction of classical extension.
The classical extension of $\rho_{AB}$ is defined by the form of $\rho_{ABE}=\sum_s p_s \rho_{AB,s} \otimes \ket{\phi_s} \bra{\phi_s}$, where $\rho_{AB}=\sum_s p_s \rho_{AB,s}$ is an arbitrary decomposition of $\rho_{AB}$, and $\{\ket{\phi_s} \}$ is a certain orthonormal basis on $E$. 
For the classical extension, the conditional mutual information~\eqref{cond_info} being equal to the average of the mutual information can be easily verified.
Thus, based on the definition, the c-squashed entanglement $E_{\rm sq}^c(\rho_{AB})$ is larger than or equal to the original squashed entanglement~\eqref{Squashed_entanglement_def}.
Further, as reported in Ref.~\cite[Sec.~6.5.2]{brandao2008entanglement}, in general, $E_{\rm sq}^c(\rho_{AB})\neq E_{\rm sq}(\rho_{AB})$ is obtained.}: 
\begin{align}
\label{def_Entanglement_form_corr_extend}
E_{\rm sq}^c(\rho_{AB}):= \inf_{\{p_s, \rho_{s,AB}\}} 
\sum_{s}\frac{p_s}{2} \mathcal{I}_{\rho_{s,AB}}(A:B), 
\end{align}
where $\inf_{\{p_s, \rho_{s,AB}\}}$ is considered for all possible decompositions of $\rho_{AB}$ such that 
$\rho_{AB}=\sum_s p_s \rho_{s,AB}$.
The difference between $\mathcal{QI}_{\rho_{AB}}(A:B)$ and $E_F(\rho_{AB})$ is 
whether the decomposed quantum states of $\rho$ are restricted to a pure state~\footnote{The difference between $E_{\rm sq}^c(\rho_{AB})$ and $E_F(\rho_{AB})$ can be significantly large. 
For example, the quantity $E_{\rm sq}^c(\rho_{AB})$ is trivially upper-bounded using the mutual information as $\mathcal{I}_{\rho_{AB}}(A:B)/2$. 
By contrast, as indicated in Ref.~\cite{PhysRevB.98.235154}, there exist quantum states~\cite{Hayden2006} such that the entanglement of formation may be considerably larger than the mutual information, that is, $E_F(\rho_{AB}) \gg \mathcal{I}_{\rho_{AB}}(A:B) \ge E_{\rm sq}^c(\rho_{AB})$.  
In Ref.~\cite[Example~9]{doi:10.1063/1.1643788}, an explicit example where $E_F(\rho_{AB}) > \mathcal{I}_{\rho_{AB}}(A:B)/2$ has been provided.}.  
Trivially, the entanglement of formation $E_F(\rho_{AB})$ is lower-bounded as
\begin{align}
\label{bound_EF_classical_squashed}
E_F(\rho_{AB}) 
&\ge E_{\rm sq}^c(\rho_{AB}). 
\end{align}

Finally, $E_{\rm sq}^c(\rho_{AB})$ is compared with the LHS in \eqref{ave_square_corre_ineq}. 
For this purpose, the following inequality reported in Ref.~\cite[Inequality~(5)]{PhysRevLett.100.070502} is utilized:
\begin{align}
\label{correlation_mutual_wolf}
\mathcal{I}_{\rho_{AB}}(A:B) \ge \frac{|C_{\rho_{AB}} (O_A,O_B)|^2}{2 \|O_A\|^2 \cdot \|O_B\|^2}.
\end{align}
The application of the above inequality to definition~\eqref{def_Entanglement_form_corr_extend} yields 
\begin{align}
E_{\rm sq}^c(\rho_{AB})
&\ge \inf_{\{p_s, \rho_{s,AB}\}} 
\sum_{s}\frac{p_s}{2} \cdot \frac{|C_{\rho_{s,AB}} (O_A,O_B)|^2}{2 \|O_A\|^2 \cdot \|O_B\|^2} \notag \\
&\ge \frac{ [\QC_{\rho_{AB}} (O_A,O_B)]^2}{4 \|O_A\|^2 \cdot \|O_B\|^2}, 
\end{align}
where ~\eqref{ave_square_corre_ineq} is used in the last inequality. 
Thus, by combining the above inequality with~\eqref{bound_EF_classical_squashed}, the main inequality~\eqref{def_Entanglement_form_corr_upperbound} is proven.
This completes the proof. $\square$

\subsection{Optimality of the obtained bounds}\label{sec:Optimality of the obtained bounds}

Herein, the optimality of the correlation length $\xi_\beta$ or $\xi'_\beta$ in Theorems~\ref{thm:quantum_correlation} and \ref{quantum_correlation_skey_type} is discussed. 
The $\beta$-dependence of the correlation length $\xi_\beta$ (i.e., $\xi_\beta \propto \beta$) is shown to be qualitatively optimal, which cannot be improved in general. This point is ensured by the correspondence of the inverse temperatures and spectral gap, as follows:
\begin{align}
\label{correspond_gap_beta}
\beta \longleftrightarrow 1/\Delta 
\end{align}
with $\Delta$ being the spectral gap between the ground and first excited states.
Consequently, the correlation length of $\orderof{\Delta^{-1}}$ in the gapped ground states~\cite{PhysRevLett.93.140402,ref:Hastings2006-ExpDec,ref:Nachtergaele2006-LR} implies the correlation length of $\orderof{\beta}$ in the thermal states.

To elaborate, first, the following inequality for the number of energy eigenstates in an arbitrary energy shell $(E-1,E]$~\cite{PhysRevB.76.035114,PhysRevA.80.052104,PhysRevLett.124.200604} is assumed:  
\begin{align}
{\cal N}_{E, 1} & \le n^{c E}, \label{density_of_state_DE}  
\end{align}
where ${\cal N}_{E,1}$ is the number of eigenstates within the energy shell of $(E-1,E]$, and $c$ is a constant of $\orderof{1}$.  
Furthermore, the energy origin is set such that the ground state's energy is equal to zero. 
Here, the above condition is satisfied in various types of quantum many-body systems~\cite{PhysRevB.76.035114}. 

Thus, under condition~\eqref{density_of_state_DE}, the quantum Gibbs states $\rho_{\beta}$ are close to the ground state $\rho_{\infty}$ in the sense that 
\begin{align}
\norm{ \rho_{\beta} - \rho_{\infty}}_1 \le {\rm const.} \times \frac{e^{-(\beta-c \log(n) )\Delta}}{\beta-c\log(n)}. \label{density_of_state_DE2}  
\end{align}
Therefore, the properties of the thermal states and the ground state are approximately the same for $\beta \approx \log(n)/\Delta$, as follows:
\begin{align}
\norm{ \rho_{\beta} - \rho_{\infty}}_1= 1/{\rm poly}(n). 
\end{align}

When the ground state is non-degenerate and gapped, the correlation function $\C_{\rho_{\infty}}(O_A,O_B)$ is expressed as~\cite{ref:Hastings2006-ExpDec,ref:Nachtergaele2006-LR}
\begin{align}
\label{clustering_gapped_gs} 
 \C_{\rho_{\infty}}(O_A,O_B) =  \QC_{\rho_{\infty}}(O_A,O_B)  = {\rm const.} \times e^{-\orderof{\Delta} R},  
\end{align}
where Eq.~\eqref{def_Quantum_corr_pure} is used for the pure state in the first equation. 
Subsequently, using the continuity bound~\eqref{continuity_main_ineq}
\begin{align}
&\QC_{\rho_{\beta}}(O_A,O_B) =  \C_{\rho_{\infty}}(O_A,O_B) - 1/{\rm poly}(n)  \notag \\
&= {\rm const.} \times e^{-\orderof{\Delta} R}  - 1/{\rm poly}(n)  \notag \\
&=  {\rm const.} \times e^{- \orderof{R}/ (\beta / \log(n)) }  - 1/{\rm poly}(n),
\end{align}
where the second equation results from the fact that $\beta \approx \log(n)/\Delta$ implies $\Delta\approx \log(n)/\beta $.
Thus, the quantum correlation starts to decay for $R\gtrsim \beta / \log(n)$; hence, the correlation length is proportional to $\beta$ at sufficiently low temperatures. 

By contrast, for the WYD skew and quantum Fisher information, there is scope for improvement in the present $\beta$-dependences, which have been assigned inequalities~\eqref{Ineq:corol:quantum_information_skey_type} and \eqref{Ineq2:corol:quantum_information_skey_type}, respectively. 
In the ground states, the WYD skew and quantum Fisher information reduce to the variance of the operator. 
For an arbitrary operator $K$ expressed in Eq.~\eqref{op_K_O_i_i}, the variance $(\Delta K)^2 = \tr\br{\rho_\infty K^2}- [\tr\br{\rho_\infty K}]^2$ is upper-bounded by~\cite{Kuwahara_2016_gs,Kuwahara_2017}:
\begin{align}
\mathcal{I}_{\rho_\infty}^{(\alpha)}(K) = (\Delta K)^2  \le {\rm const.} \times \Delta^{-1} n. 
\end{align}
The above inequality holds in infinite dimensional systems and long-range interacting systems; hence, the ($\beta$, $\Delta$) correspondence~\eqref{correspond_gap_beta} indicates an improvement in the current upper bounds as
\begin{align}
\mathcal{I}_{\rho_\beta}^{(\alpha)}(K) \le \orderof{\beta n} \quad {\rm and}  \quad \mathcal{F}_{\rho_\beta} (K) \le \orderof{\beta n},
\end{align}
which affords better bounds in dimensions greater than $1$ ($D\ge 2$).

\subsection{Beyond quantum Gibbs states} \label{sec:Beyond quantum Gibbs states}

Throughout the discussion, the equilibrium situation is considered at a finite temperature. However, when considering a non-equilibrium density matrix, the entanglement properties exhibit different properties in general~\cite{PhysRevLett.105.180501}.
Consequently, a natural question arises as to whether the current results hold for more general quantum states. 
Based on definition~\eqref{def_Quantum_corr} of the quantum correlation, concavity is satisfied, that is, 
 \begin{align}
\QC_{\rho} (O_A,O_B) \le p_1 \QC_{\rho_1} (O_A,O_B)  + p_2 \QC_{\rho_2} (O_A,O_B)  \notag
\end{align}
for an arbitrary decomposition of $\rho=p_1 \rho_1 + p_2 \rho_2$ ($p_1>0$, $p_2>0$). 
Hence, considering a quantum state in the form of 
\begin{align}
\label{def:passive_state_laplace}
\rho = \int_0^\infty a(z) e^{-z H}dz  
\end{align}
with $a(z)$ being a non-negative function, Theorem~\ref{thm:quantum_correlation} can be applied to the state $\rho$. 
Subsequently, the state $\rho$ has a finite quantum correlation length, while the entanglement clustering is also satisfied. 
A similar discussion can be also applied to the WYD skew information $\mathcal{I}_{\rho}^{(\alpha)}(K)$ and the quantum Fisher information $\mathcal{F}_{\rho} (K) $ owing to their concavities~\footnote{The concavity of the WYD skew information has been proven through the celebrated Wigner-Yanase-Dyson-Lieb theorem~\cite{LIEB1973267,Hansen2006}. 
The concavity of the quantum Fisher information can be easily proven from the convex roof expression~\eqref{def_quantum_Fisher_convex_roof}, which has been proven in~\cite{toth2013extremal,yu2013quantum}.}.
Herein, if the state $\rho$ includes extremely low-temperature states, for example, $\int_{\beta_0}^\infty a(z) \tr(e^{-z H})\approx 1$ with $\beta_0\approx \orderof{n}$, 
the state $\rho$ is similar to low-temperature Gibbs states; consequently, the quantum correlation length may become large.

As an important class of quantum states, the following density matrix is considered to be characterized by a monotonically decreasing function $F(x)$:
\begin{align}
\label{def:passive_state}
\rho = \frac{F(H)}{\tr\brr{F(H)}}, 
\end{align}
where $F(x)\ge 0$.
This class of the quantum state is referred to as the passive state~\cite{Pusz1978,Lenard1978} and plays a crucial role in quantum thermodynamics~\cite{PhysRevE.92.042147,PhysRevE.91.052133,Sparaciari2017,PhysRevX.8.021064}.
Moreover, the quantum Gibbs state trivially corresponds to the case $F(x)=e^{-\beta x}$.
Based on the Bernstein-Widder theorem~\cite{Bernstein1929,Widder1931,Qi2019}, the passive state~\eqref{def:passive_state} can be represented in the form of Eq.~\eqref{def:passive_state_laplace} if and only if the function $F(x)$ is completely monotonic, as follows:
\begin{align}
\label{def:complete_monotone}
(-1)^m \frac{d^m}{dx^m}F(x) \ge 0
\end{align}
for arbitrary $m\ge 0$. 
Therefore, for every passive state with condition~\eqref{def:complete_monotone}, structural restrictions similar to that for the quantum Gibbs state must be imposed~\footnote{Passive states usually possess different properties from those in Gibbs states. 
For example, regarding the Markov property in classical systems, only the Gibbs states satisfy the property from the Hammersley-Clifford theorem~\cite{HammersleyClifford1971}, whereas the passive states do not.}.

\section{Summary and Future works}\label{sec:Summary and Future works}

This study primarily addressed the conjecture of the exponential clustering of bipartite entanglement,   
which revealed the fundamental aspect of long-range entanglement.
The entanglement was accessed via the introduction of a novel concept, referred to as the quantum correlation $\QC_\rho(O_A,O_B)$,
which is defined by the convex roof of the standard bipartite correlation function, as in Eq.~\eqref{def_Quantum_corr}.
Consequently, as a fundamental theorem, the exponential clustering of the quantum correlation was derived, which holds at arbitrary temperatures, even at the critical point of thermal phase transition. 
Based on its definition and exploiting the fact that it uses the convex roof, quantum correlation exhibits properties similar to those of entanglement.
Subsequently, several basic statements in Sec.~\ref{sec:Quantum correlation} were derived, including the relationship between the quantum correlation and the PPT relative entanglement (Proposition~\ref{prop:quantum_correlation_negativity}). 
Further, based on the clustering theorem for the quantum correlation, entanglement clustering theorems (Corollary~\ref{corol:quantum_correlation_negativity} and Theorem~\ref{thm:quantum_correlation_negativity_1D}) for PPT relative entanglement~\eqref{Ineq:main_conj_partial_proof} were presented. 
Moreover, using similar analytical techniques, the exponential clustering of another type of quantum correlation based on the WYD skew information (Theorem~\ref{quantum_correlation_skey_type}) was derived, which yielded the fundamental limitations of the WYD skew and quantum Fisher information (Corollary~\ref{corol:quantum_information_skey_type}). Consequently, these serve as representative measures for quantum coherence and macroscopic entanglement.

Furthermore, this study expressed simple and general no-go theorems on the existence of long-range entanglement. 
On the other hand, there is still room for improvement of the present analytical techniques, and hence the obtained results may be further strengthened. 
Based on the results obtained, the strongest form of the bi-partite entanglement clustering may be expressed as follows:
\begin{align}
\label{Ineq:main_conj}
&\textbf{[The strongest conjecture]} \notag \\
&E_R(\rho_{\beta,AB}) \le \poly(|A|,|B|) e^{-R/\tilde{\xi}_\beta}
\end{align}
for an arbitrary choice of $A$ and $B$ such that $\dist_{A,B}=R$, 
where $\tilde{\xi}_\beta=\poly(\beta)$ and $\poly(x)$ denote a finite degree polynomial.  
As shown in Sec.~\ref{sec:Quantum entanglement}, from the continuity bounds, inequality~\eqref{Ineq:main_conj} yields the same upper bound for other entanglement measures.  
However, the main theorems presented in this paper did not arrived at this form of entanglement clustering, and further investigations are required to refine the current results.

In conclusion, this study unveiled a fundamental limit on the characteristic length scale, such that certain types of quantum effects can exist. 
Moreover, the present results do not depend on system details and hold at arbitrary temperatures. 
The understanding of the universal structural constraints in low-temperature physics, which must be satisfied for every quantum many-body system, still remains limited. 
Consequently, identifying these constraints is a critical task for understanding the complicated quantum many-body phases as well as developing efficient algorithms for quantum many-body simulations. 
This study is expected to introduce a novel approach to address this profound problem.

Finally, the following topics are mentioned as specific open questions:



\begin{itemize}
\item{} First, deriving a clustering theorem for the relative entanglement instead of the PPT relative entanglement.
This may be addressed by resolving conjecture~\ref{conj_entanglement_exist}.  
Subsequently, Proposition~\ref{prop:quantum_correlation_negativity} can be improved; in other words, under the condition of (almost) zero quantum correlations [i.e., Ineq.~\eqref{cond_M_Quantum_corr}], a similar inequality to \eqref{ineq:prop:quantum_correlation_negativity} may hold for the relative entanglement $E_{R}(\rho_{AB})$ instead of $E_{R}^{\PPT}(\rho_{AB})$. 
This improvement immediately yields the entanglement clustering for other popular measures, such as the entanglement of formations [see also the discussion after Ineq.~\eqref{sep_closeness_ineq}].

\item{} As a related problem, the $(|A|, |B|)$ dependence in Corollary~\ref{corol:quantum_correlation_negativity} may be improved under dimensions greater than one. 
In the present form, the independence is in the exponential form, and hence, a meaningful bound for the case of $|A|$ and $|B|$ being as large as the system size cannot be obtained. 
To improve this, as has been discussed after Corollary~\ref{corol:quantum_correlation_negativity}, considering the operator correlations $\QC_\rho(A,B)$ alone is not sufficient. 
Instead, the complete information between the two subsystems must be considered. 
However, at the current stage, the problem may be challenging as it should include an analogous difficulty to the data hiding problem in the context of the area law conjecture at zero temperature~\cite{PhysRevB.76.035114,brandao2013area,Brandao2015}. 

\item{} Third, identifying the class of quantum coherence measures~\cite{PhysRevLett.113.140401}, which are always short range at non-zero temperatures, remains an intriguing problem. 
In this study, it was shown that bi-partite entanglement cannot exist at long distances; however, as has been demonstrated in Sec.~\ref{sec:Macroscopic quantum effect v.s. quantum entanglement}, macroscopic quantum effects do not necessarily imply long-distance entanglement. 
For example, quantum discord, a well-known measure for quantum correlation~\cite{PhysRevLett.88.017901,PhysRevA.77.042303}, only decays algebraically at thermal critical points~\cite{PhysRevLett.117.130401}.  
Thus, the current results can still be expanded to include other coherence measures.

\item{} Finally, the question remains as to whether entanglement clustering can be applied to more practical problems such as the efficient simulation of quantum Gibbs states.
The clustering of entanglement imposes a strong constraint on the structure of quantum Gibbs states. Hence, it is likely that the property can be utilized to reduce computational complexity.

 \end{itemize}



\begin{acknowledgments}
T. K. was supported by the RIKEN Center for AIP, JSPS KAKENHI (Grant No. 18K13475), and JST PRESTO (Grant No. JPMJPR2116).
TK gives thanks to God for his wisdom.
K.S. was supported by JSPS Grants-in-Aid for Scientific Research (JP16H02211 and JP19H05603).
\end{acknowledgments}

\appendix

\section{Spectral decomposition of operators}

As a convenient notation, $O_\omega$ is defined for the arbitrary operator $O$ as follows~\cite{PhysRevB.76.201102}: 
\begin{align}
\label{Fourier_op_def}
O_\omega : = \sum_{i,j} \bra{E_i} O \ket{E_j} \delta (E_i-E_j -\omega) \ket{E_i}\bra{E_j},
\end{align}
where $\{\ket{E_i}\}$ and $\{E_i\}$ are the eigenstates and the corresponding eigenvalues of $H$, respectively.
The operator $O_\omega$ yields terms such as $\bra{E+\omega} O \ket{E} \ket{E+\omega}\bra{E}$. 
Based on the above definition, the following can be obtained:
\begin{align}
\label{Fourier_op_def_2}
\int_{-\infty}^\infty O_\omega d\omega  = O,\quad O_\omega = \frac{1}{2\pi} \int_{-\infty}^\infty O(t) e^{-i\omega t} dt, 
\end{align}
and 
\begin{align}
\label{O_omega_equation_simple}
&\ad_H (O_\omega) = \omega O_\omega,\quad [ \ad_H (O_\omega)]^\dagger = \omega O_\omega^\dagger, \notag \\
& [e^{-\beta H}, O_\omega ] = (1-e^{\beta \omega}) e^{-\beta H} O_\omega, 
\end{align}
where $\ad_H(\cdot) := [H, \cdot]$ is defined.

%
%
%
%
%
%
%

\section{Beyond short-range interacting spin systems} \label{sec:Beyond short/range interacting spin systems}

\subsection{Long-range interacting cases}

This section discusses the manner in which the current analyses can be generalized to systems with long-range interactions, 
where the decay of the function $J(x)$ in Eq.~\eqref{supp_def:interaction length} is given in a polynomial form: 
\begin{align}
\label{int_long_range}
J(x) \le \frac{g_0}{(x+1)^\alpha} \quad (\textrm{long-range interaction}). 
\end{align}
When $\alpha >2D$, 
the Lieb-Robinson bound~\eqref{Lieb-Robinson_main_short} can be generalized to long-range interacting systems~\cite{PhysRevLett.114.157201,PhysRevX.10.031010,PhysRevX.10.031009,chen2019finite,PhysRevLett.126.030604,tran2021optimal,tran2021liebrobinson}. 
The Lieb-Robinson bound can be obtained in the following form:
\begin{align}
\label{Lieb-Robinson_main_general}
\norm{ [O_X(t), O_Y]} \le C'  \min(|\partial X|, |\partial Y|) \frac{t (1+t)^{\eta_\alpha}   }{R^{-\tilde{\alpha}}},
\end{align}
where $\eta_\alpha$ and $\tilde{\alpha}$ depend on the spatial dimension $D$ and the decay exponent $\alpha$.
For example, a loose estimation affords $\eta_\alpha=\alpha-D-1$ and $\tilde{\alpha}=\alpha-2D$~\cite{PhysRevLett.126.030604}.
Nevertheless, a quantitatively optimal estimation of the parameters $\eta_\alpha$ and $\tilde{\alpha}$ remains unaddressed. 

Using the Lieb-Robinson bound~\eqref{Lieb-Robinson_main_general}, the main results can be generalized to long-range interacting systems.
In this case, the exponential decay becomes the power-law decay.  
Analyses using the Lieb-Robinson bound can be summarized as follows:
\begin{enumerate}

\item{} For the proof of Theorem~\ref{quantum_correlation_skey_type}, the Lieb-Robinson bound is used in~\eqref{integrate_Lieb-Robinson_QC_WY} or \eqref{integrate_QC_WY_t_le_t_0}.

\item{} For the proof of Theorem~\ref{thm:quantum_correlation}, the Lieb-Robinson bound is used in~\eqref{integrate_QC_L_1_t_le_t_0} and \eqref{Lieb-Robinson_main_short_region_claim_proof_com}.
 
\item{} For the proof of Theorem~\ref{thm:quantum_correlation_negativity_1D}, the Lieb-Robinson bound is used in~\eqref{Lieb-Robinson_1D_finite}. 

\end{enumerate}

\subsection{Disordered systems}

Other interesting systems include the disordered systems where randomness is added to the Hamiltonians. 
In such systems, the Lieb-Robinson bound can be proven to have improved, as follows~\cite{PhysRevLett.99.167201,kim2014local}:
\begin{align}
\label{Lieb-Robinson_main_MBL}
\norm{ [O_X(t), O_Y]} \le  C \min(|\partial X|, |\partial Y|) t^{\eta} e^{-\mu R},
\end{align}
where $C,\eta,\mu$ are constants of $\orderof{1}$, which depend on the system parameters.
In this case, the norm $\norm{ [O_X(t), O_Y]}$ is exponentially small with respect to the distance $R$ up to time $t\sim e^{\orderof{R}}$. 
This leads to the quantum correlation length of $\orderof{{\rm polylog}(\beta)}$ in the main theorem (i.e., Theorem~\ref{thm:quantum_correlation}, Theorem~\ref{quantum_correlation_skey_type}, and Theorem~\ref{thm:quantum_correlation_negativity_1D}).

\subsection{Quantum boson systems}

Finally, in quantum boson systems, the Hamiltonian is locally unbounded (i.e., the parameter $g$ is infinitely large, as shown in Fig.~\ref{tab:fund_para}). 
In such systems, typically, the Lieb-Robinson bound is not obtained with a finite Lieb-Robinson velocity~\cite{PhysRevLett.102.240501}. 
To extend the obtained results, the study may need to be restricted to particular classes of quantum many-body boson systems, 
such as free boson systems~\cite{cramer2008locality,Nachtergaele2009}, spin-boson models~\cite{PhysRevLett.111.230404,PhysRevLett.115.130401}, and 
Bose-Hubbard type Hamiltonians~\cite{PhysRevA.84.032309,PhysRevLett.127.070403,yin2021finite}. The establishment of the Lieb-Robinson bound in boson systems is still an active area of research.



\section{Proofs of Theorem~\ref{quantum_correlation_skey_type} and Corollary~\ref{corol:quantum_information_skey_type}} 
\label{proof_WYD_app}

In this section, Theorem~\ref{quantum_correlation_skey_type} is proven, following by Theorem~\ref{thm:quantum_correlation}.
The proof for Theorem~\ref{quantum_correlation_skey_type} is considerably simpler than that for Theorem~\ref{thm:quantum_correlation}, although the essence for both are similar. 

Theorem~\ref{quantum_correlation_skey_type} and the resulting Corollary~\ref{corol:quantum_information_skey_type} provide the upper bounds for
\begin{align}
\label{app_QC_Wigner_Yanase_alpha_average}
Q_{\rho_\beta}^{(\alpha)}(O_A,O_B) := \tr (\rho_\beta O_AO_B) -  \tr \br{ \rho_\beta^{1-\alpha} O_A \rho_\beta^{\alpha}  O_B } 
\end{align}
and
\begin{align}
\label{app_definition_skew_information}
\mathcal{I}_{\rho_\beta}^{(\alpha)}(K):= \tr (\rho_\beta K^2) -  \tr \br{ \rho_\beta^{1-\alpha} K \rho_\beta^{\alpha} K}
\end{align}
with $K= \sum_{i\in \Lambda} O_i \quad (\|O_i\|\le 1)$, respectively. 

For the convenience of the readers, the rough forms of the statements are provided.
In Theorem~\ref{quantum_correlation_skey_type}, 
\begin{align}
Q_{\rho_\beta}^{(\alpha)}(O_A,O_B)  \le C'_\beta\min(|\partial A |,|\partial B |) e^{-  R/ \xi'_\beta},
\end{align}
where the parameters are $\orderof{1}$ constants that are expressed in Eq.~\eqref{def_C'_beta_si_beta}. 
Furthermore, Corollary~\ref{corol:quantum_information_skey_type} provides the inequalities
\begin{align}
\label{app_Ineq:corol:quantum_information_skey_type}
\mathcal{I}_{\rho_\beta}^{(\alpha)}(K) \le 
\tilde{C}_\beta' \xi_\beta'^{D} n = \orderof{\beta^D} n,
\end{align}
for the WYD skew information.

\subsection{Remark on the parameter regime $\alpha \notin [0,1]$}

As is evident, in general, obtaining the same results for the parameter regime $\alpha \notin [0,1]$ is not possible. 
Mathematically, the proof in Sec.~\ref{Proof of Theorem_refquantum_correlation_skey_type} breaks down for $\alpha \notin [0,1]$, because the function $g_{\alpha,\beta}(t)$ in \eqref{def_g_alpha_fourier} no longer decays exponentially with $t$. 

For example, when $\alpha=-1$, $\mathcal{I}_{\rho}^{(-1)}(K)$ is referred to as the purity of coherence~\cite{Marvian2020}:
\begin{align}
\mathcal{I}_{\rho}^{(-1)}(K) 
&= \tr (\rho K^2) -  \tr \br{ \rho^{2}  K \rho^{-1} K}  \notag \\
&=-\sum_{j,k} \frac{\lambda_k^2-\lambda_j^2}{\lambda_j} |\bra{\lambda_j} K \ket{\lambda_k}|^2,
\end{align}
where $\rho= \sum_j \lambda_j \ket{\lambda_j}\bra{\lambda_j}$ is the spectral decomposition of $\rho$.
In general, 
\begin{align}
\label{expression_wigner_Y_D_skew} 
\mathcal{I}_{\rho}^{(\alpha)}(K) 
&= \tr (\rho K^2) -  \tr \br{ \rho_\beta^{1-\alpha}  K \rho^{\alpha} K} \notag \\
&=- \sum_{j,k} \frac{\lambda_k^{1-\alpha}-\lambda_j^{1-\alpha}}{\lambda_j^{-\alpha}} |\bra{\lambda_j} K \ket{\lambda_k}|^2.
\end{align}
For $\beta = {\rm poly}[\log(n)]$, under the same assumption as for Eq.~\eqref{density_of_state_DE}, 
the quantum Gibbs state $\rho_\beta$ satisfies
\begin{align}
\lambda_0\approx 1, \quad 
\lambda_j= e^{-\beta E_j}.
\end{align}
Hence, the quantum Gibbs state is approximately given by the ground state.
Thus, $\mathcal{I}_{\rho_\beta}^{(\alpha)}(K)$ in Eq.~\eqref{expression_wigner_Y_D_skew} includes the following terms: 
\begin{align}
\label{expression_wigner_Y_D_skew_k=0}
&\sum_{j}\br{ \frac{\lambda_0^{1-\alpha}}{\lambda_j^{-\alpha}}
+ \frac{\lambda_j^{1-\alpha}}{\lambda_0^{-\alpha}}}
 |\bra{\lambda_j} K \ket{\lambda_0}|^2 \notag \\
&\approx \sum_{j} \br{ e^{-\alpha \beta E_j} + e^{(\alpha-1) \beta E_j} } |\bra{\lambda_j} K \ket{\lambda_0}|^2.
\end{align}
For $\alpha \in [0,1]$, both $e^{-\alpha \beta E_j}$ and $e^{(\alpha-1) \beta E_j}$ decay with $E_j$, whereas for $\alpha \notin [0,1]$, either $e^{-\alpha \beta E_j}$ or $e^{(\alpha-1) \beta E_j}$ grows exponentially with $E_j$.  

Typically, only $ |\bra{\lambda_j} K \ket{\lambda_0}|^2 \lesssim e^{-{\rm const.} \times E_j}$ from Ref.~\cite{Arad_2016} can be ensured.
Hence, for $\alpha<0$ ($\alpha>1$), there exists a critical temperature $\beta_c\propto 1/(-\alpha)$ [$\beta_c\propto 1/(\alpha-1)$], such that Eq.~\eqref{expression_wigner_Y_D_skew_k=0} 
exponentially grows with the system size $n$ for $\beta > \beta_c$. 
Therefore, a meaningful upper bound $\mathcal{I}_{\rho_\beta}^{(\alpha)}(K)$ cannot be obtained without additional conditions (such as the high-temperature condition).

\subsection{Proof of Corollary~\ref{corol:quantum_information_skey_type}}
\label{proof_WYD_app_corol}

First, Corollary~\ref{corol:quantum_information_skey_type} based on Theorem~\ref{quantum_correlation_skey_type} is proven, as follows:
\begin{align}
\label{mathcal_I_rho_beta_alpha_K__upp}
\mathcal{I}_{\rho_\beta}^{(\alpha)}(K)
&=\sum_{i,j} \tr (\rho_\beta O_i O_j) -  \tr \br{\rho_\beta^{1-\alpha} O_i \rho_\beta^{\alpha} O_j}  \notag \\
&\le \sum_{i,j} C'_\beta e^{- \dist_{i,j}/ \xi'_\beta} \notag \\
&\le C'_\beta  |\Lambda| \max_{i\in \Lambda} \sum_{j\in \Lambda} e^{- \dist_{i,j}/ \xi'_\beta}  
= C'_\beta \zeta_{0,\xi'_\beta} n 
\end{align}
with
$
\zeta_{s,\xi}:= \max_{i\in \Lambda} \sum_{j\in \Lambda}  \dist_{i,j}^s e^{-\dist_{i,j}/\xi}.
$

The parameter $\zeta_{s,\xi}$ is upper-bounded by
\begin{align}
\label{upper_bound_zeta_s_xi}
\zeta_{s,\xi}  \le 1+  \gamma e^{1/\xi} \xi^{s+D} (s+D)!. 
\end{align}
Using definition~\eqref{def:parameter_gamma} for the parameter $\gamma$, the proof is straightforward, as follows:
\begin{align}
\sum_{j\in \Lambda}  \dist_{i,j}^s  e^{-\dist_{i,j}/\xi} &=1+ \sum_{x=1}^\infty \sum_{j: \dist_{i,j}=x} x^s e^{-x/\xi} \notag \\
&\le 1 +  \gamma \sum_{x=1}^\infty x^{s+D-1} e^{-x/\xi}   \notag \\
& \le1 + \gamma  \int_0^\infty x^{s+D-1} e^{-(x-1)/\xi}dx  \notag \\
&=1 + \gamma e^{1/\xi} \int_0^\infty \xi (\xi z)^{s+D-1} e^{z}dz \notag \\
&= 1+  \gamma e^{1/\xi} \xi^{s+D} (s+D)!.
\end{align}

Using~\eqref{upper_bound_zeta_s_xi} and $\xi_\beta^{'-1}\le \mu/2$, $\zeta_{0,\xi'_\beta}$ can be reduced to the form of
\begin{align}
\zeta_{0,\xi'_\beta}&= 1+  \gamma e^{1/\xi'_\beta} \xi_\beta'^{D} D! 
=\xi_\beta^{'D} \br{\xi_\beta'^{-D} + \gamma e^{1/\xi'_\beta} D!} \notag \\
&\le  \xi_\beta^{'D} \brr{(\mu/2)^D + \gamma e^{\mu/2} D! }.
\end{align}
Thereafter, on applying the above inequality to~\eqref{mathcal_I_rho_beta_alpha_K__upp}, 
the desired inequality~\eqref{app_Ineq:corol:quantum_information_skey_type} can be obtained. 
This completes the proof. $\square$

\subsection{Proof of Theorem~\ref{quantum_correlation_skey_type}} \label{Proof of Theorem_refquantum_correlation_skey_type}

Herein, the upper bound of $Q_{\rho_\beta}^{(\alpha)}$ in Eq.~\eqref{app_QC_Wigner_Yanase_alpha_average} is considered.
Before beginning the proof, first, we consider the following trivial upper bound for $Q_{\rho}^{(\alpha)}(O_A,O_B)$ for arbitrary $\rho$, as follows: 
\begin{align}
\label{trivial_upp_QC_alpha}
Q_{\rho}^{(\alpha)}(O_A,O_B) &\le \tr(\rho |O_AO_B|) + \frac{\tr(\rho |O_A|^2)+\tr(\rho |O_B|^2)}{2} \notag \\
&\le  (\|O_A\|+ \|O_B\|)^2/2=2,
\end{align}
where $\|O_A\|=\|O_B\|=1$. 
For the proof of inequality~\eqref{trivial_upp_QC_alpha}, because $\tr (\rho O_AO_B)\le \tr (\rho |O_AO_B|)$ is trivial,  the following must be proven: 
\begin{align}
\label{second_term_QC_upper}
| \tr \br{ \rho^{1-\alpha} O_A \rho^{\alpha}  O_B}| \le\frac{\tr(\rho |O_A|^2)+\tr(\rho |O_B|^2)}{2}.
\end{align}
Using the spectral decomposition of $\rho=\sum_s \lambda_s \ket{\lambda_s} \bra{\lambda_s}$, 
\begin{align}
&\abs{ \tr \br{ \rho^{1-\alpha} O_A \rho^{\alpha}  O_B } } \notag \\
&\le \sum_{s,s'} \lambda_s^{1-\alpha} \lambda_{s'}^{\alpha} 
| \bra{\lambda_s} O_A \ket{\lambda_{s'}}\bra{\lambda_{s'}} O_B \ket{\lambda_s} | \notag \\
&\le \sum_{s,s'} \lambda_s^{1-\alpha} \lambda_{s'}^{\alpha} 
\frac{ | \bra{\lambda_s} O_A \ket{\lambda_{s'}}|^2+|\bra{\lambda_{s'}} O_B \ket{\lambda_s} |^2 }{2}.
\label{second_term_QC_alpha}
\end{align}
Using the H\"older inequality
\begin{align}
&\sum_{s,s'} \lambda_s^{1-\alpha} \lambda_{s'}^{\alpha} | \bra{\lambda_s} O_A \ket{\lambda_{s'}}|^2  \notag \\
&= \sum_{s,s'} \br{ \lambda_s | \bra{\lambda_s} O_A \ket{\lambda_{s'}}|^2} ^{1-\alpha} \br{\lambda_{s'}| \bra{\lambda_s} O_A \ket{\lambda_{s'}}|^2}^{\alpha} \notag \\
&\le  \br{\sum_{s,s'}  \lambda_s | \bra{\lambda_s} O_A \ket{\lambda_{s'}}|^2}^{1-\alpha} 
\br{\sum_{s,s'} \lambda_{s'}| \bra{\lambda_s} O_A \ket{\lambda_{s'}}|^2}^{\alpha}\notag \\
&= \sum_{s,s'}  \lambda_s | \bra{\lambda_s} O_A \ket{\lambda_{s'}}|^2 = \tr( \rho |O_A|^2 ),
\label{second_term_QC_alpha_Holder}
\end{align}
where $O_AO_A^\dagger = |O_A|^2$ is used in the last equation.
Thus, on applying inequality~\eqref{second_term_QC_alpha_Holder} to \eqref{second_term_QC_alpha}, inequality~\eqref{second_term_QC_upper} is proven. Therefore, inequality~\eqref{trivial_upp_QC_alpha} is proven.

Thereafter, we consider the non-trivial upper bound presented in Theorem~\ref{quantum_correlation_skey_type}, 
which utilizes the properties of quantum Gibbs states.
When $\rho$ is a Gibbs state (i.e., $\rho=\rho_\beta=e^{-\beta H}$), $\rho_\beta^{-\alpha} O_A\rho_\beta^\alpha$ is reduced to the imaginary time evolution.
Therefore, at the first glance, the quantity~\eqref{app_QC_Wigner_Yanase_alpha_average} is not upper-bounded for low temperatures because the imaginary time evolution $e^{\beta \alpha H} O_A e^{-\beta \alpha H}$ is usually unbounded~\cite{doi:10.1063/1.4936209}. 
To prove Theorem~\ref{quantum_correlation_skey_type}, a direct treatment of the imaginary time evolution should necessarily be avoided.
Instead, the condition $\alpha \in [0,1]$ is utilized for this purpose.
However, for $\alpha \notin [0,1]$, the unboundedness of the norm of $e^{\beta \alpha H} O_A e^{-\beta \alpha H}$ cannot be avoided, which is reflected in the fact that the function $g_{\alpha,\beta}(t)$ in \eqref{def_g_alpha_fourier} converges only  for $\alpha \in [0,1]$. 

For this purpose, the imaginary time evolution is transformed in an appropriate manner. 
Using the notation of Eq.~\eqref{Fourier_op_def} 
\begin{align}
&\tr (\rho_\beta O_AO_B) -\tr (\rho_\beta^{1-\alpha} O_A \rho_\beta^{\alpha}  O_B)  \notag \\
&= \int_{-\infty}^\infty  \tr \br{
\rho_\beta O_{A,\omega}O_B - 
 \rho_\beta^{1-\alpha} O_{A,\omega} \rho_\beta^{\alpha}  O_B} d\omega .
\end{align}
Using $\rho_\beta=e^{-\beta H}$, we obtain 
\begin{align}
&\rho_\beta O_{A,\omega}- \rho_\beta^{1-\alpha} O_{A,\omega} \rho_\beta^{\alpha}
=e^{-\beta H} \br{O_{A,\omega}- e^{\alpha \beta H} O_{A,\omega} e^{-\alpha \beta H}} \notag \\
&=e^{-\beta H} \br{ 1- e^{\alpha \beta \omega} } O_{A,\omega}  =\frac{1-e^{\alpha \beta \omega}}{1-e^{\beta \omega}} [e^{-\beta H}, O_{A,\omega}], \notag
\end{align}
where Eq.~\eqref{O_omega_equation_simple} is used in the last equation.
Hence, using the identity $\tr \br{[O_A,O_B]O_3}= \tr \br{O_A[O_B,O_3]}$,
\begin{align}
&Q_{\rho_\beta}^{(\alpha)}(O_A,O_B)
= \int_{-\infty}^\infty  \frac{1-e^{\alpha \beta \omega}}{1-e^{\beta \omega}} 
\tr \br{ [e^{-\beta H}, O_{A,\omega}]  O_B} d\omega  \notag \\
&= \int_{-\infty}^\infty  \frac{1-e^{\alpha \beta \omega}}{1-e^{\beta \omega}} 
\tr \br{ e^{-\beta H} [O_{A,\omega},  O_B]} d\omega. 
\label{quantum_corre_alpha_Skew}
\end{align}

From Eq.~\eqref{Fourier_op_def_2}, 
\begin{align}
\label{fourier_transform_O_Aomega}
&\int_{-\infty}^\infty  \frac{1-e^{\alpha \beta \omega}}{1-e^{\beta \omega}}  O_{A,\omega} d\omega \notag \\
&= \int_{-\infty}^\infty  \frac{1-e^{\alpha \beta \omega}}{1-e^{\beta \omega}}   \frac{1}{2\pi} \int_{-\infty}^\infty O_A(t) e^{-i\omega t} dt d\omega  \notag \\
&= \int_{-\infty}^\infty g_{\alpha,\beta}(t) O_A(t)  dt,
\end{align}
where $g_{\alpha,\beta}(t)$ is defined by the Fourier transform of $(1-e^{\alpha \beta \omega})/(1-e^{\beta \omega})$ as
\begin{align}
\label{def_g_alpha_fourier}
&g_{\alpha,\beta}(t)  :=  \frac{1}{2\pi}  \int_{-\infty}^\infty  \frac{1-e^{\alpha \beta \omega}}{1-e^{\beta \omega}} e^{-i\omega t}d\omega  \notag \\
&= - i  \beta^{-1} \sum_{m=1}^\infty {\rm sign}(t) e^{-2\pi m |t| /\beta} (-1 + e^{-2\pi i \alpha  m  {\rm sign}(t)}),
\end{align}
where the proof of the second equation is provided in Sec.~\ref{Fourier_calculation/sec}. 
Based on the above form, the following can be obtained 
\begin{align}
\label{upp_g_alpha_t}
| g_{\alpha,\beta}(t) |  &\le 2 \beta^{-1} \sum_{m=1}^\infty e^{-2\pi m |t| /\beta}  \notag \\
&=2 \beta^{-1} \frac{e^{-2\pi  |t| /\beta}}{1- e^{-2\pi  |t| /\beta} }.
\end{align}

Further, combining Eqs.~\eqref{quantum_corre_alpha_Skew} and \eqref{fourier_transform_O_Aomega} with inequality~\eqref{upp_g_alpha_t} yields  
\begin{align}
\label{QC_ineq_integrate_Lieb-Robinson_QC_WY}
&\left |Q_{\rho_\beta}^{(\alpha)}(O_A,O_B) \right|=\left| \int_{-\infty}^\infty g_{\alpha,\beta}(t) \tr \br{ \rho_\beta [O_A(t), O_B]} dt \right| \notag \\
&\le 2 \beta^{-1}  \int_{-\infty}^\infty \frac{e^{-2\pi  |t| /\beta}}{1- e^{-2\pi  |t| /\beta} } \|  [O_A(t),  O_B ]\| dt,
\end{align}
where $\tr \br{ \rho_\beta [O_A(t), O_B]}\le \|[O_A(t), O_B]\|$ are used. 
\begin{widetext}
Subsequently, using the Lieb-Robinson bound~\eqref{Lieb-Robinson_main_short}, 
\begin{align}
\label{integrate_Lieb-Robinson_QC_WY}
2 \beta^{-1}  \int_{-\infty}^\infty \frac{e^{-2\pi  |t| /\beta}}{1- e^{-2\pi  |t| /\beta} } \|  [O_A(t), O_B ]\| dt  
\le \min(|\partial A |,|\partial B |) \brr{
\frac{4}{\pi}  \br{1+ \frac{\xi'_\beta}{R}}
+
2C  \br{\frac{2}{v\beta}+\frac{1}{\pi}} 
} e^{-R/\xi'_\beta}.
\end{align}
The proof is provided in Sec.~\ref{Proof of the inequality_integrate_Lieb-Robinson_QC_WY}. 
For $R\le \xi'_\beta/2$, the RHS in \eqref{integrate_Lieb-Robinson_QC_WY} is larger than the trivial upper bound~\eqref{trivial_upp_QC_alpha}.
Hence, $R\ge \xi'_\beta/2$ must be considered, which yields  
\begin{align}
\label{eq:orig-C24}
2 \beta^{-1}  \int_{-\infty}^\infty \frac{e^{-2\pi  |t| /\beta}}{1- e^{-2\pi  |t| /\beta} } \|  [O_A(t), O_B ]\| dt  
\le 
\min(|\partial A |,|\partial B |) \br{
\frac{12+2C}{\pi}  
+
\frac{4C}{v\beta}  
}e^{-R/\xi'_\beta}.
\end{align}
On applying the above inequality to~\eqref{QC_ineq_integrate_Lieb-Robinson_QC_WY}, Theorem~\ref{quantum_correlation_skey_type} is proven. $\square$
\end{widetext}

\subsubsection{Fourier transform of $(1-e^{\alpha \beta \omega})/(1-e^{\beta \omega})$}
\label{Fourier_calculation/sec}

\begin{figure}[tt]
\centering
{\includegraphics[clip, scale=0.4]{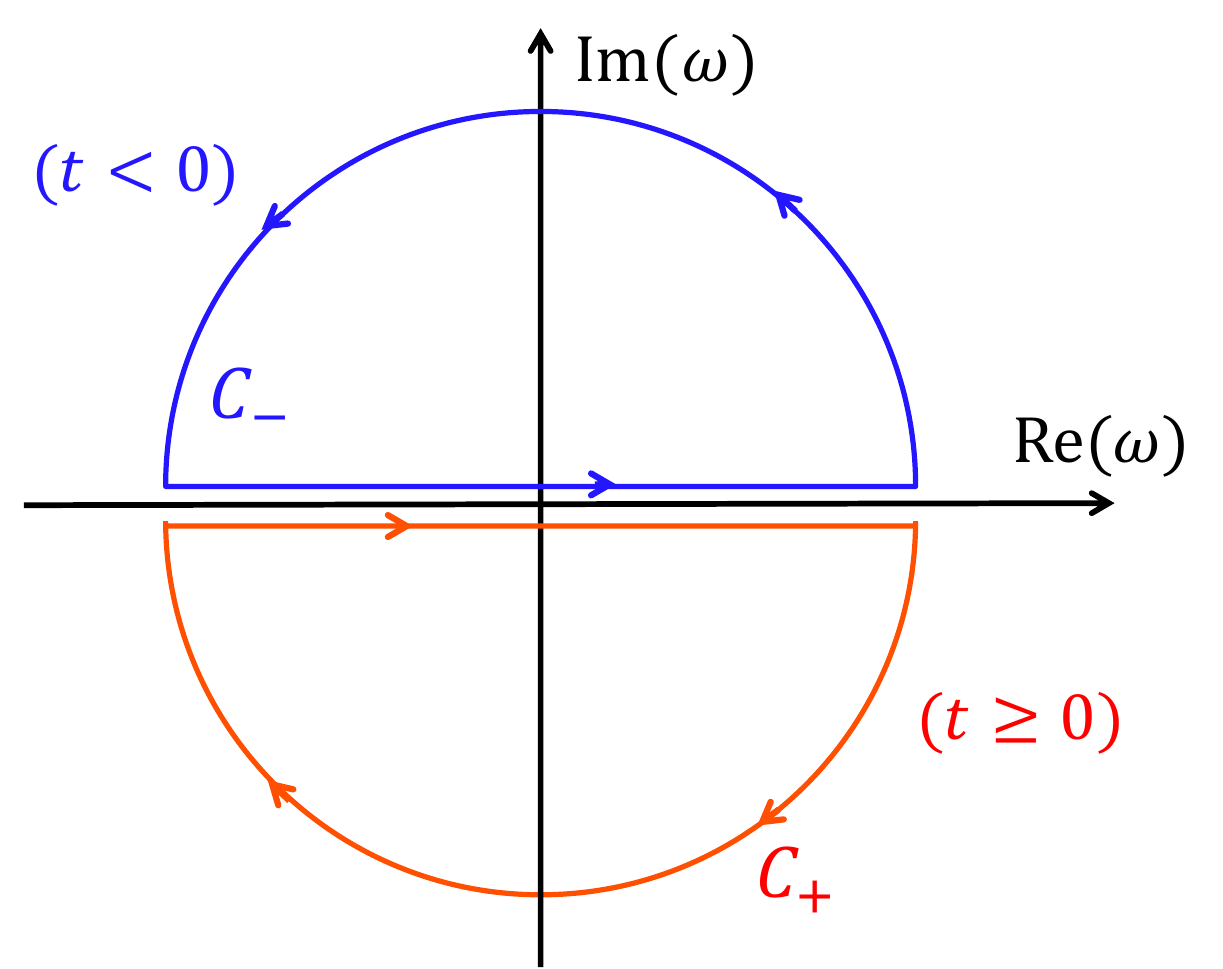}}
\caption{Schematic of the integral paths in Eq.~\eqref{integral_rewrite_fourier}. 
}
\label{fig_int}
\end{figure}

Herein, equation~\eqref{def_g_alpha_fourier} is proven. 
For this proof, the integral is rewritten as follows: 
\begin{align}
&\frac{1}{2\pi} \int_{-\infty}^\infty  \frac{1-e^{\alpha \beta \omega}}{1-e^{\beta \omega}} e^{-i\omega t}d\omega  \notag \\
&= \begin{cases}
\displaystyle \frac{1}{2\pi} \int_{C_-} \frac{1-e^{\alpha \beta \omega}}{1-e^{\beta \omega}} e^{-i\omega t}d\omega  &\for t<0, \\
\displaystyle  \frac{1}{2\pi} \int_{C_+} \frac{1-e^{\alpha \beta \omega}}{1-e^{\beta \omega}} e^{-i\omega t}d\omega  &\for t\ge0,
\end{cases} 
\label{integral_rewrite_fourier}
\end{align}
where the integral paths $C_-$ and $C_+$ are described in Fig.~\ref{fig_int}. 

First, the case of $t<0$ is considered. Then, using the residue theorem, 
\begin{align}
\label{integral_C_-_eq}
&\frac{1}{2\pi} \int_{C_-} \frac{1-e^{\alpha \beta \omega}}{1-e^{\beta \omega}} e^{-i\omega t}d\omega   \notag \\
&= i \sum_{m=1}^\infty {\rm Res}_{\omega=(2\pi i m)/\beta} \br{\frac{1-e^{\alpha \beta \omega}}{1-e^{\beta \omega}} e^{-i\omega t}},
\end{align}
where ${\rm Res}_{\omega=(2\pi i m)/\beta}$ is the residue at $\omega=(2\pi i m)/\beta$. 
Owing to 
\begin{align}
&i {\rm Res}_{\omega=(2\pi im)/\beta} \br{ \frac{1-e^{\alpha \beta \omega}}{1-e^{\beta \omega}} e^{-i\omega t}}  \notag\\
&=i \beta^{-1} e^{2\pi m t /\beta} (-1 + e^{2\pi i m \alpha}),
\end{align}
Eq.~\eqref{integral_C_-_eq} can be reduced to 
\begin{align}
\label{integral_C_-_eq_fin}
&\frac{1}{2\pi} \int_{C_-} \frac{1-e^{\alpha \beta \omega}}{1-e^{\beta \omega}} e^{-i\omega t}d\omega   \notag \\
&=i \beta^{-1} \sum_{m=1}^\infty e^{2\pi m t /\beta} (-1 + e^{2\pi i m \alpha}).
\end{align}

In the same manner, we obtain
\begin{align}
\label{integral_C_+_eq_fin}
&\frac{1}{2\pi} \int_{C_+} \frac{1-e^{\alpha \beta \omega}}{1-e^{\beta \omega}} e^{-i\omega t}d\omega   \notag\\
&=-  i \sum_{m=1}^\infty {\rm Res}_{\omega=-(2\pi im)/\beta} \br{\frac{1-e^{\alpha \beta \omega}}{1-e^{\beta \omega}} e^{-i\omega t}}  \notag \\
&=- i \beta^{-1} \sum_{m=1}^\infty e^{-2\pi m t /\beta} (-1 + e^{-2\pi i m \alpha}).
\end{align}
By combining the two cases~\eqref{integral_C_-_eq_fin} and \eqref{integral_C_+_eq_fin},
\begin{align}
&\frac{1}{2\pi} \int_{-\infty}^\infty  \frac{1-e^{\alpha \beta \omega}}{1-e^{\beta \omega}} e^{-i\omega t}d\omega \notag\\
&= - i  \beta^{-1} \sum_{m=1}^\infty {\rm sign}(t) e^{-2\pi m |t| /\beta} (-1 + e^{-2\pi i \alpha  m  {\rm sign}(t)}). \notag 
\end{align}
This completes the proof of Eq.~\eqref{def_g_alpha_fourier}. 
$\square$

\subsubsection{Proof of the inequality~\eqref{integrate_Lieb-Robinson_QC_WY}} 
\label{Proof of the inequality_integrate_Lieb-Robinson_QC_WY}

We first consider the decomposition
\begin{align}
\label{integrate_QC_WY_all_t}
&\int_{-\infty}^\infty \frac{e^{-2\pi  |t| /\beta}}{1- e^{-2\pi  |t| /\beta} } \|  [O_A(t), O_B ]\| dt   \notag \\
&= \int_{|t| > t_0}\frac{e^{-2\pi  |t| /\beta}}{1- e^{-2\pi  |t| /\beta} } \|  [O_A(t), O_B ]\| dt  \notag \\
&\quad +\int_{|t| \le t_0}\frac{e^{-2\pi  |t| /\beta}}{1- e^{-2\pi  |t| /\beta} } \|  [O_A(t), O_B ]\| dt,
\end{align}
where $t_0:=\mu R/(2v)$ is chosen.
For the first term in the RHS of \eqref{integrate_QC_WY_all_t}, from $1/(1-e^{-|x|}) \le 1+1/|x|$,
\begin{align}
\frac{e^{-2\pi  |t| /\beta}}{1- e^{-2\pi  |t| /\beta} }
\le  e^{-2\pi  |t| /\beta} \br{1+ \frac{1}{2\pi  |t| /\beta}},
\end{align}
which yields
\begin{align}
\label{integrate_QC_WY_t_>_t_0}
&\int_{|t| > t_0}\frac{e^{-2\pi  |t| /\beta}}{1- e^{-2\pi  |t| /\beta} } \|  [O_A(t), O_B ]\| dt  \notag \\
&\le 2 \int_{|t| > t_0} e^{-2\pi  |t| /\beta} \br{1+ \frac{1}{2\pi  |t| /\beta}} dt \notag \\ 
&\le \frac{2\beta}{\pi}  e^{-2\pi  t_0 /\beta}\br{1+ \frac{1}{2\pi  t_0 /\beta}} \notag \\
&= \frac{2\beta}{\pi}  e^{-\pi  \mu R / (v\beta)}\br{1+ \frac{v\beta}{\pi  \mu R}}
\le \frac{2\beta}{\pi}  e^{-R / \xi'_\beta}\br{1+ \frac{\xi'_\beta}{R}},
\end{align}
where $\|  [O_A(t), O_B ]\| \le 2 \|O_A\|\cdot \|O_B\|=2$ and $\pi  \mu/ (v\beta) \ge \xi_\beta^{'-1}=\mu/[2+(v\beta)/\pi]$ are used in the first and last inequalities, respectively.
For the second term in the RHS of \eqref{integrate_QC_WY_all_t}, the Lieb-Robinson bound~\eqref{Lieb-Robinson_main_short} is used as 
\begin{align}
 \|  [O_A(t), O_B ]\| \le C \min(|\partial A |,|\partial B |) \br{e^{v|t|}-1} e^{-\mu R},\notag 
\end{align}
which yields
\begin{align}
\label{integrate_QC_WY_t_le_t_0}
&\int_{|t| \le  t_0}\frac{e^{-2\pi  |t| /\beta}}{1- e^{-2\pi  |t| /\beta} } \|  [O_A(t), O_B ]\| dt  \notag \\
&\le C \min(|\partial A |,|\partial B |) e^{-\mu R}\times \notag \\
& \int_{|t|\le t_0} e^{-2\pi  |t| /\beta} \br{1+ \frac{1}{2\pi  |t| /\beta}}\br{e^{v|t|}-1}  dt.
\end{align}
The integral for $|t|\le t_0$ is upper-bounded as follows: 
\begin{align}
&\int_{|t|\le t_0} e^{-2\pi  |t| /\beta} \br{1+ \frac{1}{2\pi  |t| /\beta}}\br{e^{v|t|}-1}  dt  \notag \\
&\le  2\int_{0}^{t_0} e^{(v-2\pi /\beta)t}   dt + \frac{v}{\pi/\beta} \int_0^1  \int_{0}^{t_0} e^{-2\pi t /\beta}  e^{\lambda v t} dt d\lambda \notag \\
&\le \br{2+\frac{v}{\pi/\beta}}  \int_{0}^{t_0} e^{v t}   dt \le \br{\frac{2}{v}+\frac{1}{\pi/\beta}}e^{v t_0},
\end{align}
where $e^{v t}-1= v t \int_0^1 e^{\lambda v t} d\lambda$ is used in the first inequality. 
Further, the above inequality reduces inequality~\eqref{integrate_QC_WY_t_le_t_0} to
\begin{align}
\label{integrate_QC_WY_t_le_t_0_2}
&\int_{|t| \le  t_0}\frac{e^{-2\pi  |t| /\beta}}{1- e^{-2\pi  |t| /\beta} } \|  [O_A(t), O_B ]\| dt  \notag \\
&\le C   \min(|\partial A |,|\partial B |) \br{\frac{2}{v}+\frac{1}{\pi/\beta}} e^{vt_0-\mu R} \notag \\
&\le \min(|\partial A |,|\partial B |)  C  \br{\frac{2}{v}+\frac{1}{\pi/\beta}} e^{- R/\xi'_\beta},
\end{align}
where we use $t_0=\mu R/(2v)$ and $\mu/2 \ge \xi_\beta^{'-1}=\mu/[2+(v\beta)/\pi]$.

Thereafter, applying inequalities~\eqref{integrate_QC_WY_t_>_t_0} and \eqref{integrate_QC_WY_t_le_t_0_2} to Eq.~\eqref{integrate_QC_WY_all_t} yields
\begin{align}
&\int_{-\infty}^\infty \frac{e^{-2\pi  |t| /\beta}}{1- e^{-2\pi  |t| /\beta} } \|  [O_A(t), O_B ]\| dt 
\le \min(|\partial A |,|\partial B |) 
  \notag \\
&\quad \quad  \quad \quad  \   \times \brr{\frac{2\beta}{\pi}  \br{1+ \frac{\xi'_\beta}{R}}
+
C  \br{\frac{2}{v}+\frac{1}{\pi/\beta}} 
} e^{-R/\xi'_\beta}, \notag 
\end{align}
which, in turn, affords inequality~\eqref{integrate_Lieb-Robinson_QC_WY}. 
This competes the proof.  $\square$

\section{Proof of Theorem~\ref{thm:quantum_correlation}} 
\label{app_Proof of Theorem_thm:quantum_correlation}

This section presents the proof for one of the primary proposed theorems, which provides the exponential decay of the quantum correlation defined by
\begin{align}
\label{app_def_Quantum_corr}
\QC_\rho (O_A,O_B):=  \inf_{\{p_s,\rho_s\}} 
\sum_{s}p_s |\C_{\rho_s}(O_A,O_B)|. 
\end{align} 
In Theorem~\ref{thm:quantum_correlation}, the following inequality was proven: 
\begin{align}
\label{app_ineq:thm:quantum_correlation}
&\QC_\rho (O_A,O_B)\notag \\
&\le C_\beta (|\partial A| +|\partial B| )\br { 1 +  \log|AB| }  e^{-R/\xi_\beta},
\end{align}
where $\xi_\beta$ is a $\orderof{\beta}$ constant expressed as Eq.~\eqref{parameters_definitions}, and 
$C_\beta$ is obtained from $c_{\beta, 1}+c_{\beta, 2}$, with $c_{\beta, 1}$ and $c_{\beta, 2}$ defined in Eq.~\eqref{parameters_definitions}

Here, the logarithmic term $1 +  \log|AB| $ originates from the norm of $\rho^{-1/2 }  \mathcal{L}_{O_A} \rho^{1/2}$ and $\rho^{-1/2 }  \mathcal{L}_{O_B} \rho^{1/2}$ in Eq.~\eqref{commutator_imaginary_time_final}.
The explicit norm estimation is provided in Claim~\ref{claim_L_O_imag_norm}.

\subsection{Proof of Theorem~\ref{thm:quantum_correlation}}

For an arbitrary quantum state $\rho$, the spectral decomposition of $\rho$ is denoted as 
\begin{align}
\rho=\sum_s \lambda_s  \ket{\lambda_s}\bra{\lambda_s}. 
 \end{align}
In the proof, the aim is to explicitly construct a set of ensembles $\{p_m, \ket{\phi_m}\}$ such that 
\begin{align}
\rho_\beta= \sum_m p_m \ket{\phi_m}\bra{\phi_m},
 \end{align}
 which satisfies inequality~\eqref{app_ineq:thm:quantum_correlation}. 
To prove the statements, the following steps are adopted. 
In the first and second lemmas (Lemmas~\ref{lem:quantum_correlation_1} and \ref{lem:quantum_correlation_2}), 
the generic quantum states are considered and provide general statements regarding the quantum correlations. 
Thereafter, in the third, fourth, and fifth lemmas (Lemmas~\ref{lem:quantum_correlation_3}, \ref{lem:quantum_correlation_4}, and \ref{lem:quantum_correlation_5}), the property of quantum Gibbs states is utilized to provide an upper-bound to the quantum correlations. 

In the first step, the general upper bound for the quantum correlation is proven, as follows:
\begin{lemma} \label{lem:quantum_correlation_1}
For an arbitrary operator $O$, $\mathcal{L}_O$ is defined as follows: 
\begin{align}
\label{def_L_O_lambda}
\mathcal{L}_O :=  \sum_{s,s'} \frac{2\sqrt{\lambda_s \lambda_{s'}}}{\lambda_s +\lambda_{s'}} 
\bra{\lambda_s} O \ket{\lambda_{s'}} \ket{\lambda_{s}}\bra{\lambda_{s'}}. 
\end{align}
Then, for the two operators $O_A$ and $O_B$, if  
 \begin{align}
 \label{condition_lem:quantum_correlation_1}
[\mathcal{L}_{O_A}, \mathcal{L}_{O_B} ] = 0,
 \end{align}
the quantum correlation is bound from above as follows:
\begin{align}
\label{main_ineq:quantum_correlation_1}
&\QC_\rho(O_A,O_B)  \notag \\
&\le \frac{1}{4} \norm{\left[  \br{ \rho^{-1/2 }  \mathcal{L}_{O_A} \rho^{1/2} }, \br{\rho^{1/2}\mathcal{L}_{O_B} \rho^{-1/2 }} \right]}.
\end{align}
\end{lemma} 

Typically, condition~\eqref{condition_lem:quantum_correlation_1} is not satisfied. 
Further, in the second lemma, consider the case where Eq.~\eqref{condition_lem:quantum_correlation_1} holds only in an approximate sense.
Thus, the lemma can be proven as follows:
\begin{lemma} \label{lem:quantum_correlation_2}
For two arbitrary operators $O_A$ and $O_B$, if two operators $\tilde{\mathcal{L}}_{O_A}$ and $\tilde{\mathcal{L}}_{O_B}$ can be determined such that 
  \begin{align}
[\tilde{\mathcal{L}}_{O_A}, \tilde{\mathcal{L}}_{O_B}] =0 
\end{align}
and 
 \begin{align}
 \label{condition_L_01_LO2_L01O2}
&\| \mathcal{L}_{O_A} - \tilde{\mathcal{L}}_{O_A} \| \le \delta_1, \quad \| \mathcal{L}_{O_B} - \tilde{\mathcal{L}}_{O_B} \| \le \delta_2,
 \end{align}
the quantum correlation $\QC_\rho(O_A,O_B)$ is upper-bounded as follows:
\begin{align}
\label{main_inequality_lemma_2_QC}
&\QC_\rho(O_A,O_B) \le  3\delta_1+3\delta_2 \notag \\
  &+ \frac{1}{4} \norm{  
\left[  \br{ \rho^{-1/2 }  \mathcal{L}_{O_A} \rho^{1/2} }, \br{\rho^{1/2}\mathcal{L}_{O_B} \rho^{-1/2 }} \right]}.
 \end{align}
\end{lemma} 

The final task is to provide an upper-bound for the parameters $\{\delta_1, \delta_2\}$ and the norm of the commutator between $\rho^{-1/2 }  \mathcal{L}_{O_A} \rho^{1/2}$ and $\rho^{1/2}\mathcal{L}_{O_B} \rho^{-1/2 }$.  
Thus, we first consider an integral form of $\mathcal{L}_O$, which comprises the time evolution of $t\approx \beta$.
The lemma on the basic properties of the operator $\mathcal{L}_O$ is proven as follows:

\begin{lemma} \label{lem:quantum_correlation_3}
Let $\rho$ be a quantum Gibbs state as $\rho=\rho_\beta=e^{-\beta H}$.
Then, for an arbitrary operator $O$, the operator $\mathcal{L}_O$ is given as follows:
\begin{align}
\label{form_simple_mathcal_L_O}
\mathcal{L}_O  =\int_{-\infty}^\infty f_{\beta} (t) O(t) dt,
 \end{align}
where $f_{\beta} (t)$ is defined as
\begin{align}
f_{\beta}  (t) &= \frac{1}{\beta \cosh(\pi t /\beta)}.
 \end{align}
Furthermore, the norm of $\mathcal{L}_O$ is upper-bounded as follows:
\begin{align}
\label{fupper_bound_of_mathcal_L_O}
 \| \mathcal{L}_O\| \le  \|O\|. 
 \end{align}
\end{lemma} 

Because the function $f_{\beta}  (t)$ decays exponentially as $e^{-\orderof{|t|/\beta}}$, 
the operator $\mathcal{L}_O$ is approximately constructed using the time-evolved operator $O(t)$ with $t\approx \beta$. 
Consequently, the Lieb-Robinson bound is applied to prove the quasi-locality of $\mathcal{L}_O$ and construct the operators $\tilde{\mathcal{L}}_{O_A}$ and $\tilde{\mathcal{L}}_{O_B}$ in Lemma~\ref{lem:quantum_correlation_2}.
From Lemma~\ref{lem:quantum_correlation_3}, the following lemma, which provides the upper bounds for $\delta_1$ and $\delta_2$, is proven: 
\begin{lemma} \label{lem:quantum_correlation_4}
When $\rho$ is given by the quantum Gibbs state with a short-range Hamiltonian, as in \eqref{int_short_range}, 
$\delta_1$ and $\delta_2$ are upper-bounded as 
\begin{align}
\label{lemma_upperbound_delta_1_delta_2}
&\delta_1 \le   e^{\mu/(2+2v\beta/\pi)} \br{ \frac{8}{\pi} + \frac{4C}{v\beta}  }  |\partial A|  e^{-\mu R/[4(1+v\beta/\pi)]}, \notag \\
&\delta_2  \le  e^{\mu/(2+2v\beta/\pi)} \br{ \frac{8}{\pi} + \frac{4C}{v\beta}  }  |\partial B|  e^{-\mu R/[4(1+v\beta/\pi)]}. 
\end{align}
\end{lemma}

\noindent 
This lemma provides the upper bound for the first term of the RHS in inequality~\eqref{main_inequality_lemma_2_QC}, as follows:
\begin{align}
\label{upp_delta_1+_delta_2}
3\delta_1+3\delta_2
 \le c_{\beta, 1} ( |\partial A| +|\partial B|)  e^{- R/\xi_\beta},
 \end{align}
 where the definition of $c_{\beta, 1}$ and $\xi_\beta$ is used in Eq.~\eqref{parameters_definitions}. 
 
 Before detailing the estimation for the second term of the RHS of \eqref{main_inequality_lemma_2_QC}, it is shown that, for $R-2\le \xi_\beta$, the upper bound~\eqref{upp_delta_1+_delta_2} results in a trivial upper bound for $\QC_\rho(O_A,O_B)$.
Indeed, for $R-2\le \xi_\beta$,
 \begin{align}
&c_{\beta, 1} ( |\partial A| +|\partial B|)  e^{- R/\xi_\beta} 
\ge  c_{\beta, 1} e^{- R/\xi_\beta}   \notag \\
&\ge e^{- (R-2)/\xi_\beta} \frac{24}{\pi} \ge  \frac{24}{e\pi} \approx 2.8104,
 \end{align}
 which is larger than the trivial upper bound $\|O_A\|\cdot \|O_B\|=1$ (i.e., $\QC_\rho(O_A,O_B)\le 1$).
 Therefore, we consider the regime of $R-2> \xi_\beta$ in the following.

The final task involves estimating the commutator.
\begin{align}
\label{commutator_imaginary_time_final}
\norm{\left[  \br{ \rho^{-1/2 }  \mathcal{L}_{O_A} \rho^{1/2} }, \br{\rho^{1/2}\mathcal{L}_{O_B} \rho^{-1/2 }} \right]}.
 \end{align}
Herein, the quasi-locality of $\rho^{-1/2 }  \mathcal{L}_{O_A} \rho^{1/2}$ must be characterized. 
For $\rho=e^{-\beta H}$, it is obtained from the imaginary time-evolution of $\mathcal{L}_{O_A}$. 
For a large $\beta$, the unboundedness of the imaginary time evolution usually occurs~\cite{doi:10.1063/1.4936209}.
Notably, owing to the specialty of $\mathcal{L}_{O_A} $, such an unboundedness can be avoided and the following lemma can be proven: 
\begin{widetext}
\begin{lemma} \label{lem:quantum_correlation_5}
The norm of the commutator~\eqref{commutator_imaginary_time_final} is upper-bounded by
\begin{align}
\label{upper_bound_commutator_imaginary_time_final}
&\norm{\left[  \br{ \rho^{-1/2 }  \mathcal{L}_{O_A} \rho^{1/2} }, \br{\rho^{1/2}\mathcal{L}_{O_B} \rho^{-1/2 }} \right]}   \notag \\
&\le 3 e^{2/\xi_\beta}  \brr{\frac{8}{\pi} \br{1+ \frac{\xi_\beta}{R-2}} 
+
4C\br{\frac{1}{\pi}+\frac{1}{v\beta}}
 }     e^{-R/\xi_\beta} \notag \\
&\times \bigl \{ |\partial A| \brr{2+ \log (1+\beta \| \ad_H(O_B)\| ) }
+|\partial B| \brr{2+ \log (1+\beta \| \ad_H(O_A)\| ) } \bigr\}   \notag \\
&\le  e^{2/\xi_\beta}  \br{\frac{48+12C}{\pi}
+
\frac{12C}{v\beta}}
     e^{-R/\xi_\beta} \notag \\
&\times \bigl \{ |\partial A| \brr{2+ \log (1+\beta \| \ad_H(O_B)\| ) }
+|\partial B| \brr{2+ \log (1+\beta \| \ad_H(O_A)\| ) } \bigr\},
 \end{align}
 where $R-2> \xi_\beta$ is used in the second inequality. 
\end{lemma} 

To estimate the upper bound of $\| \ad_H(O_A)\|$ ($\| \ad_H(O_B)\|$), 
consider the norm of a commutator $\ad_H(O_X)$ ($\|O_X\|=1$) for a general operator $O_X$, which is upper-bounded using \eqref{supp_def:Ham} as follows:
\begin{align}
\| \ad_H(O_X) \| \le \sum_{i\in X}\sum_{Z: Z\ni i} \| \ad_{h_Z}(O_X) \| 
\le 2 \sum_{i\in X}\sum_{Z: Z\ni i} \| h_Z\| \cdot \|O_X \| \le 2 g |X|.
 \end{align}
Hence, using $\log(1+x y) \le \log(1+y) + \log(x)$ for $x\ge 1$ and $y\ge 0$, 
\begin{align}
 &|\partial A| \Bigl( 2+ \log (1+\beta \| \ad_H(O_B)\| )\Bigr)  +|\partial B| \Bigl( 2+ \log (1+\beta \| \ad_H(O_A)\| )\Bigr)  \notag \\
 & \le 
  ( |\partial A|+|\partial B|)\Bigl(  2+ \log (1+2g\beta  |AB| )\Bigr) \notag \\
  &\le  ( |\partial A|+|\partial B|) \br{\frac{2+ \log (1+2g\beta)+\log  |AB| }{ \log  |AB| +1 }  } (\log  |AB|  +1)\notag \\
  &\le ( |\partial A|+|\partial B|) [ 2+ \log (1+2g\beta)]  (\log  |AB|  +1).
 \end{align}
Thus, combining the above inequality with \eqref{upper_bound_commutator_imaginary_time_final}, 
an upper-bound is provided for the second term of the RHS in inequality~\eqref{main_inequality_lemma_2_QC} by
 \begin{align}
 \label{commutator_L_O_AL_O_B_imag_fin}
\frac{1}{4} \norm{  
\left[  \br{ \rho^{-1/2 }  \mathcal{L}_{O_A} \rho^{1/2} }, \br{\rho^{1/2}\mathcal{L}_{O_B} \rho^{-1/2 }} \right]}  
\le c_{\beta,2}(|\partial A| +|\partial B| ) ( 1 +  \log|AB| ) e^{-R/\xi_\beta},
 \end{align}
 where the definitions of $c_{\beta,2}$ in Eq.~\eqref{parameters_definitions} are used. 
 
Thus, by applying inequalities~\eqref{upp_delta_1+_delta_2} and \eqref{commutator_L_O_AL_O_B_imag_fin} to Lemma~\ref{lem:quantum_correlation_2}, the desired inequality~\eqref{app_ineq:thm:quantum_correlation} can be obtained. 
This completes the proof of Theorem~\ref{thm:quantum_correlation}. $\square$

\end{widetext}

\subsection{Proof of Lemma~\ref{lem:quantum_correlation_1}}

In this proof, a technique similar to that outlined in Ref.~\cite{yu2013quantum} is employed.  
Let $\{\ket{\psi_m}\}$ be a set of orthonormal quantum states.
Define the unitary matrix $U$, which provides the quantum states: $\{\ket{\psi_m}\}$ in the base of $\{\ket{\lambda_s}\}_s$:
 \begin{align}
\ket{\psi_m}= \sum_s U_{m,s} \ket{\lambda_{s}}.
\end{align}
Then, by defining the ensemble $\{p_m, \ket{\phi_m}\}$ as  
 \begin{align}
\ket{\phi_m}=\frac{1}{\sqrt{p_m}} \sqrt{\rho}\ket{\psi_m}, \quad p_m = \bra{\psi_m} \rho \ket{\psi_m},
\end{align}
Then, density operator $\rho$ is rewritten as 
 \begin{align}
\rho = \sum_m p_m  \ket{\phi_m}\bra{\phi_m}. 
\end{align}
In general, $\{\ket{\phi_m}\}$ are not orthogonal to each other (i.e., $\bra{\phi_m} \phi_{m'} \rangle \neq 0$). 
For this decomposition, the quantum correlation $\QC_\rho(O_A,O_B)$ is upper-bounded by
 \begin{align}
 \label{QC_upper_bound_lemma_cond}
\QC_\rho(O_A,O_B) \le \sum_m p_m  |\C_{\ket{\phi_m}}(O_A,O_B) |, 
\end{align}
where $\C_{\ket{\phi_m}}(O_A,O_B)$ has been defined as a standard correlation function, that is, $\C_{\ket{\phi_m}}(O_A,O_B)=\bra{\phi_m} O_A O_B \ket{\phi_m} -\bra{\phi_m} O_A \ket{\phi_m} \bra{\phi_m} O_B \ket{\phi_m}$. 
Our task is to identify a good set $\{\ket{\psi_m}\}$ such that $\{\ket{\phi_m}\}$ has a weak correlation with $O_A$ and $O_B$.

For an arbitrary operator $O$, 
 \begin{align}
 \label{Eq_phi_m_O_phi_m}
\bra{\phi_m}O\ket{\phi_m}& = \sum_{s,s'}  \frac{U_{m,s'}U^\ast_{m,s}}{p_m} \sqrt{\lambda_{s}\lambda_{s'}} 
\bra{\lambda_{s}}O\ket{\lambda_{s'}} \notag \\
&=\sum_{s,s'}  \frac{U_{m,s'}U^\ast_{m,s}}{p_m} \frac{\lambda_s + \lambda_s'}{2} 
\bra{\lambda_{s}} \mathcal{L}_{O} \ket{\lambda_{s'}}  \notag \\
&= \sum_{s,s'}  \frac{U_{m,s'}U^\ast_{m,s}}{p_m} \frac{1}{2}  
\bra{\lambda_{s}} \{\rho, \mathcal{L}_{O} \} \ket{\lambda_{s'}} \notag \\
&=\frac{1}{2p_m}  
\bra{\psi_m} \{\rho, \mathcal{L}_{O} \} \ket{\psi_m},
 \end{align}
where definition~\eqref{def_L_O_lambda} is used for $\mathcal{L}_{O}$ from the second to third equations.
Here, the definition is shown again for the convenience of the reader:
\begin{align}
\label{def_L_O_lambda_re}
\mathcal{L}_O :=  \sum_{s,s'} \frac{2\sqrt{\lambda_s \lambda_{s'}}}{\lambda_s +\lambda_{s'}} 
\bra{\lambda_s} O \ket{\lambda_{s'}} \ket{\lambda_{s}}\bra{\lambda_{s'}}. 
\end{align} 
 
Herein, $\{\ket{\psi_m}\}$ are chosen as the simultaneous eigenstates of $\mathcal{L}_{O_A}$ and $\mathcal{L}_{O_B}$.
Note that such a choice is possible because of condition~\eqref{condition_lem:quantum_correlation_1}, that is, $[\mathcal{L}_{O_A},\mathcal{L}_{O_B}]=0$. 
We then obtain, from Eq.~\eqref{Eq_phi_m_O_phi_m},
 \begin{align}
  \label{Eq_phi_m_O_AO_B_phi_m}
\bra{\phi_m}O_A\ket{\phi_m}&=\frac{1}{2p_m}  
\bra{\psi_m} \{\rho, \mathcal{L}_{O_A} \} \ket{\psi_m}  \notag \\
&= \frac{\alpha_{1,m}}{p_m}  \bra{\psi_m} \rho \ket{\psi_m}  = \alpha_{1,m}
 \end{align}
 and $\bra{\phi_m}O_B\ket{\phi_m}= \alpha_{2,m}$, 
where $\alpha_{1,m}$ and $\alpha_{2,m}$ are defined as the $m$th eigenvalues of $\mathcal{L}_{O_A}$ and $\mathcal{L}_{O_B}$, respectively. 
We, therefore, obtain
 \begin{align}
  \label{Eq_phi_m_O_AO_B_phi_m_fin}
\bra{\phi_m}O_A\ket{\phi_m}\bra{\phi_m}O_B\ket{\phi_m}  = \alpha_{1,m}\alpha_{2,m}
 \end{align}
for an arbitrary $m$.

We next consider $\bra{\phi_m}O_A O_B\ket{\phi_m}$.
Then, from Eq.~\eqref{Eq_phi_m_O_phi_m},
\begin{align}
\bra{\phi_m}O_A O_B\ket{\phi_m}=\frac{1}{2p_m}  
\bra{\psi_m} \{\rho, \mathcal{L}_{O_AO_B} \} \ket{\psi_m}.
 \end{align}
Further, based on the equation, if $\mathcal{L}_{O_AO_B}=\mathcal{L}_{O_A}\mathcal{L}_{O_B}$ can be obtained, 
$\bra{\phi_m}O_A O_B\ket{\phi_m}=\alpha_{1,m}\alpha_{2,m}$ can also be easily proven in the same manner as for Eq.~\eqref{Eq_phi_m_O_AO_B_phi_m}.
However, the difficulty lies in the fact that, in general, $\mathcal{L}_{O_AO_B}\neq \mathcal{L}_{O_A}\mathcal{L}_{O_B}$; hence, a different approach is required.

For this purpose, first consider 
\begin{align}
\bra{\phi_m}O\ket{\psi_{m'}}& = \sum_{s,s'}  \frac{U_{m',s'}U^\ast_{m,s}}{\sqrt{p_m}} \sqrt{\lambda_{s}} 
\bra{\lambda_{s}}O\ket{\lambda_{s'}} \notag \\
&= \sum_{s,s'}  \frac{U_{m',s'}U^\ast_{m,s}}{\sqrt{p_m}} \sqrt{\lambda_{s}\lambda_{s'}} 
\bra{\lambda_{s}}O  \rho^{-1/2} \ket{\lambda_{s'}} \notag \\
&=\sum_{s,s'}  \frac{U_{m',s'}U^\ast_{m,s}}{\sqrt{p_m}} \frac{\lambda_s + \lambda_s'}{2} 
\bra{\lambda_{s}} \mathcal{L}_{O\rho^{-1/2}}  \ket{\lambda_{s'}}  \notag \\
&=\frac{1}{2\sqrt{p_m}}  
\bra{\psi_m} \{\rho, \mathcal{L}_{O} \rho^{-1/2}\} \ket{\psi_{m'}},
 \end{align} 
where $\mathcal{L}_{O\rho^{-1/2}}  = \mathcal{L}_{O} \rho^{-1/2}$ is used 
from definition~\eqref{def_L_O_lambda_re}. 
\begin{widetext}
Subsequently, 
\begin{align}
\label{phi_m_expect_O_AO_B}
\bra{\phi_m}O_AO_B\ket{\phi_m}&=\sum_{m'} \bra{\phi_m}O_A\ket{\psi_{m'}} \bra{\psi_{m'}} O_B\ket{\phi_m}  \notag \\
&= \frac{1}{4p_m} \sum_{m'} \bra{\psi_m}\{\rho, \mathcal{L}_{O_A} \rho^{-1/2}\} \ket{\psi_{m'}} \bra{\psi_{m'}} \{\rho, \rho^{-1/2} \mathcal{L}_{O_B} \}  \ket{\psi_m}   \notag \\
&=\frac{1}{4p_m} \bra{\psi_m}\{\rho, \mathcal{L}_{O_A} \rho^{-1/2}\}  \{\rho, \rho^{-1/2} \mathcal{L}_{O_B} \}  \ket{\psi_m},  
 \end{align}
where $ \sum_{m'} \ket{\psi_{m'}} \bra{\psi_{m'}} =1$ is used. 
Thus, Eq.~\eqref{phi_m_expect_O_AO_B} is further reduced to 
\begin{align}
\label{phi_m_expect_O_AO_B_expansion}
\bra{\phi_m}O_AO_B\ket{\phi_m}
&=\frac{1}{4p_m} 
\bra{\psi_m}( \rho \mathcal{L}_{O_A} \rho^{-1/2}+\mathcal{L}_{O_A} \rho^{1/2} )(\rho^{1/2} \mathcal{L}_{O_B} +\rho^{-1/2} \mathcal{L}_{O_B}\rho)  \ket{\psi_m} \notag \\
&=\frac{1}{4p_m} 
\bra{\psi_m} (\rho \mathcal{L}_{O_A}\mathcal{L}_{O_B} + \mathcal{L}_{O_A} \rho  \mathcal{L}_{O_B}  
+ \mathcal{L}_{O_A} \mathcal{L}_{O_B}\rho +\rho \mathcal{L}_{O_A} \rho^{-1}\mathcal{L}_{O_B}\rho )
  \ket{\psi_m}.
 \end{align}
 Using $\mathcal{L}_{O_A} \ket{\psi_m}=\alpha_{1,m} \ket{\psi_m}$ and $\mathcal{L}_{O_B} \ket{\psi_m}=\alpha_{2,m} \ket{\psi_m}$, the above equation can be reduced to 
\begin{align}
\label{phi_m_expect_O_AO_B_2}
\bra{\phi_m}O_AO_B\ket{\phi_m}
&=\frac{1}{4p_m}  \bra{\psi_m} \br{ \rho\alpha_{1,m} \alpha_{2,m} + \alpha_{1,m} \rho \alpha_{2,m}
+ \alpha_{1,m} \alpha_{2,m} \rho +  \rho \mathcal{L}_{O_A} \rho^{-1}  \mathcal{L}_{O_B} \rho } \ket{\psi_m} \notag \\
&= \frac{3}{4}\alpha_{1,m} \alpha_{2,m} 
+\frac{1}{4p_m} \bra{\psi_m}  \rho \mathcal{L}_{O_A} \rho^{-1}  \mathcal{L}_{O_B} \rho  \ket{\psi_m},
\end{align}
 where $\bra{\psi_m} \rho \ket{\psi_m}=p_m$.

The remaining task entails estimating the error as 
\begin{align}
\bra{\psi_m}  \rho \mathcal{L}_{O_A} \rho^{-1}  \mathcal{L}_{O_B} \rho  \ket{\psi_m} - p_m \alpha_{1,m} \alpha_{2,m}.
 \end{align} 
To obtain this, consider 
\begin{align}
\label{phi_m_expect_O_AO_B_3}
\bra{\psi_m}  \rho \mathcal{L}_{O_A} \rho^{-1}  \mathcal{L}_{O_B} \rho  \ket{\psi_m}
&= \bra{\psi_m}  \rho^{1/2}  \br{\rho^{1/2}\mathcal{L}_{O_A} \rho^{-1/2 }} \br{ \rho^{-1/2 }  \mathcal{L}_{O_B} \rho^{1/2} }
\rho^{1/2 }   \ket{\psi_m} \notag \\
&= \bra{\psi_m}  \rho^{1/2}  \br{ \rho^{-1/2 }  \mathcal{L}_{O_B} \rho^{1/2} } \br{\rho^{1/2}\mathcal{L}_{O_A} \rho^{-1/2 }} 
\rho^{1/2 }   \ket{\psi_m}  \notag \\
&\quad + \bra{\psi_m}  \rho^{1/2} 
\left[  \br{ \rho^{-1/2 }  \mathcal{L}_{O_A} \rho^{1/2} }, \br{\rho^{1/2}\mathcal{L}_{O_B} \rho^{-1/2 }} \right]
 \rho^{1/2 }   \ket{\psi_m}  \notag \\
 &= p_m \alpha_{1,m} \alpha_{2,m} + p_m \bra{\phi_m}  
\left[  \br{ \rho^{-1/2 }  \mathcal{L}_{O_A} \rho^{1/2} }, \br{\rho^{1/2}\mathcal{L}_{O_B} \rho^{-1/2 }} \right]
  \ket{\phi_m},
\end{align} 
where $\mathcal{L}_{O_A} \ket{\psi_m}=\alpha_{1,m}  \ket{\psi_m}$ and $\mathcal{L}_{O_B} \ket{\psi_m}=\alpha_{2,m}  \ket{\psi_m}$ are used from the second to third equations.
Thus, by applying Eq.~\eqref{phi_m_expect_O_AO_B_3} to Eq.~\eqref{phi_m_expect_O_AO_B_2},
\begin{align}
\left| \bra{\phi_m}O_AO_B\ket{\phi_m}- \alpha_{1,m} \alpha_{2,m} \right| 
\le \frac{1}{4} \norm{\left[  \br{ \rho^{-1/2 }  \mathcal{L}_{O_A} \rho^{1/2} }, \br{\rho^{1/2}\mathcal{L}_{O_B} \rho^{-1/2 }} \right]}. 
\end{align}
Therefore, by combining the above inequality and Eq.~\eqref{Eq_phi_m_O_AO_B_phi_m_fin} 
with \eqref{QC_upper_bound_lemma_cond}, inequality~\eqref{main_ineq:quantum_correlation_1} is proven.
This completes the proof. $\square$
\end{widetext}


\subsection{Proof of Lemma~\ref{lem:quantum_correlation_2}}

The approach used in this proof is similar to that for the proof of Lemma~\ref{lem:quantum_correlation_1}. 
Herein, $\{\ket{\psi_m}\}$ are chosen as the simultaneous eigenstates of $\tilde{\mathcal{L}}_{O_A}$ and $\tilde{\mathcal{L}}_{O_B}$, instead of $\mathcal{L}_{O_A}$ and $\mathcal{L}_{O_B}$:
\begin{align}
\tilde{\mathcal{L}}_{O_A} \ket{\psi_m}=\tilde{\alpha}_{1,m}  \ket{\psi_m}, \quad 
\tilde{\mathcal{L}}_{O_B} \ket{\psi_m}=\tilde{\alpha}_{2,m}  \ket{\psi_m}.
\end{align}
Then, the same inequality as \eqref{QC_upper_bound_lemma_cond} is obtained:
 \begin{align}
 \label{QC_upper_bound_lemma_cond_2nd}
\QC_\rho(O_A,O_B) \le \sum_m p_m  |\C_{\ket{\phi_m}}(O_A,O_B) |, 
\end{align}

We begin by estimating $\bra{\phi_m} O_A \ket{\phi_m} \bra{\phi_m} O_B \ket{\phi_m}$. 
Using Eq.~\eqref{Eq_phi_m_O_phi_m}, 
 \begin{align}
 \label{Eq_phi_m_O_phi_m_2nd_Lemma}
\bra{\phi_m}O_A\ket{\phi_m}
&=\frac{1}{2p_m}  
\bra{\psi_m} \{\rho, \mathcal{L}_{O_A} \} \ket{\psi_m}  \notag \\ 
&=\tilde{\alpha}_{1,m}+ \frac{1}{2p_m}  
\bra{\psi_m} \{\rho, \delta \mathcal{L}_{O_A} \} \ket{\psi_m},
 \end{align}
 where $\delta \mathcal{L}_{O_A}:= \mathcal{L}_{O_A} - \tilde{\mathcal{L}}_{O_A}$. 
In the same manner, $\bra{\phi_m}O_B\ket{\phi_m}
= \tilde{\alpha}_{2,m}+ \frac{1}{2p_m}  
\bra{\psi_m} \{\rho, \delta \mathcal{L}_{O_B} \} \ket{\psi_m}$. 
Thus, 
 \begin{align}
 \label{prod_O_Aex_O_Bex_phi_m}
&| \bra{\phi_m} O_A \ket{\phi_m} \bra{\phi_m} O_B \ket{\phi_m} - \tilde{\alpha}_{1,m}\tilde{\alpha}_{2,m}|   \notag \\
&\le \frac{1}{2p_m}| \bra{\psi_m} \{\rho, \delta \mathcal{L}_{O_A}+\delta \mathcal{L}_{O_B} \} \ket{\psi_m} |,
\end{align}
which yields 
 \begin{align}
&\sum_m p_m | \bra{\phi_m} O_A \ket{\phi_m} \bra{\phi_m} O_B \ket{\phi_m} - \tilde{\alpha}_{1,m}\tilde{\alpha}_{2,m}|  \notag \\
&\le \frac{1}{2} \sum_m | \bra{\psi_m} \{\rho, \delta \mathcal{L}_{O_A}+\delta \mathcal{L}_{O_B} \} \ket{\psi_m}|. 
\end{align}

For an arbitrary operator $O$, $|\bra{\psi_m} O\ket{\psi_m}|\le \bra{\psi_m}\ |O|\ \ket{\psi_m}$; hence, 
  \begin{align}
  \label{sum_m_tilde_delta_1_m_le_2}
&\sum_m | \bra{\psi_m} \{\rho, \delta \mathcal{L}_{O_A}+\delta \mathcal{L}_{O_B} \} \ket{\psi_m}|   \notag \\
&\le \sum_m \bra{\psi_m}\ |\{\rho, \delta \mathcal{L}_{O_A}+\delta \mathcal{L}_{O_B} \}|\ \ket{\psi_m} \notag \\
&\le \| \{ \rho, \delta \mathcal{L}_{O_A}\} \|_1  + \| \{ \rho, \delta \mathcal{L}_{O_B}\} \|_1  \notag \\
&\le 2\|\rho\|_1 \cdot (\| \delta \mathcal{L}_{O_A}\| + \| \delta \mathcal{L}_{O_B}\|)\le 2(\delta_1 + \delta_2),
 \end{align}
 where $\tr (|O|)=\|O\|_1$, $\|O+O'\|_1\le \|O\|_1+\|O'\|_1$ and $\|OO'\|_1\le \|O\|_1 \cdot \|O'\|$ are used for the arbitrary operators $O$ and $O'$, respectively. 
 Further, applying inequality~\eqref{sum_m_tilde_delta_1_m_le_2} to \eqref{prod_O_Aex_O_Bex_phi_m} yields
 \begin{align}
 \label{prod_O_Aex_O_Bex_phi_m_fin}
\sum_m p_m | \bra{\phi_m} O_A \ket{\phi_m} \bra{\phi_m} O_B \ket{\phi_m} - \tilde{\alpha}_{1,m}\tilde{\alpha}_{2,m}|  
\le \delta_1 + \delta_2. 
\end{align}

\begin{widetext}
Next the error that originates from $\bra{\phi_m}O_AO_B\ket{\phi_m}$ is estimated. 
Consider the same equation as Eq.~\eqref{phi_m_expect_O_AO_B_expansion}:
\begin{align}
\label{phi_m_expect_O_AO_B_Lemma_2nd}
\bra{\phi_m}O_AO_B\ket{\phi_m}
&=\frac{1}{4p_m} 
\bra{\psi_m} (\rho \mathcal{L}_{O_A}\mathcal{L}_{O_B} + \mathcal{L}_{O_A} \rho  \mathcal{L}_{O_B}  
+ \mathcal{L}_{O_A} \mathcal{L}_{O_B}\rho +\rho \mathcal{L}_{O_A} \rho^{-1}\mathcal{L}_{O_B}\rho )
  \ket{\psi_m} \notag\\
&=\frac{1}{4p_m}  \bra{\psi_m} (\rho \mathcal{L}_{O_A}\mathcal{L}_{O_B} + \mathcal{L}_{O_A} \rho  \mathcal{L}_{O_B}  
+ \mathcal{L}_{O_A} \mathcal{L}_{O_B}\rho +\mathcal{L}_{O_B} \rho  \mathcal{L}_{O_A}  )
  \ket{\psi_m} \notag \\
&\quad   +  \frac{1}{4} \bra{\phi_m}  
\left[  \br{ \rho^{-1/2 }  \mathcal{L}_{O_A} \rho^{1/2} }, \br{\rho^{1/2}\mathcal{L}_{O_B} \rho^{-1/2 }} \right]
  \ket{\phi_m},
 \end{align}
Where, in the second equation, Eq.~\eqref{phi_m_expect_O_AO_B_3} is used as follows:
\begin{align}
\bra{\psi_m}  \rho \mathcal{L}_{O_A} \rho^{-1}  \mathcal{L}_{O_B} \rho  \ket{\psi_m}
&= \bra{\psi_m}  \rho^{1/2}  \br{ \rho^{-1/2 }  \mathcal{L}_{O_B} \rho^{1/2} } \br{\rho^{1/2}\mathcal{L}_{O_A} \rho^{-1/2 }} 
\rho^{1/2 }   \ket{\psi_m}  \notag \\
&\quad + \bra{\psi_m}  \rho^{1/2} 
\left[  \br{ \rho^{-1/2 }  \mathcal{L}_{O_A} \rho^{1/2} }, \br{\rho^{1/2}\mathcal{L}_{O_B} \rho^{-1/2 }} \right]
 \rho^{1/2 }   \ket{\psi_m}  \notag \\
 &=\bra{\psi_m} \mathcal{L}_{O_B} \rho \mathcal{L}_{O_A} \ket{\psi_m}   + p_m \bra{\phi_m}  
\left[  \br{ \rho^{-1/2 }  \mathcal{L}_{O_A} \rho^{1/2} }, \br{\rho^{1/2}\mathcal{L}_{O_B} \rho^{-1/2 }} \right]
  \ket{\phi_m}.
\end{align} 

Further, in Eq.~\eqref{phi_m_expect_O_AO_B_Lemma_2nd}, 
 \begin{align}
\bra{\psi_m} \rho \mathcal{L}_{O_A}\mathcal{L}_{O_B}\ket{\psi_m}  
&= \bra{\psi_m}\rho \mathcal{L}_{O_A}  (\tilde{\alpha}_{2,m}+ \delta \mathcal{L}_{O_B}) \ket{\psi_m}   \notag \\
&=\bra{\psi_m} \rho ( \tilde{\alpha}_{1,m} \tilde{\alpha}_{2,m}+ \delta \mathcal{L}_{O_A} \tilde{\alpha}_{2,m} +   \mathcal{L}_{O_A}  \delta \mathcal{L}_{O_B}) \ket{\psi_m}  \notag \\
&=\bra{\psi_m} \rho \tilde{\alpha}_{1,m} \tilde{\alpha}_{2,m}+ \rho \delta \mathcal{L}_{O_A} \mathcal{L}_{O_B} + \rho  \mathcal{L}_{O_A} \delta \mathcal{L}_{O_B} - \rho \delta  \mathcal{L}_{O_A} \delta \mathcal{L}_{O_B} \ket{\psi_m}.
\end{align}
In a similar manner,
 \begin{align}
&\bra{\psi_m} \mathcal{L}_{O_A} \rho  \mathcal{L}_{O_B} \ket{\psi_m}  
=\bra{\psi_m}   \tilde{\alpha}_{1,m}\rho \tilde{\alpha}_{2,m}+ \delta \mathcal{L}_{O_A}\rho \mathcal{L}_{O_B}
+  \mathcal{L}_{O_A}\rho\delta \mathcal{L}_{O_B} -  \delta\mathcal{L}_{O_A}\rho\delta \mathcal{L}_{O_B}
 \ket{\psi_m}, \notag \\
&\bra{\psi_m} \mathcal{L}_{O_A} \mathcal{L}_{O_B}\rho  \ket{\psi_m}  
=\bra{\psi_m}\tilde{\alpha}_{1,m} \tilde{\alpha}_{2,m} \rho 
+  \delta \mathcal{L}_{O_A}\mathcal{L}_{O_B}\rho
+ \mathcal{L}_{O_A} \delta \mathcal{L}_{O_B}\rho 
-  \delta \mathcal{L}_{O_A} \delta\mathcal{L}_{O_B}\rho
 \ket{\psi_m}, \notag \\
&\bra{\psi_m} \mathcal{L}_{O_B} \rho  \mathcal{L}_{O_A}    \ket{\psi_m}   
=\bra{\psi_m}   \tilde{\alpha}_{1,m} \rho \tilde{\alpha}_{2,m}+ \mathcal{L}_{O_B}\rho \delta \mathcal{L}_{O_A} 
+ \delta \mathcal{L}_{O_B}\rho \mathcal{L}_{O_A}
- \delta \mathcal{L}_{O_B}\rho \delta \mathcal{L}_{O_A} 
\ket{\psi_m}.
\end{align}
Using the above equations, Eq.~\eqref{phi_m_expect_O_AO_B_Lemma_2nd} is reduced to
\begin{align}
\bra{\phi_m}O_AO_B\ket{\phi_m}&= 
\tilde{\alpha}_{1,m} \tilde{\alpha}_{2,m}+
\frac{1}{4p_m}  \bra{\psi_m} 
( \rho  \mathcal{L}_{O_A} \delta\mathcal{L}_{O_B} +  \delta \mathcal{L}_{O_A}\rho \mathcal{L}_{O_B}
+\delta \mathcal{L}_{O_A} \mathcal{L}_{O_B}\rho+ \mathcal{L}_{O_B}\rho \delta\mathcal{L}_{O_A} )
  \ket{\psi_m} \notag \\
  &+
\frac{1}{4p_m}  \bra{\psi_m} 
( \rho \mathcal{L}_{O_A} \delta \mathcal{L}_{O_B} +   \mathcal{L}_{O_A}\rho \delta\mathcal{L}_{O_B}
+\mathcal{L}_{O_A} \delta \mathcal{L}_{O_B}\rho+\delta \mathcal{L}_{O_B}\rho \mathcal{L}_{O_A} )
  \ket{\psi_m} \notag \\
&  -
\frac{1}{4p_m}  \bra{\psi_m} 
( \rho \delta\mathcal{L}_{O_A} \delta \mathcal{L}_{O_B} +   \delta\mathcal{L}_{O_A}\rho \delta\mathcal{L}_{O_B}
+\delta\mathcal{L}_{O_A} \delta \mathcal{L}_{O_B}\rho+\delta \mathcal{L}_{O_B}\rho \delta \mathcal{L}_{O_A} )
  \ket{\psi_m} \notag \\
&\quad   +  \frac{1}{4} \bra{\phi_m}  
\left[  \br{ \rho^{-1/2 }  \mathcal{L}_{O_A} \rho^{1/2} }, \br{\rho^{1/2}\mathcal{L}_{O_B} \rho^{-1/2 }} \right]
  \ket{\phi_m},
\end{align}
where $\bra{\psi_m} \rho  \ket{\psi_m}=p_m$. 
Thus, 
\begin{align}
&\sum_m p_m | \bra{\phi_m}O_AO_B\ket{\phi_m} -  \tilde{\alpha}_{1,m} \tilde{\alpha}_{2,m}| \notag \\
&\le  \br{  \norm {\delta \mathcal{L}_{O_A}}\cdot \norm{\mathcal{L}_{O_B}}+ 
 \norm {\mathcal{L}_{O_A}}\cdot \norm{\delta \mathcal{L}_{O_B}}+\norm {\delta \mathcal{L}_{O_A}}\cdot \norm{\delta\mathcal{L}_{O_B}}  } 
 + \frac{1}{4} \norm{  
\left[  \br{ \rho^{-1/2 }  \mathcal{L}_{O_A} \rho^{1/2} }, \br{\rho^{1/2}\mathcal{L}_{O_B} \rho^{-1/2 }} \right]},
 \end{align}
where analyses similar to those for inequality~\eqref{sum_m_tilde_delta_1_m_le_2} are used.  
Using condition~\eqref{condition_L_01_LO2_L01O2} and $\norm {\mathcal{L}_{O_A}}\le \|O_A\|=1$, which is proven as inequality~\eqref{fupper_bound_of_mathcal_L_O} in Lemma~\ref{lem:quantum_correlation_3}, the inequality of 
\begin{align}
\label{prod_O_AO_Bex_phi_m_fin}
&\sum_m p_m | \bra{\phi_m}O_AO_B\ket{\phi_m} -  \tilde{\alpha}_{1,m} \tilde{\alpha}_{2,m}|  
\le  \delta_1+\delta_2 + \delta_1\delta_2
  + \frac{1}{4} \norm{  
\left[  \br{ \rho^{-1/2 }  \mathcal{L}_{O_A} \rho^{1/2} }, \br{\rho^{1/2}\mathcal{L}_{O_B} \rho^{-1/2 }} \right]}.
 \end{align}
is obtained. Further, by combining inequalities~\eqref{prod_O_Aex_O_Bex_phi_m_fin} and \eqref{prod_O_AO_Bex_phi_m_fin}, 
 \begin{align}
&\sum_m p_m  | \bra{\phi_m}O_AO_B\ket{\phi_m}  - \bra{\phi_m}O_A\ket{\phi_m} \bra{\phi_m}O_B\ket{\phi_m}|  \notag \\
=& \sum_m p_m  \left | \bra{\phi_m}O_AO_B\ket{\phi_m} - \tilde{\alpha}_{1,m}\tilde{\alpha}_{2,m}  - \bra{\phi_m}O_A\ket{\phi_m} \bra{\phi_m}O_B\ket{\phi_m}  + \tilde{\alpha}_{1,m}\tilde{\alpha}_{2,m} \right |   \notag \\
\le & \sum_m p_m  \br{ \left | \bra{\phi_m}O_AO_B\ket{\phi_m} - \tilde{\alpha}_{1,m}\tilde{\alpha}_{2,m} \right |  + p_m  \left |   \bra{\phi_m}O_A\ket{\phi_m} \bra{\phi_m}O_B\ket{\phi_m} - \tilde{\alpha}_{1,m}\tilde{\alpha}_{2,m} \right | }  \notag \\
\le &2\delta_1+2\delta_2 + \delta_1\delta_2
  + \frac{1}{4} \norm{  
\left[  \br{ \rho^{-1/2 }  \mathcal{L}_{O_A} \rho^{1/2} }, \br{\rho^{1/2}\mathcal{L}_{O_B} \rho^{-1/2 }} \right]}.
\end{align}
When $\delta_1\ge 1/2$ or $\delta_2\ge 1/2$, the upper bound is worse than the trivial bound $1$, and hence, the inequality is meaningful only for $\delta_1\le 1/2$ and $\delta_2\le 1/2$, which yields $\delta_1\delta_2\le \delta_1+\delta_2$. 
 Thus, by applying the above inequality to \eqref{QC_upper_bound_lemma_cond_2nd}, the main inequality~\eqref 
{main_inequality_lemma_2_QC} is proven.
This completes the proof.   $\square$
\end{widetext}


\subsection{Proof of Lemma~\ref{lem:quantum_correlation_3}}

First, the eigenvalues $\{\lambda_s\}$ and the eigenstates $\{\ket{\lambda_s}\}$ are rewritten as 
\begin{align}
\lambda_s = e^{-\beta E_s}, \quad \ket{\lambda_s}= \ket{E_s},
 \end{align}
 where $H \ket{E_s} = E_s \ket{E_s}$. 
 Then, for an arbitrary operator $O$, definition~\eqref{def_L_O_lambda} provides 
\begin{align}
\label{mathcal_O_express_1}
\mathcal{L}_O &=  \sum_{s,s'} \frac{2\sqrt{e^{-\beta (E_s-E_s')}}}{1 +e^{-\beta (E_s-E_s')} } 
 \bra{E_s} O \ket{E_{s'}} \ket{E_{s}}\bra{E_{s'}}  \notag \\
 &=\int_{-\infty}^\infty \frac{2\sqrt{e^{-\beta \omega}}}{1 +e^{-\beta \omega} } O_\omega d\omega,
 \end{align}
 where the notation of Eq.~\eqref{Fourier_op_def} is used. 
 
Using Eq.~\eqref{Fourier_op_def_2}, the above form is reduced to 
\begin{align}
\label{mathcal_O_express_2}
\mathcal{L}_O 
&=\int_{-\infty}^\infty \frac{2\sqrt{e^{-\beta \omega}}}{1 +e^{-\beta \omega} } \frac{1}{2\pi} \int_{-\infty}^\infty O(t) e^{-i\omega t} dt d\omega \notag \\
& =\int_{-\infty}^\infty f_{\beta} (t) O(t) dt,
 \end{align}
with 
\begin{align}
f_{\beta} (t) :=\frac{1}{2\pi} \int_{-\infty}^\infty \frac{2\sqrt{e^{-\beta \omega}}}{1 +e^{-\beta \omega} } e^{-i\omega t} d\omega.
 \end{align}
Further, by following the same analysis as in Sec.~\ref{Fourier_calculation/sec}, it can be proven that $f_\beta(t)$ is given by
\begin{align}
&f_\beta (t)  \notag \\
&= \begin{cases}
&\displaystyle  i \sum_{m=1}^\infty {\rm Res}_{\omega=(2\pi i m - i\pi )/\beta} \br{\frac{2\sqrt{e^{-\beta \omega}}}{1 +e^{-\beta \omega} } e^{-i\omega t} }   \\
&{\rm for} \quad t<0, \\
&\displaystyle  - i \sum_{m=1}^\infty {\rm Res}_{\omega=(-2\pi i m + i\pi )/\beta} \br{\frac{2\sqrt{e^{-\beta \omega}}}{1 +e^{-\beta \omega} } e^{-i\omega t} }   \\
&{\rm for} \quad  t\ge0,
\end{cases}  \notag \\
&= \begin{cases}
\displaystyle  - \sum_{m=1}^\infty\frac{2(-1)^m}{\beta} e^{\pi (2m-1) t/\beta }    &\for t<0, \\
\displaystyle  - \sum_{m=1}^\infty \frac{2(-1)^m}{\beta} e^{-\pi (2m-1) t/\beta }     &\for t\ge0,
\end{cases} \notag \\ 
&=\sum_{m=1}^\infty\frac{2(-1)^m}{\beta} e^{-\pi (2m-1) |t|/\beta } = \frac{1}{\beta \cosh(\pi |t| /\beta)}. \notag 
 \end{align}
This completes the proof of Eq.~\eqref{form_simple_mathcal_L_O}.

The proof of inequality~\eqref{fupper_bound_of_mathcal_L_O} is simply given as follows.
Owing to $f_\beta (t) \ge 0$, 
\begin{align}
\label{upper_bound_L_0_f_t_int}
\| \mathcal{L}_O \|
&\le \int_{-\infty}^\infty f_{\beta} (t) \| O(t)\| dt  \le \| O\| \int_{-\infty}^\infty f_{\beta} (t)  dt.
 \end{align}
Using the inverse Fourier transform  
\begin{align}
 \int_{-\infty}^\infty  f_{\beta} (t)  e^{i\omega t} dt =\frac{2\sqrt{e^{-\beta \omega}}}{1 +e^{-\beta \omega} }  
 \end{align}
with $\omega=0$, the inequality~\eqref{upper_bound_L_0_f_t_int} is reduced to \eqref{fupper_bound_of_mathcal_L_O}. 
This completes the proof. $\square$


\subsection{Proof of Lemma~\ref{lem:quantum_correlation_4}}
 \begin{figure}[tt]
\centering
\includegraphics[clip, scale=0.3]{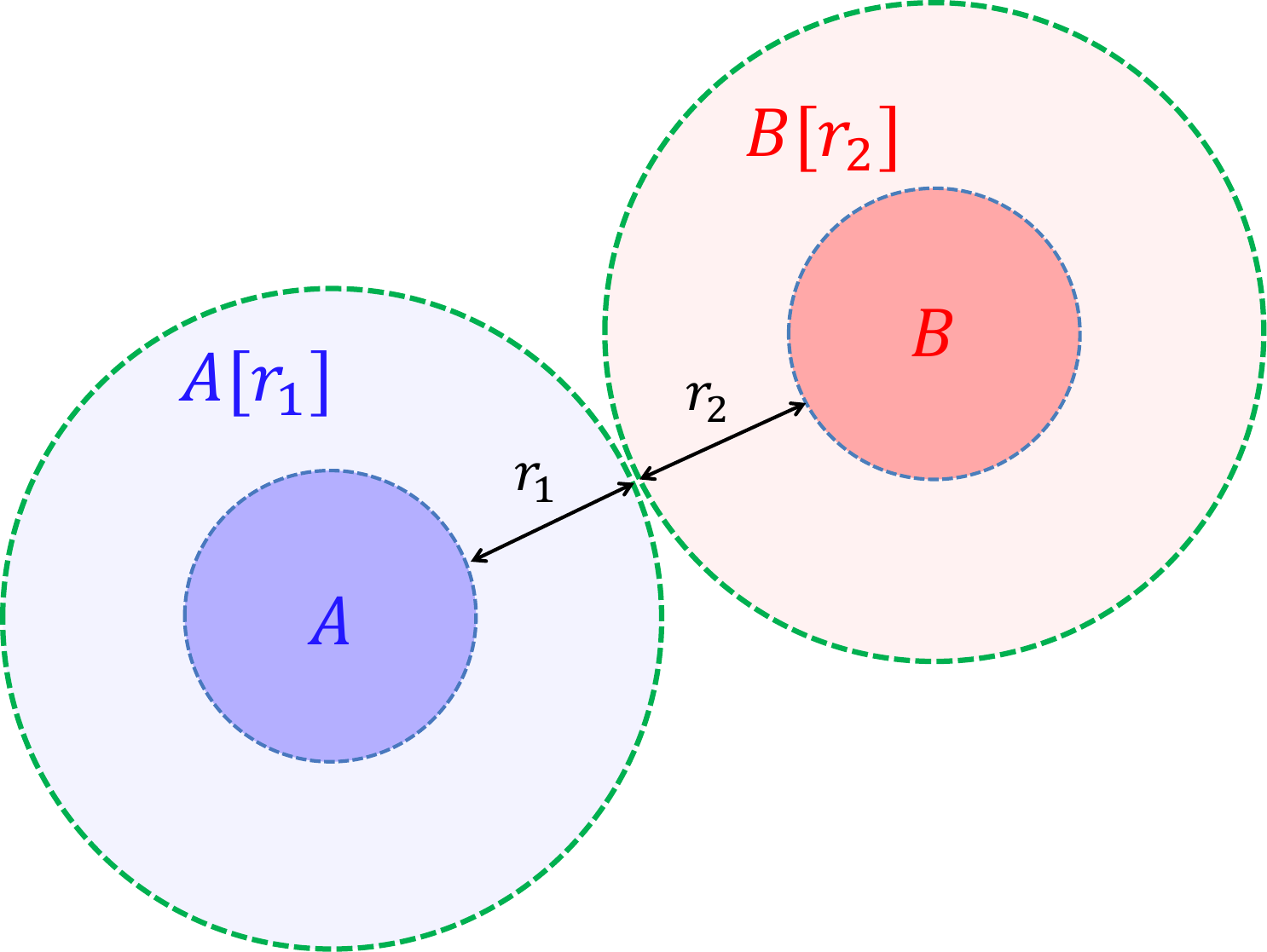}
\caption{Approximations of $\mathcal{L}_{O_A}$ and $\mathcal{L}_{O_B}$.
To obtain the approximations $\tilde{\mathcal{L}}_{O_A}$ and $\tilde{\mathcal{L}}_{O_B}$, which commute with each other, 
$\mathcal{L}_{O_A}$ and $\mathcal{L}_{O_B}$ are approximated onto the extended regions $A[r_1]$ and $B[r_2]$ ($r_1+r_2<R$), respectively.  In Eqs.~\eqref{tilde_mathcal_L_O_A} and \eqref{tilde_mathcal_L_O_B}, the explicit forms of $\mathcal{L}_{O_A}$ and $\mathcal{L}_{O_B}$ are presented.  
}
\label{fig:extended_region}
\end{figure}

First, consider the explicit construction of $\tilde{\mathcal{L}}_{O_A}$ and $\tilde{\mathcal{L}}_{O_B}$, such that $[\tilde{\mathcal{L}}_{O_A}, \tilde{\mathcal{L}}_{O_B}] =0 $. 
For this purpose, Eq.~\eqref{form_simple_mathcal_L_O} is used in Lemma~\ref{lem:quantum_correlation_3}, and 
the time-evolved operator $O_A(t) $ is approximated on $A[r_1]$ (see Fig.~\ref{fig:extended_region}), which yields  
\begin{align}
\label{tilde_mathcal_L_O_A}
\tilde{\mathcal{L}}_{O_A}  =\int_{-\infty}^\infty f_{\beta} (t) O_A(t,A[r_1]) dt,
 \end{align}
where the notation of $O_A(t,A[r_1])$ has been provided in Eq.~\eqref{supp_def:W_X_local_approx}, and $r_1$ is chosen appropriately. 
Note that $\tilde{\mathcal{L}}_{O_A}$ is now supported on the subset $A[r_1]$.
In the same manner, $\tilde{\mathcal{L}}_{O_A}$ is defined as 
\begin{align}
\label{tilde_mathcal_L_O_B}
\tilde{\mathcal{L}}_{O_B}  =\int_{-\infty}^\infty f_{\beta} (t) O_B(t,B[r_2]) dt.
 \end{align}
Thus, if we set $r_1 + r_2 < \dist_{A,B}=R$, $[\tilde{\mathcal{L}}_{O_A}, \tilde{\mathcal{L}}_{O_B}]=0$ is obtained.
Therefore, in the following discussions, $r_1=r_2=\lceil R/2 \rceil-1$ is chosen. 

Using Eq.~\eqref{tilde_mathcal_L_O_A}, $\delta_1$ can be estimated as 
\begin{align}
\delta_1  \le \int_{-\infty}^\infty f_{\beta} (t) \|  O_A(t) -  O_A(t,A[r_1]) \|  dt.
 \end{align}
 For the estimation of the integral, an approach similar to that in Sec.~\ref{Proof of the inequality_integrate_Lieb-Robinson_QC_WY} is used.
 First, 
 \begin{align}
\label{integrate_QC_L_1_all_t}
&\int_{-\infty}^\infty f_{\beta} (t) \|  O_A(t) -  O_A(t,A[r_1]) \|  dt  \notag \\
&= \int_{|t| > t_0}f_{\beta} (t) \|  O_A(t) -  O_A(t,A[r_1]) \|  dt  \notag \\
&+\int_{|t| \le t_0}f_{\beta} (t) \|  O_A(t) -  O_A(t,A[r_1]) \|  dt,
\end{align}
 where $t_0:=\mu r_1/(2v)$. 
Owing to
 \begin{align}
& f_{\beta} (t) = \frac{1}{\beta \cosh(\pi |t| /\beta) }  \le \frac{2}{\beta}  e^{-\pi |t| /\beta}, \notag \\
& \|  O_A(t) -  O_A(t,A[r_1]) \|  \le 2\|O_A\|=2,
\end{align}
the first term is upper-bounded as 
 \begin{align}
\label{integrate_QC_L_1_t_>_t_0}
&\int_{|t| > t_0}f_{\beta} (t) \|  O_A(t) -  O_A(t,A[r_1]) \|  dt  \notag \\
&\le \frac{4}{\beta} \int_{|t| > t_0}e^{-\pi |t| /\beta} dt \le \frac{8}{\pi}  e^{-\pi \mu r_1/(2v\beta)}.
\end{align}
The quantity $\|  O_A(t) -  O_A(t,A[r_1]) \|$ is upper-bounded using the Lieb-Robinson bound~\eqref{Lieb-Robinson_main_short_region},
and hence, the second term is upper-bounded as 
\begin{align}
\label{integrate_QC_L_1_t_le_t_0}
&\int_{|t| \le t_0}f_{\beta} (t) \|  O_A(t) -  O_A(t,A[r_1]) \|  dt  \notag \\
&\le \frac{2}{\beta}  \int_{|t| \le t_0} e^{-\pi  |t| /\beta}  C |\partial A| \br{e^{v|t|}-1} e^{-\mu r_1} dt \notag\\
&\le \frac{4C}{\beta}  |\partial A|  \int_{0}^{t_0}  e^{v|t|} e^{-\mu r_1} dt \notag\\
&\le \frac{4C}{v\beta}   |\partial A|  e^{-\mu r_1 + vt_0} =  \frac{4C}{v\beta}   |\partial A|  e^{-\mu r_1/2}.
\end{align}
Further, applying inequalities~\eqref{integrate_QC_L_1_t_>_t_0} and \eqref{integrate_QC_L_1_t_le_t_0}, 
Eq.~\eqref{integrate_QC_L_1_all_t} is reduced to 
 \begin{align}
\delta_1 & \le \int_{-\infty}^\infty f_{\beta} (t) \|  O_A(t) -  O_A(t,A[r_1]) \|  dt  \notag \\
&\le \br{ \frac{8}{\pi} + \frac{4C}{v\beta}  }  |\partial A|  e^{-\min[\mu r_1/2, \pi \mu r_1/(2v\beta)]}  \notag \\
&\le \br{ \frac{8}{\pi} +\frac{4C}{v\beta}  }  |\partial A|  e^{- \mu r_1/(2+2v\beta/\pi)},
\end{align}
where $|\partial A|\ge 1$ is used in the second inequality. 
In the same manner, 
 \begin{align}
\delta_2&\le \int_{-\infty}^\infty f_{\beta} (t) \|  O_B(t) -  O_B(t,B[r_2]) \|  dt  \notag \\
&\le \br{ \frac{8}{\pi} + \frac{4C}{v\beta}  }  |\partial B| e^{- \mu r_2/(2+2v\beta/\pi)}. 
\end{align}
Thus, applying $r_1=r_2=\lceil R/2\rceil -1 $, inequality~\eqref{lemma_upperbound_delta_1_delta_2} is proven. 
This completes the proof. $\square$


\subsection{Proof of Lemma~\ref{lem:quantum_correlation_5}}

First, consider the integral expression of $\rho^{\pm 1/2 }  \mathcal{L}_{O} \rho^{\mp 1/2}$ for an arbitrary operator $O$. 
Using 
\begin{align}
e^{\pm \beta H/2} O_{\omega}e^{\mp \beta H/2}=e^{\pm \beta \omega/2}  O_{\omega},
 \end{align}
based on Eq.~\eqref{mathcal_O_express_1}, we obtain 
\begin{align}
\rho^{\pm 1/2 }  \mathcal{L}_{O} \rho^{\mp 1/2}
 =\int_{-\infty}^\infty \frac{2\sqrt{e^{-\beta \omega}}}{1 +e^{-\beta \omega} } e^{\pm \beta \omega/2} O_{\omega} d\omega.
 \end{align}
Using Eq.~\eqref{Fourier_op_def_2}, the above equation is reduced to 
\begin{align}
\label{mathcal_O_imaginary_express_2}
&\rho^{\pm 1/2 }  \mathcal{L}_{O} \rho^{\mp 1/2}  \notag \\
&=\int_{-\infty}^\infty \frac{2\sqrt{e^{-\beta\omega}}}{1 +e^{-\beta \omega} } e^{\pm \beta \omega/2} \frac{1}{2\pi} \int_{-\infty}^\infty O( t) e^{-i\omega t} dt d\omega   \notag \\
&=\int_{-\infty}^\infty g_{\beta,\pm} (t) O(t) dt,
 \end{align}
where $g_\beta(t)$ is defined as 
\begin{align}
g_{\beta,\pm } (t)= \frac{1}{2\pi} \int_{-\infty}^\infty \frac{2\sqrt{e^{-\beta\omega}}}{1 +e^{-\beta \omega} } e^{\pm \beta \omega/2}  e^{-i\omega t} d\omega. 
 \end{align}
Further, 
 \begin{align}
g_{\beta,\pm } (t)
&=\frac{1}{2\pi} \int_{-\infty}^\infty \left[ \pm \tanh(\beta \omega/2) + 1 \right]e^{-i\omega t}  d\omega  \notag \\
&= \delta(t)  \pm g_\beta(t),
 \end{align}
where $\delta(t)$ is the delta function and $g_\beta(t)$ is the Fourier transform of $\tanh(\beta \omega/2)$. 
 
 \begin{widetext}
As in Sec.~\ref{Fourier_calculation/sec}, herein, 
\begin{align}
g_\beta(t) &= \begin{cases}
\displaystyle  i \sum_{m=1}^\infty {\rm Res}_{\omega=(2\pi i m - i\pi )/\beta} \br{\tanh(\beta \omega/2)  e^{-i\omega t} }   &\for t<0, \\
\displaystyle  - i \sum_{m=1}^\infty {\rm Res}_{\omega=(-2\pi i m + i\pi )/\beta} \br{\tanh(\beta \omega/2) e^{-i\omega t} }   &\for t\ge0,
\end{cases}  \notag \\
&= \begin{cases}
\displaystyle  i\sum_{m=1}^\infty\frac{2}{\beta} e^{\pi (2m-1) t/\beta }    &\for t<0, \\
\displaystyle  -i \sum_{m=1}^\infty \frac{2}{\beta} e^{-\pi (2m-1) t/\beta }     &\for t\ge0,
\end{cases} \notag \\ 
&=\frac{-2 i }{\beta}{\rm sign}(t)\sum_{m=1}^\infty e^{-\pi (2m-1) |t|/\beta } =-i \frac{{\rm sign}(t)}{\beta \sinh(\pi |t| /\beta)} 
= \frac{-i}{\beta \sinh(\pi t /\beta)}.
 \end{align}
Consequently,
\begin{align}
\label{mathcal_O_imaginary_express_fin}
\rho^{\pm 1/2 }  \mathcal{L}_{O} \rho^{\mp 1/2} 
&=O \pm \int_{-\infty}^\infty g_{\beta} (t) O(t) dt. 
 \end{align}

For the proof of the lemma, the following two claims must be proven:
\begin{claim} \label{claim_L_O_imag_norm}
Let $O$ be an arbitrary operator supported on a subset $X\subset \Lambda$.
Then, the norm of $\rho^{\pm 1/2 }  \mathcal{L}_{O} \rho^{\mp 1/2}$ is upper-bounded as 
\begin{align}
\label{ineq:claim_L_O_imag_norm}
\norm{ \rho^{\pm 1/2 }  \mathcal{L}_{O} \rho^{\mp 1/2} } 
&\le  \|O\|  \log \br{1+\frac{\beta \| \ad_H(O)\|}{\|O\|}  } 
+ 2\|O\|.
 \end{align}
\end{claim}
\begin{claim} \label{claim_L_O1_O2_commutator_norm}
Let $O$ be the operator defined in Claim~\ref{claim_L_O_imag_norm}.
Then, for $\|O\|=1$, the operator $\rho^{\pm 1/2 }  \mathcal{L}_{O} \rho^{\mp 1/2}$ is approximated on $X[r]$ with an error of 
\begin{align}
&\norm{ \rho^{\pm 1/2 }  \mathcal{L}_{O} \rho^{\mp 1/2}- \br{\rho^{\pm 1/2 }  \mathcal{L}_{O} \rho^{\mp 1/2}}_{X[r]} } \le  |\partial X| \brr{\frac{8}{\pi} \br{1+ \frac{\xi_\beta}{2r}} 
+
4C \br{\frac{1}{\pi}+\frac{1}{v\beta}}
 }    e^{-2r /\xi_\beta},
 \end{align}
 where $\br{ \rho^{\pm 1/2 }  \mathcal{L}_{O} \rho^{\mp 1/2}}_{X[r]}$ is supported on $X[r]$ and chosen appropriately.  
\end{claim}

Using these claims, an upper-bound for the norm of~\eqref{commutator_imaginary_time_final} can be provided.
Let us approximate 
\begin{align}
&\mathfrak{O}_1:=\rho^{-1/2 }  \mathcal{L}_{O_A} \rho^{1/2} \approx \mathfrak{O}_{1,A[r_1]}, \notag \\
&\mathfrak{O}_2:=\rho^{1/2 }  \mathcal{L}_{O_B} \rho^{-1/2} \approx \mathfrak{O}_{2,B[r_2]},
 \end{align}
where $r_1 + r_2 < R$. 
Then, from $[\mathfrak{O}_{1,A[r_1]},\mathfrak{O}_{2,B[r_2]}]=0$, 
 \begin{align}
\norm{ [\mathfrak{O}_1,\mathfrak{O}_2]  } 
&= 
 \norm{ [\mathfrak{O}_1-\mathfrak{O}_{1,A[r_1]}, \mathfrak{O}_2]  + [\mathfrak{O}_{1,A[r_1]}, \mathfrak{O}_2-\mathfrak{O}_{2,B[r_2]}] } \notag \\
 &\le  2\norm{ \delta \mathfrak{O}_1} \cdot  \norm{ \mathfrak{O}_2} 
+ 2\norm{ \delta \mathfrak{O}_2 } \cdot  \norm{ \mathfrak{O}_1} 
   +2\norm{ \delta \mathfrak{O}_1} \cdot  \norm{ \delta \mathfrak{O}_2 }, 
 \end{align}
where $\delta \mathfrak{O}_1:= \mathfrak{O}_1-\mathfrak{O}_{1,A[r_1]}$ 
and $\delta \mathfrak{O}_2:= \mathfrak{O}_2-\mathfrak{O}_{2,B[r_2]}$ are defined, and 
$\norm{ \mathfrak{O}_{1,A[r_1]} } \le \norm{ \mathfrak{O}_1} + \norm{\delta \mathfrak{O}_1}$ 
is used in the inequality. 
For $\norm{ \delta \mathfrak{O}_s}> \norm{\mathfrak{O}_s}$ ($s=1,2$), the above inequality is worse than the trivial inequality, that is, $\norm{ [\mathfrak{O}_1,\mathfrak{O}_2]  } \le  2 \norm{\mathfrak{O}_1} \cdot  \norm{ \mathfrak{O}_2} $.
Hence, only $\norm{ \delta \mathfrak{O}_s}\le \norm{\mathfrak{O}_s}$ is considered, which yields 
 \begin{align}
\norm{ [\mathfrak{O}_1,\mathfrak{O}_2]  } 
\le  3 \br{ \norm{ \delta \mathfrak{O}_1} \cdot  \norm{ \mathfrak{O}_2} 
+ \norm{ \delta \mathfrak{O}_2 } \cdot  \norm{ \mathfrak{O}_1} }. 
 \end{align}
By choosing $r_1=r_2=\lceil R/2 \rceil-1$ and applying Claims~\ref{claim_L_O_imag_norm} and \ref{claim_L_O1_O2_commutator_norm}, the main inequality~\eqref{upper_bound_commutator_imaginary_time_final} is obtained as follows:
 \begin{align}
\norm{ [\mathfrak{O}_1,\mathfrak{O}_2]  } 
\le  &3 e^{2/\xi_\beta}  \brr{\frac{8}{\pi} \br{1+ \frac{\xi_\beta}{R-2}} 
+
4C\br{\frac{1}{\pi}+\frac{1}{v\beta}}
 }     e^{-R/\xi_\beta} \notag \\
&\times \bigl \{ |\partial A| \brr{2+ \log (1+\beta \| \ad_H(O_B)\| ) }
+|\partial B| \brr{2+ \log (1+\beta \| \ad_H(O_A)\| ) } \bigr\},
 \end{align}
 where $\|O_A\|=\|O_B\|=1$.
This completes the proof of Lemma~\ref{lem:quantum_correlation_5}.

\subsubsection{Proof of Claim~\ref{claim_L_O_imag_norm}}
From the integral expression~\eqref{mathcal_O_imaginary_express_fin},
\begin{align}
\label{start_ineq_0}
\norm{\rho^{\pm 1/2 }  \mathcal{L}_{O} \rho^{\mp 1/2} } 
&\le \|O\| +  \norm{\int_{-\infty}^\infty g_{\beta} (t) O(t) dt}. 
 \end{align}
In a standard approach, the following is used 
\begin{align}
\norm{\int_{-\infty}^\infty g_{\beta} (t) O(t) dt} \le \|O\| \int_{-\infty}^\infty | g_{\beta} (t)| dt.
 \end{align}
However, the integral of $|g_\beta(t)|$ does not converge because $|g_\beta(t)| \propto 1/t$ for $t\ll 1$. 

Thus, to obtain a refined bound, $O(t)$ is parameterized as $O(\lambda t)$ using the parameter $\lambda$.
Subsequently, 
\begin{align}
O(t)& =O+ \int_0^1 \frac{d}{d\lambda}  O(\lambda t) d\lambda
=O+i t  \int_0^1 \ad_H(O) (\lambda t) d\lambda,
 \end{align}
which yields
\begin{align}
\norm{\int_{-\infty}^\infty g_{\beta} (t) O(t) dt} 
&\le
\norm{\int_{|t| > \delta t} g_{\beta} (t) O(t) dt} 
+ \norm{\int_{|t| \le \delta t} g_{\beta} (t)O   dt  +\int_{|t| \le \delta t}  i t  \int_0^1g_{\beta} (t)  \ad_H(O) (\lambda t) d\lambda  dt}  \notag \\
&\le 
2\|O\| \int_{t > \delta t} \frac{1}{\beta \sinh(\pi  t /\beta)}   dt 
+ 2 \| \ad_H(O)\|  \int_0^{\delta t }   \frac{t}{\beta \sinh(\pi  t /\beta)}   dt  \notag \\
&\le \frac{-2\|O\|  }{\pi} \log\brr{ \tanh \br{\frac{\pi \delta t}{2\beta} }} 
+ \frac{2\| \ad_H(O)\| }{\pi} \delta t \le  \frac{2\|O\|  }{\pi} \log \br{1+\frac{2\beta}{\pi \delta t} } 
+ \frac{2\| \ad_H(O)\| }{\pi} \delta t,
\end{align}
where $\int_{|t| \le \delta t} g_{\beta} (t) dt=0$, $1/\sinh(x) \le 1/x$, and $-\log[\tanh(x)] \le \log(1+1/x)$ are used in the second, third, and fourth inequalities, respectively. Note that $g_{\beta} (-t)=-g_{\beta} (t)$. 
Thus, by choosing $\delta t= \|O\|/\| \ad_H(O)\|$, 
\begin{align}
\label{end_inequality_claim_0}
\norm{\int_{-\infty}^\infty g_{\beta} (t) O(t) dt} 
&\le  \frac{2\|O\|  }{\pi} \log \br{1+\frac{2\beta \| \ad_H(O)\|}{\pi \|O\|}  } 
+ \frac{2\|O\| }{\pi}.
\end{align}
Therefore, by combining inequalities~\eqref{start_ineq_0} and \eqref{end_inequality_claim_0} with $2/\pi \le 1$, inequality~\eqref{ineq:claim_L_O_imag_norm} is proven.
$\square$

\subsubsection{Proof of Claim~\ref{claim_L_O1_O2_commutator_norm}}

As in the proof of Lemma~\ref{lem:quantum_correlation_4}, we consider a similar approximation to the one in Eq.~\eqref{tilde_mathcal_L_O_A}. 
Using the integral expression~\eqref{mathcal_O_imaginary_express_fin}, we obtain 
\begin{align}
\br{\rho^{\pm 1/2 }  \mathcal{L}_{O} \rho^{\mp 1/2}}_{X[r]} 
:=O \pm \int_{-\infty}^\infty g_{\beta} (t) O(t, X[r]) dt, 
\end{align}
which yields 
\begin{align}
\norm{ \rho^{\pm 1/2 }  \mathcal{L}_{O} \rho^{\mp 1/2}- \br{\rho^{\pm 1/2 }  \mathcal{L}_{O} \rho^{\mp 1/2}}_{X[r]} }
\le  \int_{-\infty}^\infty | g_{\beta} (t)| \cdot \norm{ O(t, X[r]) - O(t) } dt.  
 \end{align}
Using $1/\sinh(x) \le 2 e^{-x}(1+1/x)$ ($x\ge 0$),
\begin{align}
|g_\beta(t) | = \frac{1}{\beta \sinh(\pi |t| /\beta)} 
\le \frac{2e^{-\pi |t| /\beta}}{\beta} \br{1+ \frac{1}{\pi |t|/\beta}}.
 \end{align}
In addition, as per the Lieb-Robinson bound~\eqref{Lieb-Robinson_main_short_region}, 
\begin{align}
\label{Lieb-Robinson_main_short_region_claim_proof_com}
\norm{ O(t)- O_X(t,X[r])} \le   \min \br{ C |\partial X| \br{e^{v|t|}-1} e^{-\mu r}, 2 }.
\end{align}

Subsequently, analyses similar to those for \eqref{integrate_QC_WY_t_>_t_0}, \eqref{integrate_QC_WY_t_le_t_0}, and  
\eqref{integrate_QC_WY_t_le_t_0_2} can be applied.
For $t_0= \mu r/(2v)$, we obtain
\begin{align}
&\int_{-\infty}^\infty | g_{\beta} (t)| \cdot \norm{ O(t, X[r]) - O(t) } dt  \notag \\
&\le  \int_{|t| > t_0} \frac{2e^{-\pi |t| /\beta}}{\beta} \br{1+ \frac{1}{\pi |t|/\beta}}  \cdot 2 dt  + 
\int_{|t| \le t_0} \frac{2e^{-\pi |t| /\beta}}{\beta} \br{1+ \frac{1}{\pi |t|/\beta}}  \cdot C |\partial X| \br{e^{v|t|}-1} e^{-\mu r} dt  \notag \\
&\le \frac{8e^{-\pi t_0 /\beta}}{\pi} \br{1+ \frac{1}{\pi t_0/\beta}}   
+ \frac{4C}{\beta}  |\partial X|   \br{\frac{1}{v}+ \frac{1}{\pi/\beta}}  e^{-\mu r+vt_0}  \notag \\
&\le |\partial X| \brr{\frac{8}{\pi} \br{1+ \frac{2v\beta}{\pi \mu r }}  e^{-\pi \mu r /(2v\beta)}
+
4C  \br{\frac{1}{\pi}+\frac{1}{v\beta}}  e^{-\mu r/2}
 } \le  |\partial X| \brr{\frac{8}{\pi} \br{1+ \frac{\xi_\beta}{2r}} 
+
4C \br{\frac{1}{\pi}+\frac{1}{v\beta}}
 }    e^{-2r /\xi_\beta}, \notag 
 \end{align}
 where the definition of $\xi_\beta:=4/\mu(1+v\beta/\pi)$ is used in the last inequality.
This completes the proof of Claim~\ref{claim_L_O1_O2_commutator_norm}. $\square$

\end{widetext}

\section{Proof of Proposition~\ref{prop:quantum_correlation_negativity}}
\label{App_Proof of Proposition_prop:quantum_correlation_negativity}

Herein, the proof of Proposition~\ref{prop:quantum_correlation_negativity}, which connects the PPT relative entanglement and quantum correlation, is presented.
When the quantum correlation satisfies 
\begin{align}
\label{cond_M_Quantum_corr_app}
\QC_{\rho_{AB}}(O_A,O_B)\le\epsilon\|O_A\|\cdot\|O_B\|
\end{align}
for two arbitrary operators $O_A$ and $O_B$, Proposition~\ref{prop:quantum_correlation_negativity} yields 
\begin{align}
 \label{app_ineq:prop:quantum_correlation__}
&E_{R}^{\PPT}(\rho_{AB})\le 4\mathcal{D}_{AB} \bar{\delta} \log(1/\bar{\delta}) 
\le 4\mathcal{D}_{AB} \bar{\delta}^{1/2}   , 
\end{align}
where $\bar{\delta} := 4\epsilon \min(\mathcal{D}_{A},\mathcal{D}_{B})$.

\subsection{Proof}

In inequality~\eqref{app_ineq:prop:quantum_correlation__}, if $\bar{\delta}> 1/\mathcal{D}_{AB}$, the upper bound is worse than the trivial bound, i.e., $E_{R}^{\PPT}(\rho_{AB})\le \log[\min(\mathcal{D}_{A},\mathcal{D}_{B})] \le (1/2)\log(\mathcal{D}_{AB})$. Hence, only the case of $\bar{\delta}\le 1/\mathcal{D}_{AB}$ is considered.

The eigenstates of $\rho^{T_A}_{AB}$ with negative eigenvalues are defined as $\{ \ket{\eta_i} \}_{i=1}^{M_0}$. 
Then, the proof of Proposition~\ref{prop:quantum_correlation_negativity} is immediately obtained via the following lemma:

\begin{lemma}\label{lemma_Min_eigenvalue_T_A}
For the quantum state $\rho_{AB}$ given in Prop.~\ref{prop:quantum_correlation_negativity}, the minimum negative eigenvalue of $\rho_{AB}^{T_A}$ satisfies 
\begin{align}
\label{lemma_ineq_Min_eigenvalue_T_A}
\delta := - \min_{i\in[M_0]}  \bra{\eta_i} \rho_{AB}^{T_A}\ket{\eta_i}\le 4\epsilon \min(\mathcal{D}_{A},\mathcal{D}_{B})=\bar{\delta},
\end{align}
where the parameter $\epsilon$ has been defined in \eqref{cond_M_Quantum_corr_app}. 
\end{lemma}

To prove inequality~\eqref{ineq:prop:quantum_correlation_negativity}, a quantum state $\tilde{\sigma}_{AB}$ is defined as follows:
\begin{align}
\label{tilde_sigma_AB_def}
\tilde{\sigma}_{AB} =
(1- \mathcal{D}_{AB} \bar{\delta} ) \rho_{AB} + \bar{\delta} \cdot \hat{1}_{AB}, 
\end{align}
where $\tr(\tilde{\sigma}_{AB})=1$ since $\tr(\bar{\delta} \cdot \hat{1}_{AB}) = \mathcal{D}_{AB} \bar{\delta}$.  
Because of the definition of $\delta $ in \eqref{lemma_ineq_Min_eigenvalue_T_A}, we have $\tilde{\sigma}_{AB}^{T_A} \succeq 0$ (i.e., $\tilde{\sigma}_{AB} \in \PPT$). 
We then obtain 
\begin{align}
\label{upp_AB_PPT_E_proof}
E_{R}^{\PPT}(\rho_{AB}) \le S(\rho_{AB}||\tilde{\sigma}_{AB} ).
\end{align}
Subsequently, using the continuity bound on the relative entropy~\cite[Theorem~3]{doi:10.1063/1.2044667} (or Ref.~\cite{DONALD1999257}), 
\begin{align}
\label{rho_AB_tilde_sigma_AB}
S(\rho_{AB}||\tilde{\sigma}_{AB} ) \le & \delta_{AB} \log(\mathcal{D}_{AB}) -\delta_{AB} \log (\delta_{AB}) \notag \\
&-\delta_{AB} \log [\lambda_{\min} (\tilde{\sigma}_{AB})] 
\end{align}
under the assumption of $\delta_{AB} \le 1/e$, 
where $\delta_{AB}:= \|\rho_{AB}-\tilde{\sigma}_{AB}\|_1$ and $\lambda_{\min} (\tilde{\sigma}_{AB})$ are defined as the minimum eigenvalues of $\tilde{\sigma}_{AB}$. 
Based on definition~\eqref{tilde_sigma_AB_def}, $\lambda_{\min} (\tilde{\sigma}_{AB}) \ge \bar{\delta}$ and 
\begin{align}
\delta_{AB} \le 2 \mathcal{D}_{AB} \bar{\delta}.
\end{align}
First, the case of $2 \mathcal{D}_{AB} \bar{\delta} \le 1/e$, that is, $\bar{\delta}\le 1/(2e\mathcal{D}_{AB})$, is considered. 
Then, $-\delta_{AB} \log (\delta_{AB}) \le - 2 \mathcal{D}_{AB} \bar{\delta} \log(2 \mathcal{D}_{AB} \bar{\delta})$, and hence, inequality~\eqref{rho_AB_tilde_sigma_AB} reduces to
\begin{align}
\label{upp_AB_PPT_E_proof_fin}
S(\rho_{AB}||\tilde{\sigma}_{AB} )& \le -2 \mathcal{D}_{AB} \bar{\delta} \log(2\bar{\delta}^2) \notag \\
&\le -4\mathcal{D}_{AB} \bar{\delta} \log(\bar{\delta}).
\end{align}
In the case of $\bar{\delta}> 1/(2e\mathcal{D}_{AB})$, the RHS of the above inequality is larger than the trivial upper bound $(1/2)\log(\mathcal{D}_{AB})$. 
Therefore, by combining inequality~\eqref{upp_AB_PPT_E_proof_fin} with~\eqref{upp_AB_PPT_E_proof}, the main inequality~\eqref{ineq:prop:quantum_correlation_negativity} is proven. 
This completes the proof. $\square$

\subsection{Proof of Lemma~\ref{lemma_Min_eigenvalue_T_A}}
 
The next task is to estimate 
 \begin{align}
 \min_{i} \bra{\eta_i} \rho_{AB}^{T_A}\ket{\eta_i} =  \inf_{\ket{\eta}} \tr \br{ \rho_{AB}^{T_A} P_{\eta}}
 \end{align}
under the assumption of~\eqref{cond_M_Quantum_corr_app}, where $P_{\eta} :=\ket{\eta} \bra{\eta}$. 
Therefore, first,
 \begin{align}
\tr \br{ \rho_{AB}^{T_A} P_{\eta} } 
&= \tr \br{ \rho_{AB} P_{\eta}^{T_A}}  \notag\\
&= \tr \br{ \rho_{AB} P_{\eta}}  + \tr \brr{ \rho_{AB} ( P_{\eta}^{T_A} - P_{\eta}) }, \notag 
 \end{align}
is rewritten, and the second term is subsequently proven to be approximately equal to zero for an arbitrary quantum state $\ket{\eta}$. 
Because the eigenvalues of $\rho_{AB}^{T_A}$ do not depend on the choice of basis~\cite{PhysRevLett.77.1413},   
the basis that yields the Schmidt decomposition of  $\ket{\eta}$ is selected, as follows:
 \begin{align}
 \label{Schmidt_eta}
\ket{\eta} = \sum_{s=1}^{\mathcal{D}_A} \nu_s \ket{s_A, s_B}, \quad \sum_s |\nu_s|^2=1,
  \end{align}
where we assume $\mathcal{D}_A < \mathcal{D}_B$ without a loss of generality. 

To verify this point, we first consider the qubit case, that is, $\mathcal{D}_A =\mathcal{D}_B=2$. 
Consider the proof of the following lemma:
\begin{lemma}\label{two_qubits_cases_qc}
When $\mathcal{D}_A =\mathcal{D}_B=2$, we have 
\begin{align}
\label{upper_bound_qubits}
\left | \tr \brr{ \rho_{AB} ( P_{\eta}^{T_A} - P_{\eta}) } \right | \le 2 \epsilon,
\end{align}
where the parameter $\epsilon$ is given in \eqref{cond_M_Quantum_corr_app}. 
\end{lemma}

To generalize the results of two qubits to two qudit systems, consider 
 \begin{align}
&P_{\eta}^{T_A} - P_{\eta}  \notag \\
&= \sum_{s,s': s\neq s'}^{\mathcal{D}_A} \nu_s\nu_{s'} (- \ket{s_A, s_B}\bra{s'_A, s'_B} +  \ket{s'_A, s_B}\bra{s_A, s'_B} )  \notag 
  \end{align}
and 
 \begin{align}
 &\nu_s\nu_{s'} (- \ket{s_A, s_B}\bra{s'_A, s'_B} +  \ket{s'_A, s_B}\bra{s_A, s'_B} )  + {\rm h.c.} \notag \\
 &=(\nu_s^2+\nu_{s'}^2) \br{\ket{\eta_{s,s'}} \bra{\eta_{s,s'}}^{T_A} -\ket{\eta_{s,s'}} \bra{\eta_{s,s'}}   },
\end{align}
where $\ket{\eta_{s,s'}}:= (\nu_s^2+\nu_{s'}^2)^{-1/2} ( \nu_s \ket{s_A, s_B} + \nu_{s'} \ket{s'_A, s'_B} )$. 
Now, the quantum state $\ket{\eta_{s,s'}}$ is reduced to a quantum state with two qubits. 
Thus, from Lemma~\ref{two_qubits_cases_qc}
\begin{align}
\abs{\tr \brr{ \rho_{AB} \br{\ket{\eta_{s,s'}} \bra{\eta_{s,s'}}^{T_A} -\ket{\eta_{s,s'}} \bra{\eta_{s,s'}}   }  } } \le 2 \epsilon,
\end{align}
which yields 
\begin{align}
&\left |  \tr \brr{ \rho_{AB} ( P_{\eta}^{T_A} - P_{\eta}) }  \right |  \notag\\ 
& =  \sum_{1\le s<s'\le \mathcal{D}_A} 
(\nu_s^2+\nu_{s'}^2)  \notag \\
&\quad \quad \times \abs{ \tr \brr{ \rho_{AB} \br{\ket{\eta_{s,s'}} \bra{\eta_{s,s'}}^{T_A} -\ket{\eta_{s,s'}} \bra{\eta_{s,s'}}   }  } }  \notag \\
&\le 2\epsilon  \sum_{1\le s<s'\le \mathcal{D}_A}  (\nu_s^2+\nu_{s'}^2)   \le 4\epsilon \mathcal{D}_A.
\end{align}

Consequently, 
\begin{align}
\tr \br{ \rho_{AB} P_{\eta}^{T_A} }  \ge \tr \br{ \rho_{AB} P_{\eta} } -4\epsilon \mathcal{D}_A 
\ge -4\epsilon \mathcal{D}_A,
\end{align}
where $\tr \br{ \rho_{AB} P_{\eta} }\ge 0$ is used in the second inequality.
Further, using the above inequality, 
 \begin{align}
\inf_{\ket{\eta}} \tr \br{ \rho_{AB}^{T_A} P_{\eta}} \ge  -4\epsilon \mathcal{D}_A.
 \end{align}
When $\mathcal{D}_B\le \mathcal{D}_A$, the above lower bound is replaced by $\inf_{\ket{\eta}} \tr \br{ \rho_{AB}^{T_A} P_{\eta}} \ge  -4\epsilon \mathcal{D}_B$.
Therefore, the parameter $\delta$ ($= - \min_i  \bra{\eta_i} \rho_{AB}^{T_A}\ket{\eta_i}$) is upper-bounded by
  \begin{align}
\delta \le 4\epsilon \min( \mathcal{D}_A, \mathcal{D}_B).
 \end{align}
Using this, inequality~\eqref{negativity_upper_bound_start} is reduced to the main inequality~\eqref{ineq:prop:quantum_correlation_negativity}.
This completes the proof. $\square$

\subsubsection{Proof of Lemma~\ref{two_qubits_cases_qc}}

When $\mathcal{D}_A =\mathcal{D}_B=2$, an arbitrary operator $O_{AB}$ is described in the form of 
 \begin{align}
O_{AB} = \sum_{P=x,y,z} ( J_P \hat{\sigma}_{1,P} \hat{\sigma}_{2,P} + h_{1,P} \hat{\sigma}_{1,P} + h_{2,P} \hat{\sigma}_{2,P} )
 \end{align}
by appropriately choosing the bases (see Ref.~\cite[Lemma~1]{PhysRevA.83.062311} for example), where $A=\{1\}$ and $B=\{2\}$ and $\{\hat{\sigma}_x,\hat{\sigma}_y,\hat{\sigma}_z\}$ are the Pauli matrices. 
Then, the partial transpose $T_A$ only changes $\hat{\sigma}_{1,y} \to -\hat{\sigma}_{1,y}$, and hence, 
 \begin{align}
O_{AB} - O_{AB}^{T_A} 
&=2( J_y \hat{\sigma}_{1,y} \hat{\sigma}_{2,y} + h_{1,y} \hat{\sigma}_{1,y} ) \notag \\
&= 2 \hat{\sigma}_{1,y} \otimes  ( J_y  \hat{\sigma}_{2,y} + h_{1,y} ).
 \end{align}
In this manner, the following can be expressed: 
\begin{align}
P_{\eta}^{T_A} - P_{\eta} = \Phi_{A} \otimes \Phi_{B}, 
\end{align}
where $\|\Phi_{A}\| \le 2$ and $\| \Phi_{B} \| =1$ can be realized owing to $\| P_{\eta}^{T_A} - P_{\eta}\|\le 2$. 
Subsequently, based on condition~\eqref{cond_M_Quantum_corr_app} and inequality~\eqref{QC_prop_2_ineq} in Lemma~\ref{lema:basic_property_QC}, 
\begin{align}
\QC_{\rho_{AB}} (\Phi_{A}, \Phi_{B} )\le \QC_\rho (\Phi_{A}, \Phi_{B} ) \le \epsilon \|\Phi_A\|\cdot \|\Phi_B\|, \notag 
\end{align}
which yields
\begin{align}
\label{inequality_quantum_correlation_used}
&\abs{ \tr \brr{ \rho_{AB} ( P_{\eta}^{T_A} - P_{\eta}) } }
=\abs{  \tr \br{ \rho_{AB} \Phi_{A} \otimes \Phi_{B}  }  } \notag\\
&\le \abs{  \sum_s p_s  \tr (\rho_{s,A} \Phi_{A}) \tr (\rho_{s,B} \Phi_{B}) } \notag \\ 
&+\abs{\sum_s p_s \br{ \tr (\rho_{s,AB} \Phi_{A} \Phi_{B}) }-  \tr (\rho_{s,A} \Phi_{A}) \tr (\rho_{s,B} \Phi_{B}) } \notag \\
&\le \sum_s p_s  \abs{ \tr (\rho_{s,A}\otimes \rho_{s,B} \Phi_{A}\otimes \Phi_{B})}  +  
\QC_{\rho_{AB}} (\Phi_{A}, \Phi_{B} )   \notag \\
&\le \sum_s p_s   \abs{\tr \brr{\rho_{s,A}\otimes \rho_{s,B} (P_{\eta}^{T_A} - P_{\eta} ) }}  +2 \epsilon,
\end{align}
where $\{\rho_{s,A}\}_s$ and $\{\rho_{s,B}\}_s$ are the reduced density matrices of $\{\rho_{s,AB}\}_s$, which are appropriately chosen such that they yield $\QC_{\rho_{AB}} (\Phi_{A}, \Phi_{B} )$.

%
%

The aim is to prove 
\begin{align}
\label{two_qubit_cases_P_eta_purpose}
\tr \brr{\rho_{A}\otimes \rho_{B} (P_{\eta}^{T_A} - P_{\eta} ) }  = 0
\end{align}
for arbitrary $\rho_A$ and $\rho_B$. 
Let $u_A$ and $u_B$ be unitary matrices that diagonalize $\rho_A$ and $\rho_B$, respectively. 
Then,  
\begin{align}
\label{two_qubit_cases_P_eta}
&\tr \brr{\rho_{A}\otimes \rho_{B} (P_{\eta}^{T_A} - P_{\eta} ) }  \notag \\
&= \tr \brr{\tilde{\rho}_{A}\otimes \tilde{\rho}_{B}  ( u_A\otimes u_B)  (P_{\eta}^{T_A} - P_{\eta} ) ( u_A\otimes u_B)^\dagger  }  \notag \\
&= \tr \brr{\tilde{\rho}_{A}\otimes \tilde{\rho}_{B}  (\tilde{P}_{\eta}^{\dagger_A} - \tilde{P}_{\eta} ) }, 
\end{align}
where $\tilde{\rho}_{A}:=u_A\rho_A u_A^\dagger$, $\tilde{\rho}_{B}:=u_B\rho_B u_B^\dagger$, 
$\tilde{P}_{\eta} :=  ( u_A\otimes u_B)  P_{\eta} ( u_A\otimes u_B)^\dagger$.
Note that, by using the form~\eqref{Schmidt_eta}, $P_{\eta}^{T_A}=P_{\eta}^{\dagger_A}$ is true, with $\dagger_A$ being the partial conjugate transpose. This yields 
\begin{align}
 ( u_A\otimes u_B)  P_{\eta}^{\dagger_A}( u_A\otimes u_B)^\dagger  
 = \tilde{P}_{\eta}^{\dagger_A}.
\end{align}
In Eq.~\eqref{two_qubit_cases_P_eta}, only the diagonal terms of $(\tilde{P}_{\eta}^{\dagger_A} - \tilde{P}_{\eta})$ contribute to the value, as $\tilde{\rho}_{A}\otimes \tilde{\rho}_{B}$ is a diagonal matrix. 
It is evident that all the diagonal terms in $(\tilde{P}_{\eta}^{\dagger_A} - \tilde{P}_{\eta})$ are equal to zero, and hence, it can be concluded that Eq.~\eqref{two_qubit_cases_P_eta} reduces to Eq.~\eqref{two_qubit_cases_P_eta_purpose}. 
Thus, by applying Eq.~\eqref{two_qubit_cases_P_eta_purpose} to inequality~\eqref{inequality_quantum_correlation_used},
the main inequality~\eqref{upper_bound_qubits} is obtained.
This completes the proof. $\square$

\section{Proof of Theorem~\ref{thm:quantum_correlation_negativity_1D}} 
\label{Theorem_thm_quantum_correlation_PPT_rela_1D}

This section presents the proof of Theorem~\ref{thm:quantum_correlation_negativity_1D}, where the following inequality has been obtained for one-dimensional quantum Gibbs states:
\begin{align}
\label{ineq_quantum_correlation_negativity_1D}
E_{R}^{\PPT} ( \rho_{\beta,AB})
&\le \bar{C}_\beta \log(\mathcal{D}_{AB})  e^{-R/ [6 \log(d_0) \xi^2_\beta] + 7gk\beta},
%
\end{align}
where $\bar{C}_\beta:=  24\br{\tilde{C}_\beta + 16d_0^4C_\beta }^{1/2}$, with $C_\beta$ and $\tilde{C}_\beta$
defined in Eqs.~\eqref{parameters_definitions} and \eqref{thm_ineq:lemma_quantum_belief_para_tilde_C}, respectively. 
Here, the assumption of a finite interaction length has been imposed for Hamiltonian $H$.

\subsection{Proof} 

 \begin{figure*}[tt]
\centering
\includegraphics[clip, scale=0.5]{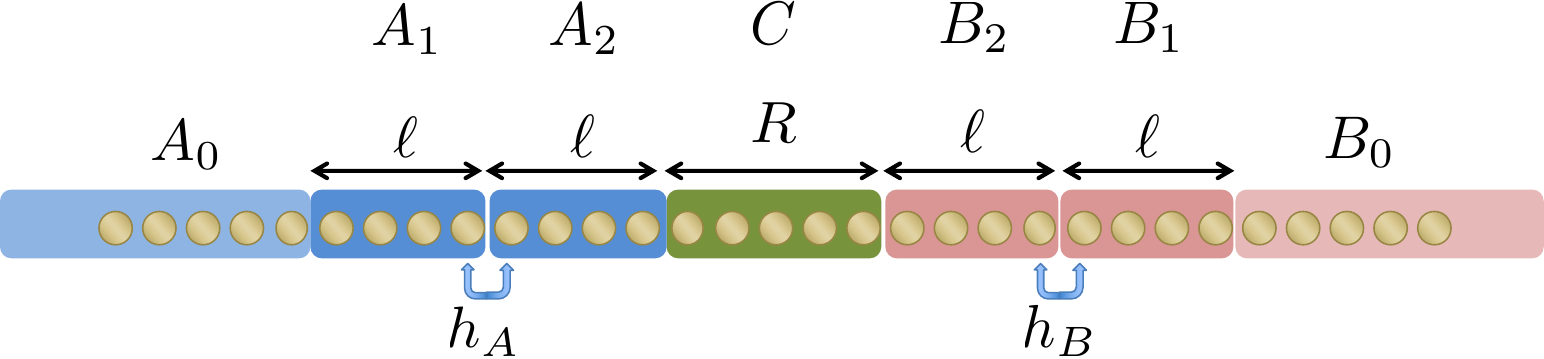}
\caption{ For the proof, subsets $A$ and $B$ are decomposed into three pieces.
The decomposed subsets, i.e., $A_1$, $A_2$, $B_2$, and $B_1$, are considered such that they have the same cardinality, that is, $|A_1|=|A_2|=|B_2|=|B_1|=\ell$. 
The interactions between the subsystems $A_1$ and $A_2$ ($B_1$ and $B_2$) are denoted as $h_{\partial A_1}$ ($h_{\partial B_1}$). 
Then, in the Hamiltonian $H-h_{\partial A_1}-h_{\partial B_1}$, the regions $A_0A_1$, $A_2CB_2$, and $B_1B_0$ are decoupled.   
Consequently, using the quantum belief propagation, it is proven that regions $A_0$ and $B_0$ do not influence the entanglement value. Then, the entanglement between $A$ and $B$ is characterized by the entanglement between $A_1A_2$ and $B_1B_2$. Further, because the size of these regions is $2\ell$, the dependence on the Hilbert space dimension in  \eqref{ineq_quantum_correlation_negativity} is significantly improved.
}
\label{fig:1D_proof}
\end{figure*}

For the proof, first, the subsystems $A$ and $B$ are decomposed as follows (Fig.~\ref{fig:1D_proof}):
\begin{align}
A= A_0 \sqcup A_1 \sqcup A_2,\quad B=B_0 \sqcup B_1 \sqcup B_2,  
\end{align}
where $|A_1|=|A_2|=|B_0|=|B_1|=\ell$. 
Let $h_{\partial A_1}$ ($h_{\partial B_1}$) denote the interactions between $A_1$ and $A_2$ ($B_1$ and $B_2$):
\begin{align}
\label{def_h_A_h_B_partial}
&h_{\partial A_1}= \sum_{Z: Z\cap A_1\neq \emptyset, Z\cap A_2\neq \emptyset} h_Z, \notag \\ 
&h_{\partial B_1}= \sum_{Z: Z\cap B_1\neq \emptyset, Z\cap B_2\neq \emptyset} h_Z.
\end{align}

Then, the quantum Gibbs state $\rho_\beta$ can be described as 
\begin{align}
\label{QB_formal}
\rho_\beta = \Phi e^{-\beta(H-h_{\partial A_1}-h_{\partial B_1})}  \Phi^\dagger,
\end{align}
where $\Phi$ is an appropriate operator. 
It can be proven that $\Phi$ is afforded by a quasi-local operator and approximated by $\Phi_{A_1,A_2} \otimes \Phi_{B_1,B_2}$,
which is formulated by the following lemma:

\begin{lemma} \label{lemma_quantum_belief}
The operator $\Phi$ in Eq.~\eqref{QB_formal} is approximated as follows:
\begin{align}
\label{ineq:lemma_quantum_belief}
&\tilde{\Phi}= \Phi_{A_1,A_2} \otimes \Phi_{B_1,B_2} \quad {\rm s.t.}  \notag \\
&\norm{ \rho_\beta - \br{ \tilde{\Phi} e^{-\beta(H-h_{\partial A_1}-h_{\partial B_1})} \tilde{\Phi}^\dagger} }_1 \notag \\
&\le 
\tilde{C}_\beta e^{-2\ell/\xi_\beta+14g k \beta}  =: \delta_{1,\ell},
\end{align}
where the correlation length $\xi_\beta$ has been defined in Eq.~\eqref{parameters_definitions}, and 
\begin{align}
\label{ineq:lemma_quantum_belief_para_tilde_C}
&\tilde{C}_\beta :=  1280 \br{\frac{5+2C e^{\mu k}}{\pi^2} + \frac{2C e^{\mu k}}{\pi v\beta}  }^2.
\end{align}
Further, 
\begin{align}
\label{ineq:lemma_quantum_belief_2}
\norm{ \tilde{\Phi}  } \le  e^{2 g k \beta }.
\end{align}
\end{lemma}

In the following, the main inequality~\eqref{ineq_quantum_correlation_negativity_1D} is proven based on the above lemma. 
For this purpose, $\tilde{\rho}_\beta $ and $\tilde{Z}$ are defined as follows:
\begin{align}
\label{def:tilde_rho_beta}
&\tilde{\rho}_\beta = \frac{ e^{-\beta(H-h_{\partial A_1}-h_{\partial B_1})} }{\tilde{Z}}, \notag \\
& \tilde{Z}:=\tr\br{e^{-\beta(H-h_{\partial A_1}-h_{\partial B_1})}},
\end{align}
Because 
\begin{align}
e^{-\beta(H-h_{\partial A_1}-h_{\partial B_1})}= e^{-\beta (H_{A_0A_1} + H_{A_2CB_2} +H_{B_1B_0})},
\end{align}
we obtain $\tilde{\rho}_{\beta,AB}$ in the form of
\begin{align}
\label{eq_negativity_belief_pro_2_1}
&\tilde{\rho}_{\beta,AB} = \tilde{\rho}_{A_0A_1} \otimes \tilde{\rho}_{A_2B_2} \otimes  \tilde{\rho}_{B_0B_1},
\end{align}
where $\tilde{\rho}_{A_0A_1}$, $\tilde{\rho}_{A_2B_2}$, and $\tilde{\rho}_{B_0B_1}$ are normalized, respectively.
Here, $\tilde{\delta}$ for $\tilde{\rho}_{A_2B_2}$ is defined in the same manner as for \eqref{lemma_ineq_Min_eigenvalue_T_A}, whereas 
$\tilde{\sigma}_{A_2B_2}$ is defined as 
\begin{align}
\label{def_tilde_sigma_AB_1}
&\tilde{\sigma}_{A_2B_2}=
\tilde{\rho}_{A_2B_2} +  \tilde{\delta} \cdot \hat{1}_{A_2B_2}.
\end{align}
Using the above $\tilde{\sigma}_{A_2B_2}$, $\tilde{\sigma}_{AB}$ is defined as 
\begin{align}
\label{def_tilde_sigma_AB_2}
&\tilde{\sigma}_{AB}:= \frac{\tilde{Z}}{Z_{\tilde{\sigma}}} \tilde{\Phi} \tilde{\rho}_{A_0A_1} \otimes \tilde{\sigma}_{A_2B_2}  \otimes  \tilde{\rho}_{B_0B_1} \tilde{\Phi}^\dagger  \notag \\
&=  \frac{\tilde{Z} }{Z_{\tilde{\sigma}}} \br{\tilde{\Phi}  \tilde{\rho}_{\beta,AB}\tilde{\Phi}^\dagger  
+  \tilde{\delta}  \cdot \tilde{\Phi}  \tilde{\rho}_{A_0A_1} \otimes \hat{1}_{A_2B_2} \otimes  \tilde{\rho}_{B_0B_1} \tilde{\Phi}^\dagger   },
\end{align}
where $Z_{\tilde{\sigma}}$ is the normalization factor used to realize $\tr ( \tilde{\sigma}_{AB})=1$.

$\tilde{\sigma}_{AB}^{T_A} \succeq 0$ can be proven as follows. 
Because $\tilde{\sigma}_{A_2B_2}^{T_A} \succeq 0$,  we obtain 
\begin{align}
\label{eq:orig-F13}
\br{\tilde{\rho}_{A_0A_1} \otimes \tilde{\sigma}_{A_2B_2}  \otimes  \tilde{\rho}_{B_0B_1}}^{T_A} \succeq 0,
\end{align}
Hence, by representing the spectral decomposition of the above operator as 
 \begin{align}
\label{eq:orig-F14}
\tilde{\rho}_{A_0A_1} \otimes \tilde{\sigma}_{A_2B_2}  \otimes  \tilde{\rho}_{B_0B_1} = 
\sum_{i} \tilde{\lambda}_i \ket{\tilde{\lambda}_i } \bra{\tilde{\lambda}_i }
\end{align}
with $\tilde{\lambda}_i \ge 0$, the following is obtained:
 \begin{align}
&\br{ \tilde{\Phi} \tilde{\rho}_{A_0A_1} \otimes \tilde{\sigma}_{A_2B_2}  \otimes  \tilde{\rho}_{B_0B_1} \tilde{\Phi}^\dagger}^{T_A}  \notag \\
&=   \sum_{i} \tilde{\lambda}_i \br{ \Phi_{A_1,A_2}^\ast \otimes \Phi_{B_1,B_2}} \ket{\tilde{\lambda}_i } \bra{\tilde{\lambda}_i }\br{\Phi_{A_1,A_2}^{T_A} \otimes \Phi_{B_1,B_2}^\dagger}  \notag \\
&\succeq 0,
\end{align}
which yields the inequality $\tilde{\sigma}_{AB}^{T_A} \succeq 0$ from definition~\eqref{def_tilde_sigma_AB_2}.

In the following calculations, the aim is to estimate the upper bound of $\norm{\tilde{\sigma}_{AB}- \rho_{\beta,AB}}_1$.
We have 
 \begin{align}
  \label{Ineq_for_the_term_tilde_sigma_start}
&\norm{\tilde{\sigma}_{AB}- \rho_{\beta,AB}}_1  \notag\\
&\le \norm{\tilde{Z} \tilde{\Phi}  \tilde{\rho}_{\beta,AB} \tilde{\Phi}^\dagger - \rho_{\beta,AB}}_1 
+ \norm{\tilde{Z} \tilde{\Phi}  \tilde{\rho}_{\beta,AB} \tilde{\Phi}^\dagger - \tilde{\sigma}_{AB}}_1.
\end{align}
For the first term, because $\tilde{\Phi}$ is supported on $A_1A_2 \cup B_1B_2$,
 \begin{align}
 \label{Ineq_for_the_first_term_tilde_sigma}
&\norm{\tilde{Z} \tilde{\Phi}  \tilde{\rho}_{\beta,AB} \tilde{\Phi}^\dagger - \rho_{\beta,AB}}_1  
= \norm{\tr_C \br{\tilde{Z}  \tilde{\Phi}  \tilde{\rho}_{\beta} \tilde{\Phi}^\dagger - \rho_{\beta}  }   }_1 \notag \\
&\le \norm{ \rho_\beta - \br{ \tilde{\Phi} e^{-\beta(H-h_{\partial A_1}-h_{\partial B_1})} \tilde{\Phi}^\dagger} }_1 
\le \delta_{1,\ell},
\end{align}
where $\delta_{1,\ell} $ has been defined in Lemma~\ref{lemma_quantum_belief}. 
For the second term, based on definition~\eqref{def_tilde_sigma_AB_2}, 
 \begin{align}
  \label{Ineq_for_the_sec_term_tilde_sigma}
&\norm{ \tilde{Z}\tilde{\Phi}  \tilde{\rho}_{\beta,AB} \tilde{\Phi}^\dagger - \tilde{\sigma}_{AB}}_1  \notag \\
&\le \norm{ \br{1-\frac{1}{Z_{\tilde{\sigma}}}} \tilde{Z}  \tilde{\Phi}  \tilde{\rho}_{\beta,AB} \tilde{\Phi}^\dagger }_1   \notag \\
&\quad +  \frac{\tilde{\delta} \tilde{Z}}{Z_{\tilde{\sigma}}} \norm{\tilde{\Phi}  \tilde{\rho}_{A_0A_1} \otimes \hat{1}_{A_2B_2} \otimes  \tilde{\rho}_{B_0B_1} \tilde{\Phi}^\dagger }_1 \notag \\
&\le \abs{1-\frac{1}{Z_{\tilde{\sigma}}}} \br{1+\delta_{1,\ell}  }
+ \frac{\tilde{\delta} \tilde{Z}}{Z_{\tilde{\sigma}}}  \norm{\tilde{\Phi}}^2,
\end{align}
where inequality~\eqref{Ineq_for_the_first_term_tilde_sigma} is used with $\|\rho_{\beta,AB}\|_1=1$ for deriving the first term of the RHS.

The remaining task entails estimating the parameters $\tilde{Z}$, $\tilde{\delta}$, and $Z_{\tilde{\sigma}}$.
Consider the proof of the following inequalities:
\begin{align}
\label{tilde_Z_tilde_Delta_Z_tilde_sigma}
&\tilde{Z} \le e^{4 g k \beta }, \quad 
\tilde{\delta} \le  \delta_{2,R}, \quad \delta_{2,R}:= 16 C_\beta e^{-R/\xi_\beta + 2\ell\log(d_0)}, \notag \\
&\frac{1}{Z_{\tilde{\sigma}}} \le 1+ 2\bar{\delta}_{\ell,R}, \quad \bar{\delta}_{\ell,R}:=\delta_{1,\ell} +\delta_{2,R}  d_0^{2\ell} e^{8gk \beta}, 
\end{align}
where the case of $\bar{\delta}_{\ell,R} \le 1/2$ is considered. 
In the case of $\bar{\delta}_{\ell,R} >1/2$, the desired inequality~\eqref{Ineq_for_the_term_tilde_sigma_fin_1_0} below is trivially true because, in this case, it becomes worse than the trivial bound $\norm{\tilde{\sigma}_{AB}- \rho_{\beta,AB}}_1\le 2$. 

{~}\\ 

\textit{Proof of inequalities in~\eqref{tilde_Z_tilde_Delta_Z_tilde_sigma}.}
The first inequality in~\eqref{tilde_Z_tilde_Delta_Z_tilde_sigma} for the partition function $\tilde{Z}$ can be immediately derived using the Golden--Thompson inequality:\begin{align}
\label{eq_negativity_belief_pro_8}
\tilde{Z}&=\tr \br{ e^{-\beta(H-h_{\partial A_1}-h_{\partial B_1})} }  \notag \\
&\le  \tr \br{ e^{-\beta H}  e^{\beta(h_{\partial A_1}+h_{\partial B_1})} }   \notag \\
&\le \tr \br{ e^{-\beta H}}  e^{\beta(\| h_{\partial A_1}\|+\|h_{\partial B_1}\|)} \le e^{4 g k \beta },
\end{align}
where we use $\tr (e^{-\beta H})=1$, and the norm of $\| h_{\partial A_1}\|+\|h_{\partial B_1}\|$ is upper-bounded in \eqref{norm_upp_h_partial A_1_ineq}. 

In addition, for $\tilde{\delta}$, Lemma~\ref{lemma_Min_eigenvalue_T_A} is applied with 
Theorem~\ref{thm:quantum_correlation} to $\tilde{\rho}_{A_2B_2}$, which yields the second inequality in~\eqref{tilde_Z_tilde_Delta_Z_tilde_sigma}:
\begin{align}
\label{eq:orig-F21}
\tilde{\delta} &\le 4\min(\mathcal{D}_{A_2},\mathcal{D}_{B_2}) \times C_\beta (|\partial A_2| +|\partial B_2| ) \notag \\
&\quad \times \br { 1 +  \log|A_2B_2| }  e^{-R/\xi_\beta}  \notag \\
&\le 16 C_\beta e^{-R/\xi_\beta + 2\ell \log(d_0)}= \delta_{2,R},
\end{align}
where we use $|A_2|=|B_2|=\ell$, $|\partial A_2| = |\partial B_2|=2$, and $1 +  \log |A_2B_2| =  1 +  \log (2\ell) \le d_0^{\ell}$ for $d_0\ge 2$. 

Finally, from Eq.~\eqref{def_tilde_sigma_AB_2}, 
\begin{align}
&Z_{\tilde{\sigma}}  \notag \\
&=\tr \br{\tilde{Z} \tilde{\Phi}  \tilde{\rho}_{\beta,AB}\tilde{\Phi}^\dagger  
+  \tilde{\delta}  \cdot \tilde{Z} \cdot \tilde{\Phi}  \tilde{\rho}_{A_0A_1} \otimes \hat{1}_{A_2B_2} \otimes  \tilde{\rho}_{B_0B_1} \tilde{\Phi}^\dagger    } \notag \\ 
&\ge   \norm{\rho_{\beta,AB}}_1-\norm{\tilde{Z} \tilde{\Phi}  \tilde{\rho}_{AB} \tilde{\Phi}^\dagger - \rho_{\beta,AB}}_1\notag \\
&\quad -\tilde{\delta} \cdot \tilde{Z} \cdot \norm{\tilde{\Phi} }^2 \mathcal{D}_{A_2B_2} \notag \\
&\ge 1- \delta_{1,\ell} - \delta_{2,R}  d_0^{2\ell} e^{8gk \beta} = 1- \bar{\delta}_{\ell,R},
\end{align}
where, in the last inequality, $\mathcal{D}_{A_2B_2}=d_0^{2\ell}$, $\tilde{Z}\le e^{4 g k \beta }$, and $\norm{ \tilde{\Phi}  } \le  e^{2 g k \beta }$ are used in \eqref{ineq:lemma_quantum_belief_2}. 
Further, using $1/(1-x) \le 1+2x$ for $0\le x\le 1/2$, the third inequality in \eqref{tilde_Z_tilde_Delta_Z_tilde_sigma} can be proven from the above inequality. 
This completes the proof of the inequalities in \eqref{tilde_Z_tilde_Delta_Z_tilde_sigma}. $\square$

{~} \\

\begin{widetext}
Combining inequalities~\eqref{Ineq_for_the_sec_term_tilde_sigma} and \eqref{tilde_Z_tilde_Delta_Z_tilde_sigma} yields
 \begin{align}
  \label{Ineq_for_the_sec_term_tilde_sigma_after_f}
&\norm{\tilde{Z} \tilde{\Phi}  \tilde{\rho}_{\beta,AB} \tilde{\Phi}^\dagger - \tilde{\sigma}_{AB}}_1  \le  2\bar{\delta}_{\ell,R}  \br{1+\delta_{1,\ell}  } 
+  \delta_{2,R}  e^{8gk \beta} \br{1+ 2\bar{\delta}_{\ell,R}}.
\end{align}
Then, on applying inequalities~\eqref{Ineq_for_the_first_term_tilde_sigma} and \eqref{Ineq_for_the_sec_term_tilde_sigma_after_f} to~\eqref{Ineq_for_the_term_tilde_sigma_start}, we obtain 
 \begin{align}
  \label{Ineq_for_the_term_tilde_sigma_fin_1_0}
\norm{\tilde{\sigma}_{AB}- \rho_{\beta,AB}}_1  &\le 2\bar{\delta}_{\ell,R}^2 + 3\bar{\delta}_{\ell,R}  \le  4\bar{\delta}_{\ell,R},
\end{align}
where $\bar{\delta}_{\ell,R}\le 1/2$ is used for the second inequality. 
Subsequently, on choosing $\ell=\lceil R/(6 \log(d_0) \xi_{\beta}) \rceil$,
\begin{align}
\label{upp_delta_upp_delta}
\bar{\delta}_{\ell,R}=\delta_{1,\ell} +\delta_{2,R}  d_0^{2\ell} e^{8gk \beta} &= \tilde{C}_\beta e^{-2\ell / \xi_\beta + 14gk\beta} 
+16C_\beta e^{-R/\xi_\beta + 4 \ell \log(d_0)  + 8gk \beta} \notag \\
&\le \br{\tilde{C}_\beta + 16d_0^4C_\beta } e^{-R/ [3 \log(d_0) \xi^2_\beta] + 14gk\beta}  =: \bar{\delta}_{AB}.
\end{align}
\end{widetext}

Finally, to apply the continuity bound~\eqref{rho_AB_tilde_sigma_AB}, 
$\lambda_{\min}(\tilde{\sigma}_{AB})$ must be controlled. 
For this purpose, we consider 
 \begin{align}
\label{eq:orig-F26}
\tilde{\sigma}_{AB}' = (1-\bar{\delta}_{AB})\tilde{\sigma}_{AB} +\bar{\delta}_{AB} \mathcal{D}_{AB}^{-1}\hat{1}_{AB},
\end{align}
which yields $\lambda_{\min}(\tilde{\sigma}_{AB}' )\ge \bar{\delta}_{AB} \mathcal{D}_{AB}^{-1}$. 
Note that $\tilde{\sigma}_{AB}' \in \PPT$. 
Then, 
 \begin{align}
  \label{Ineq_for_the_term_tilde_sigma_fin_1}
\norm{\tilde{\sigma}_{AB}'- \rho_{\beta,AB}}_1 
&\le 4 \bar{\delta}_{AB} +\norm{\tilde{\sigma}_{AB}' -\tilde{\sigma}_{AB}}_1  \notag \\
&\le  6\bar{\delta}_{AB}.
\end{align}
Inequality~\eqref{rho_AB_tilde_sigma_AB} on relative entropy yields 
\begin{align}
  \label{Ineq_for_the_term_tilde_sigma_fin_3}
S(\rho_{\beta,AB}||\tilde{\sigma}'_{AB} )& \le  6\bar{\delta}_{AB} \log(\mathcal{D}_{AB}) -6\bar{\delta}_{AB} \log (6\bar{\delta}_{AB})  \notag \\
&\quad -6\bar{\delta}_{AB} \log [\lambda_{\min} (\tilde{\sigma}'_{AB})]  \notag \\
& \le 12\bar{\delta}_{AB}  \log\br{ \mathcal{D}_{AB} \bar{\delta}_{AB}^{-1}}    \notag \\
&\le 24 \sqrt{\bar{\delta}_{AB}} \log(\mathcal{D}_{AB}),
\end{align}
where $x\log(z/x) \le 2 \sqrt{x} \log (z)$ is used for $0\le x \le 2$ and $z\ge 2$. 
Because $E_{R}^{\PPT}(\rho_{AB}) \le S(\rho_{\beta,AB}||\tilde{\sigma}'_{AB} ) $, 
the main inequality~\eqref{ineq_quantum_correlation_negativity_1D} is proven by 
applying the definition of $\bar{\delta}_{AB}$ in \eqref{upp_delta_upp_delta} to~\eqref{Ineq_for_the_term_tilde_sigma_fin_3}. 
This completes the proof. $\square$

\subsubsection{Proof of Lemma~\ref{lemma_quantum_belief}}

Using the quantum belief propagation~\cite{PhysRevB.86.245116}, $\Phi$ is described as follows:
\begin{align}
\label{Def_belief_quantum}
&\Phi := \mathcal{T} e^{\int_0^{1} \phi (\tau) d\tau},  \notag \\
&\phi(\tau):= -\frac{\beta}{2}  \int_{-\infty}^\infty F_\beta(t) [ h_{\partial A_1}(H_\tau,t) + h_{\partial B_1}(H_\tau,t)] dt,  \notag \\
&H_\tau:= H-(1-\tau) h_{\partial A_1}- (1-\tau)h_{\partial B_1},
\end{align}
where $\mathcal{T}$ is the time ordering operator, $h_{\partial A_1}(H_\tau,t)= e^{iH_\tau t} h_{\partial A_1} e^{-iH_\tau t}$. $F_\beta(t)$ is defined as 
\begin{align}
F_\beta(t) = \frac{1}{2\pi}  \int_{-\infty}^\infty 
\tilde{F}_\beta(\omega) e^{-i\omega t} d\omega, \quad 
\tilde{F}(\omega):= \frac{\tanh(\beta \omega/2)}{\beta\omega/2}. \notag 
\end{align}
The explicit form of $F_\beta(t)$ can be calculated as follows~\cite[Eq.~(103) in Supplementary Information]{Anshu_2021}:
\begin{align}
\label{F_beta_t_def}
F_\beta(t) &= \frac{2}{\beta \pi} \log \left (\frac{e^{\pi |t|/\beta }+1}{e^{\pi |t|/\beta }-1} \right) 
\le \frac{4/(\beta \pi)}{e^{\pi |t|/\beta }-1}  \notag \\
&\le  \frac{4}{\beta\pi} e^{-\pi |t|/\beta } \br{1+ \frac{1}{\pi |t|/\beta}},
\end{align}
where $\log \br{\frac{e^{x}+1}{e^{x}-1}} \le 2/(e^x-1)$ and $1/(e^x-1)\le e^{-x}(1+x^{-1})$ are used for $x\ge0$.

Herein, an approximation is adopted as follows:
\begin{align}
\label{Def_belief_quantum_approx}
&\tilde{\Phi}:= \mathcal{T} e^{\int_0^{1} \tilde{\phi}(\tau) d\tau},  \notag \\
&\tilde{\phi} (\tau):= -\frac{\beta}{2}  \int_{-\infty}^\infty  F_\beta(t)  \bigl[ h_{\partial A_1}(H_\tau,t, A_1A_2)     \notag \\
&\quad \quad\quad\quad\quad\quad\quad \quad  + h_{\partial B_1}(H_\tau,t,B_1B_2) \bigr] dt,
\end{align}
where the notations of Eq.~\eqref{supp_def:W_X_local_approx} are used.
Here, $h_{\partial A_1}(H_\tau,t, A_1A_2)$ and $h_{\partial B_1}(H_{\tau},t, B_1B_2)$ are supported on $A_1A_2$ and $B_1B_2$, respectively.
Because $[h_{\partial A_1}(H_\tau,t, A_1A_2), h_{\partial B_1}(H_{\tau},t, B_1B_2)]=0$, $\tilde{\Phi}$ is given in the form of
\begin{align}
&\tilde{\Phi} = \Phi_{A_1,A_2} \otimes \Phi_{B_1,B_2}.
\end{align}

Consider the norm of $\Phi - \tilde{\Phi}$, which is upper-bounded as 
\begin{align}
\label{Phi_minus/tilde_phi}
\norm{ \Phi - \tilde{\Phi}} \le e^{\int_0^{1}\br{ \norm{ \phi (\tau)} +\norm{ \tilde{\phi}(\tau)}  } d\tau}
\int_0^1  \norm{ \phi (\tau) - \tilde{\phi}(\tau)}   d\tau,
\end{align}
where the analysis in Ref.~\cite[Claim~25]{PhysRevX.11.011047} is used.
To estimate the RHS of \eqref{Phi_minus/tilde_phi}, first consider
\begin{align}
\label{Phi_minus/tilde_phi_first}
&\int_0^{1}\br{ \norm{ \phi (\tau)} +\norm{ \tilde{\phi}(\tau)}  } d\tau  \notag \\
&\le \beta(\|h_{\partial A_1}\|+\|h_{\partial B_1}\|) 
\int_{-\infty}^\infty F_\beta(t)dt  \notag \\
&= \beta(\|h_{\partial A_1}\|+\|h_{\partial B_1}\|),
\end{align}
where $\| h_{\partial A_1}(H_\tau,t, A_1A_2)\| \le   \| h_{\partial A_1}(H_\tau,t)\| = \|h_{\partial A_1}\|$ and 
$\int_{-\infty}^\infty F_\beta(t)dt= \tilde{F}(0)=1$ are used.
Second, using the Lieb-Robinson bound~\eqref{Lieb-Robinson_main_short_region}, 
\begin{align}
\label{Lieb-Robinson_1D_finite}
&\norm{  h_{\partial A_1}(H_\tau,t) -  h_{\partial A_1}(H_\tau,t, A_1A_2)}  \notag \\
&\le   \|h_{\partial A_1}\| \min\br{ 2, 2C \br{e^{v|t|}-1} e^{-\mu (\ell-k)}},
\end{align}
where $| \partial {\rm Supp}(h_{\partial A_1}) |=2$ is used on a 1D chain, and
it is assumed that $h_{\partial A_1}$ has the interaction length $k$ (i.e., $| {\rm Supp}(h_{\partial A_1}) |\le 2k$).  
Note that $\norm{  h_{\partial A_1}(H_\tau,t) -  h_{\partial A_1}(H_\tau,t, A_1A_2)}$ is trivially smaller than $2\|h_{\partial A_1}\|$.

\begin{widetext}

Consequently, on combining the above inequality with Eqs.~\eqref{Def_belief_quantum} and \eqref{Def_belief_quantum_approx}, we obtain
\begin{align}
\norm{ \phi (\tau) - \tilde{\phi}(\tau)}\le 
\frac{\beta(\|h_{\partial A_1}\|+\|h_{\partial B_1}\|) }{2}  \int_{-\infty}^\infty F_\beta(t) \min \br{ 2, 2 C \br{e^{v|t|}-1} e^{-\mu (\ell-k)}}dt. 
\end{align}
Given the form of $F_\beta(t)$ in Eq.~\eqref{F_beta_t_def}, the same calculations as in Sec.~\ref{Proof of the inequality_integrate_Lieb-Robinson_QC_WY} can be applied. Thus, for $t_0= \mu \ell/(2v)$, 
\begin{align}
\label{Phi_minus/tilde_phi_second}
\frac{\norm{ \phi (\tau) - \tilde{\phi}(\tau)}}{\beta (\|h_{\partial A_1}\| + \|h_{\partial B_1}\| )}
& \le \int_{t_0}^{\infty}  \frac{4}{\beta\pi} e^{-\pi t/\beta } \br{1+ \frac{1}{\pi t/\beta}} \cdot 2 dt 
+  \int_0^{t_0}\frac{4}{\beta\pi} e^{-\pi t/\beta } \br{1+ \frac{1}{\pi t/\beta}}  \cdot 2 C \br{e^{vt}-1} e^{-\mu (\ell-k)} dt  \notag \\
&\le \frac{8}{\pi^2}  \br{1+ \frac{1}{\pi t_0/\beta}} e^{-\pi t_0/\beta }  
+ \frac{8C e^{\mu k}}{\beta\pi} e^{-\mu \ell} \int_0^{t_0}\br{1+ \frac{1}{\pi t/\beta}}  \br{e^{vt}-1}  dt  \notag  \\
&\le \frac{8}{\pi^2}  \br{1+ \frac{1}{\pi t_0/\beta}} e^{-\pi t_0/\beta }  
+ \frac{8C e^{\mu k}}{\beta\pi} e^{-\mu \ell} \br{\frac{e^{v t_0}}{v}+  \frac{e^{v t_0}}{\pi/\beta} }  \notag \\
&=\frac{8}{\pi^2}  \br{1+ \frac{2\beta v}{\pi \mu \ell}} e^{-\pi \mu \ell/(2 \beta v) }  
+ \frac{8C e^{\mu k}}{\beta\pi} \br{\frac{1}{v}+  \frac{\beta}{\pi} }  e^{-\mu \ell/2} \notag \\
&\le \brr{\frac{8}{\pi^2}  \br{1+ \frac{\xi_\beta}{2\ell}} 
+ \frac{8C e^{\mu k}}{\pi} \br{\frac{1}{\pi}+\frac{1}{v\beta} } } e^{-2\ell/\xi_\beta},
\end{align}
where the definition of $\xi_\beta :=\frac{4}{\mu} \br{ 1 + \frac{v\beta}{\pi}}$ is used. 
\end{widetext}
Owing to inequality~\eqref{Phi_minus/tilde_phi_first}, the LHS of \eqref{Phi_minus/tilde_phi_second} is trivially smaller than $1$. 
By contrast, for $\ell \le \xi_\beta/3$, the RHS of \eqref{Phi_minus/tilde_phi_second} is larger than $20e^{-2/3}/\pi^2$, which is worse than the trivial upper bound. 
Hence, only the case of $\ell \ge \xi_\beta/3$ is considered, which reduces~\eqref{Phi_minus/tilde_phi_second} to 
 \begin{align}
\label{Phi_minus/tilde_phi_second'}
&\frac{\norm{ \phi (\tau) - \tilde{\phi}(\tau)}}{\beta (\|h_{\partial A_1}\| + \|h_{\partial B_1}\| )} \notag \\
&\le \br{\frac{20+8C e^{\mu k}}{\pi^2} + \frac{8C e^{\mu k}}{\pi v\beta}  } e^{-2\ell/\xi_\beta},
\end{align}

From Eq.~\eqref{supp_def:Ham}, the upper bound can be obtained as
\begin{align}
\label{norm_upp_h_partial A_1_ineq}
\|h_{\partial A_1}\| \le \sum_{i\in {\rm Supp}(h_{\partial A_1})} \sum_{Z: Z\ni i}  \|h_Z\| &\le |{\rm Supp}(h_{\partial A_1})| g  \notag \\
&\le 2g k,
\end{align}
which reduces inequalities~\eqref{Phi_minus/tilde_phi_first} and \eqref{Phi_minus/tilde_phi_second'} to 
\begin{align}
\label{Phi_minus/tilde_phi_third}
&\int_0^{1}\br{ \norm{ \phi (\tau)} +\norm{ \tilde{\phi}(\tau)}  } d\tau \le 4gk \beta, \notag \\
&\norm{ \phi (\tau) - \tilde{\phi}(\tau)}  \notag \\
&\le 4gk \beta \br{\frac{20+8C e^{\mu k}}{\pi^2} + \frac{8C e^{\mu k}}{\pi v\beta}  } e^{-2\ell/\xi_\beta},
\end{align}
respectively.
Further, by applying the above inequalities to \eqref{Phi_minus/tilde_phi}, the following is obtained: 
\begin{align}
\label{uppe_norm_Phi_tilde_Phi}
\norm{ \Phi - \tilde{\Phi} } \le 
16 gk \beta  e^{ 4 g k \beta}  \br{\frac{5+2C e^{\mu k}}{\pi^2} + \frac{2C e^{\mu k}}{\pi v\beta}  } e^{-2\ell/\xi_\beta}.
\end{align}

Finally, consider the norm of
\begin{align}
&\rho_\beta-  \tilde{\Phi} e^{-\beta(H-h_{\partial A_1}-h_{\partial B_1})} \tilde{\Phi}^\dagger \notag \\
&=\rho_\beta  -  \tilde{\Phi}  \Phi^{-1}\rho_\beta  (\tilde{\Phi}\Phi^{-1})^{\dagger}  \notag \\
&=(1- \tilde{\Phi}\Phi^{-1}  )  \rho_\beta  \brr{1-(\tilde{\Phi}\Phi^{-1})^{\dagger} }  \notag \\
&+  \tilde{\Phi}  \Phi^{-1}  \rho_\beta  \brr{1-(\tilde{\Phi}\Phi^{-1})^{\dagger}} 
    + (1-\tilde{\Phi}  \Phi^{-1} ) \rho_\beta (\tilde{\Phi}\Phi^{-1})^{\dagger}, \notag 
\end{align}
where Eq.~\eqref{QB_formal}, that is, $e^{-\beta(H-h_{\partial A_1}-h_{\partial B_1})} = \Phi^{-1}  \rho_\beta (\Phi^\dagger)^{-1} $, is used.
Subsequently, using the above equation, 
\begin{align}
\label{final_norm_inequality_rho_beta_belief0_2}
&\norm{ \rho_\beta - \br{ \tilde{\Phi} e^{-\beta(H-h_{\partial A_1}-h_{\partial B_1})} \tilde{\Phi}^\dagger} }_1\notag \\
&\le \norm{ \rho_\beta}_1 \norm{ 1- \tilde{\Phi} \Phi^{-1} }
\br{\norm{ 1- \tilde{\Phi} \Phi^{-1} } + 2 \norm{ \tilde{\Phi} \Phi^{-1} } }  \notag \\
&\le  3 \norm{ 1- \tilde{\Phi} \Phi^{-1} }^2 +  2 \norm{ 1- \tilde{\Phi} \Phi^{-1} },
\end{align}
where the triangle inequality is employed to obtain $\norm{\tilde{\Phi} \Phi^{-1} } \le \norm{1- \tilde{\Phi} \Phi^{-1} }+1$.
Based on the inequality of $\|\Phi^{-1}\|\le e^{2 g k \beta }$, which is derived in the same manner as~\eqref{Phi_minus/tilde_phi_first}, we obtain 
\begin{align}
\label{final_norm_inequality_rho_beta_belief_3}
&\norm{1 -\tilde{\Phi}\Phi^{-1} }  \le \norm{\Phi^{-1}} \cdot \norm{\Phi -\tilde{\Phi}}   \notag \\
&\le 16 gk \beta  e^{6g k \beta}  \br{\frac{5+2C e^{\mu k}}{\pi^2} + \frac{2C e^{\mu k}}{\pi v\beta}  } e^{-2\ell/\xi_\beta} \notag \\
&\le 16 e^{7g k \beta} \br{\frac{5+2C e^{\mu k}}{\pi^2} + \frac{2C e^{\mu k}}{\pi v\beta}  } e^{-2\ell/\xi_\beta} 
\end{align}
from inequality~\eqref{uppe_norm_Phi_tilde_Phi}, where $xe^{6x} \le e^{7x}$ is used for $x\ge 0$. 

Therefore, by combining inequalities~\eqref{final_norm_inequality_rho_beta_belief0_2} and \eqref{final_norm_inequality_rho_beta_belief_3}, inequality~\eqref{ineq:lemma_quantum_belief} can be obtained as follows:
\begin{align}
&\norm{ \rho_\beta - \br{ \tilde{\Phi} e^{-\beta(H-h_{\partial A_1}-h_{\partial B_1})} \tilde{\Phi}^\dagger} }_1\notag \\
&\le 
1280 \br{\frac{5+2C e^{\mu k}}{\pi^2} + \frac{2C e^{\mu k}}{\pi v\beta}  }^2 e^{-2\ell/\xi_\beta+14g k \beta}. \notag 
\end{align}  


Finally, on the norm $\norm{\tilde{\Phi} }$, considering \eqref{Phi_minus/tilde_phi_first},
\begin{align}
\norm{\tilde{\Phi} } \le 
e^{\int_0^{1}  \norm{ \tilde{\phi}(\tau)}  d\tau } \le 
e^{\frac{\beta}{2}(\|h_{\partial A_1}\|+\|h_{\partial B_1}\|) } \le e^{2 g k \beta }, \notag 
\end{align}
which yields inequality~\eqref{ineq:lemma_quantum_belief_2}.
This completes the proof. $\square$

\section{Remark on entanglement negativity} \label{sec:Remark on the entanglement negativity}

The PPT relative entanglement in Eq.~\eqref{def_Entanglement_PPT_relative} is relevant to another definition of quantum entanglement.
Herein, consider entanglement negativity, which is given by~\cite{PhysRevA.65.032314}
\begin{align}
\label{def_Entanglement_Negativity_corr}
E_N(\rho_{AB}) := \log \|\rho_{AB}^{T_A}\|_1.
\end{align}
Using Proposition~\ref{prop:quantum_correlation_negativity}, the following corollary is obtained:
\begin{corol} \label{corol:quantum_correlation_negativity_2}
Let $\rho$ be an arbitrary quantum state such that 
\begin{align}
\label{cond_M_Quantum_corr_2}
\QC_\rho (O_A,O_B) \le \epsilon \|O_A\| \cdot \|O_B\| 
\end{align}
for two arbitrary operators $O_A$ and $O_B$; then,
\begin{align}
 \label{ineq:prop:quantum_correlation_negativity_2}
E_N(\rho_{AB}) \le \|\rho_{AB}^{T_A}\|_1-1\le 8\epsilon \min( \mathcal{D}_A,\mathcal{D}_B) \mathcal{D}_{AB},
\end{align}
where the first inequality is trivially derived from $\log(1+x) \le x$ for $x\ge 0$. 
Recall that $\mathcal{D}_{AB}$ is the Hilbert space dimension in the region $AB$. 
Thus, by applying Theorem~\ref{thm:quantum_correlation} to inequality~\eqref{ineq:prop:quantum_correlation_negativity_2}, 
an inequality similar to \eqref{ineq_quantum_correlation_negativity} can be derived.
\end{corol}

\noindent
\textit{Proof of Corollary~\ref{corol:quantum_correlation_negativity_2}.} 
First, because $\tr(\rho_{AB}^{T_A})=1$, 
\begin{align}
\label{negativity_upper_bound_start}
\|\rho_{AB}^{T_A}\|_1 =1+ \sum_{i=1}^{M_0} 2 | \bra{\eta_i} \rho_{AB}^{T_A}\ket{\eta_i} | 
&\le 1+2 M_0 \cdot \delta  \notag \\
&\le 1+ 2 \mathcal{D}_{AB} \cdot \delta
\end{align}
with $\delta := - \min_i  \bra{\eta_i} \rho_{AB}^{T_A}\ket{\eta_i}$, 
where $M_0 \le \mathcal{D}_{AB}$. 
Here, the value $M_0$ can be as large as $(\mathcal{D}_{A}-1)(\mathcal{D}_{B}-1)$, in general (see Ref.~\cite{PhysRevA.87.054301}). 
Thus, using the upper bound on $\delta$ in Lemma~\ref{lemma_Min_eigenvalue_T_A}, 
inequality~\eqref{ineq:prop:quantum_correlation_negativity_2} is proven.
This completes the proof. $\square$

{~} \\

By contrast, an inequality similar to \eqref{ineq_quantum_correlation_negativity_1D} cannot be derived for 1D quantum Gibbs states if entanglement negativity is considered. 
This is explained as follows.
As shown in Lemma~\ref{lemma_quantum_belief}, the following was derived: 
\begin{align}
\label{ineq:lemma_quantum_belief_standard2}
\norm{ \rho_\beta- \tilde{\Phi} e^{-\beta(H-h_{\partial A_1}-h_{\partial B_1})} \tilde{\Phi} }_1
\le  e^{-\ell/\orderof{\beta} + \orderof{\beta}},
\end{align}
where $\tilde{\Phi}$ has been supported on $A_1A_2 \cup B_1B_2$. 
Thus, it is concluded that, for $\ell \gtrsim \beta^2$, 
 \begin{align}
\rho_\beta \approx \tilde{\Phi} e^{-\beta(H-h_{\partial A_1}-h_{\partial B_1})} \tilde{\Phi}. 
\end{align}
The primary difficulty is that entanglement negativity cannot satisfy a convenient continuity inequality. 
In Ref.~\cite[Ineq.~(16)]{PhysRevB.102.235110}, it has been proven that, for arbitrary quantum states $\rho_{AB}$ and $\rho'_{AB}$,
\begin{align}
&|E_N(\rho_{AB})- E_N(\rho'_{AB})| \notag \\
&\le\log \br{1+ \sqrt{\mathcal{D}_{AB}}\|\rho_{AB}-\rho_{AB}'\|_2}  \notag \\
&\le\log \br{1+ \sqrt{\mathcal{D}_{AB}}\|\rho_{AB}-\rho_{AB}'\|_1}.
\end{align}
Hence, even for $\|\rho - \rho'\|_1 = e^{-\orderof{n^z}}$ ($0<z<1$), the difference in entanglement negativity can be significantly large~\footnote{For example, consider the case in which $\rho_{AB}=\ket{0^{\otimes n}}\bra{0^{\otimes n}}$, with $AB=\Lambda$ and $\rho_{AB}'=(1-\epsilon)\rho_{AB}+\epsilon \rho_{\rm r} $. Here, the state $\rho_{\rm r}=\ket{\rm r}\bra{\rm r}$ is a random pure state on $\Lambda$, which is orthogonal to $\ket{0^{\otimes n}}$. We then obtain $E_N(\rho_{AB})=0$ and 
$E_N(\rho_{AB})=\log\br{1-\epsilon+\epsilon \bigl \|  \rho_{\rm r}^{T_A}\bigr\|_1}$. 
Here, $\bigl \|  \rho_{\rm r}^{T_A}\bigr\|_1=e^{\Theta(n)}$~\cite{PhysRevA.81.052312}, and hence, provided $\epsilon =e^{-\orderof{n^z}}$ ($0<z<1$), $| E_N(\rho_{AB}) - E_N(\rho'_{AB})| \propto n$ is obtained.
}. Therefore, error estimation~\eqref{ineq:lemma_quantum_belief_standard2} cannot be utilized for this purpose.

Adopting the same steps as those for Sec.~\ref{Theorem_thm_quantum_correlation_PPT_rela_1D}, 
\begin{align}
\norm{  \br{\rho_\beta - \tilde{\Phi} e^{-\beta \br{H-h_{\partial A_1}-h_{\partial B_1}}} \tilde{\Phi}^\dagger}^{T_A} }_1 \notag
\end{align}
needs to be calculated instead of
\begin{align}
\norm{ \rho_\beta- \tilde{\Phi} e^{-\beta \br{H-h_{\partial A_1}-h_{\partial B_1}} } \tilde{\Phi} }_1  \notag 
\end{align}
to obtain a meaningful upper bound for entanglement negativity. 
However, in general, the partial-transpose operation can significantly increase the operator norm, that is, 
$\|O^{T_A}\|_1 \le \min(\mathcal{D}_A, \mathcal{D}_{A^\co}) \|O\|_1$, as shown in Ref.~\cite{tomiyama1983transpose,Ando2008}. 
Owing to this difficulty, the possibility of deriving a statement similar to~Theorem~\ref{thm:quantum_correlation_negativity_1D} for entanglement negativity~\eqref{def_Entanglement_Negativity_corr} remains unclear. 
However, it is expected to be proven for entanglement negativity by employing an analysis similar to that in Ref.~\cite{PhysRevE.93.022128}.

\section{Quantum Fisher information matrix}\label{sec:Relation to quantum Fisher information matrix}

Here, the definition~\eqref{def_Quantum_corr} for the quantum correlation $\QC_\rho(O_A,O_B)$ proposed is compared 
with the quantum Fisher information matrix.   
First, it should be noted that the quantum Fisher information can be defined in the form of the convex roof of the variance.  
If $\rho$ is a pure state, the quantum Fisher information $\mathcal{F}_{\rho}(K)$ simply reduces to the variance of $K$:
\begin{align}
\mathcal{F}_{\rho}(K) = 4 \left(  \bra{\psi}K^2 \ket{\psi} - \bra{\psi}K\ket{\psi}^2  \right),
\end{align}
where $\rho=\ket{\psi} \bra{\psi}$. 
For the general state $\rho$, the quantum Fisher information is known to be equal to the convex roof of the variance~\cite{toth2013extremal,yu2013quantum}:
\begin{align}
\label{def_quantum_Fisher_convex_roof}
\mathcal{F}_{\rho}(K) =4  \inf_{\{p_s,\ket{\psi_s}\}} \sum_{s}p_s \left(  \bra{\psi_s}K^2 \ket{\psi_s} - \bra{\psi_s}K\ket{\psi_s}^2  \right),
\end{align}
where minimization is considered for all possible decompositions of $\rho$, such that $\rho= \sum_{s}p_s \ket{\psi_s}\bra{\psi_s}$ with $p_s> 0$.
Thus, the quantum Fisher information shows a certain similarity to the quantum correlation $\QC_\rho(O_A,O_B)$.

To view this similarity in more detail, consider the following quantum Fisher information matrix~\cite{Liu_2019}: 
\begin{align}
\label{def_quantum_Fisher_mat}
\mathcal{F}_{\rho}(O_i,O_j) = \sum_{s,s'} \frac{2(\lambda_s - \lambda_{s'})^2}{\lambda_s + \lambda_{s'}} 
\bra{\lambda_s} O_i \ket{\lambda_{s'}} \bra{\lambda_{s'}} O_j \ket{\lambda_{s}}. 
\end{align}
Herein, 
\begin{align}
\mathcal{F}_{\rho}(K) = \sum_{i,j} \mathcal{F}_{\rho}(O_i,O_j).  
\end{align}
The quantum Fisher information matrix has been used in the multiparameter quantum estimation theory~\cite{Helstrom1969,Liu_2019,PhysRevLett.121.130503,PRXQuantum.2.020308}.
Then, the question remains as to whether it can be associated with the convex roof of certain observables in the analogy of Eq.~\eqref{def_quantum_Fisher_convex_roof}.

The partial answer to this question is yes. The quantum Fisher information matrix is relevant to the following quantity $\QC^\ast_\rho(O_A,O_B)$, which is weaker than \eqref{def_Quantum_corr}:
\begin{align}
\label{def_Quantum_corr_ave}
\QC^\ast_\rho(O_A,O_B):= \inf_{\{p_s,\rho_s\}} \left| \sum_{s}p_s \C_{\rho_s}(O_A,O_B) \right|,
\end{align}
which is the minimization of the absolute value of the average correlation.
Based on the above quantity, the following statement can be proven, which is similar to Lemma~\ref{lem:quantum_correlation_1}:
\begin{lemma} \label{lem:quantum_Fisher_matrix}
For two arbitrary operators $O_A$ and $O_B$, if 
 \begin{align}
 \label{cond_L_o_1o_2}
[\mathcal{L}_{O_A}, \mathcal{L}_{O_B}] =0,
\end{align}
the quantity $\QC^\ast_\rho(O_A,O_B)$ is upper-bounded in Eq.~\eqref{def_Quantum_corr_ave}, as follows:
\begin{align}
\label{main_inequality_lemma_2_QC_Fisher}
\QC^\ast_\rho(O_A,O_B) \le  \frac{1}{4} | \mathcal{F}_{\rho}(O_A,O_B) |. 
\end{align}
Here, the operator $\mathcal{L}_O$ has been defined in Eq.~\eqref{def_L_O_lambda}.  
If condition~\eqref{cond_L_o_1o_2} holds only approximately (i.e.,  $[\mathcal{L}_{O_A}, \mathcal{L}_{O_B}] \approx 0$), a similar modification to Lemma~\ref{lem:quantum_correlation_2} is required.  
\end{lemma}

{\bf Remark.}
For the quantity $\QC^\ast_\rho(O_A,O_B)$ in \eqref{def_Quantum_corr_ave}, 
at the first glance, no meaningful constraints on the entanglement structure can be observed, as $\C_{\rho_s}(O_A,O_B) $ can have a negative value.
In other words, even if $\QC^\ast_\rho(O_A,O_B)$ is equal to zero, $\QC_\rho(O_A,O_B)$ may still be large. 
However, the same statement as Lemma~\ref{lem:QC_Entanglement_condition_2} can be proven for $\QC^\ast_\rho(O_A,O_B)$ on the Peres-Horodecki separability criterion (i.e., the PPT condition):
\begin{lemma} \label{PPT_Fisher_Matrix}
Consider the proof for the following statement:
\begin{align}
\label{QC_ast_PPT_cond}
&\textrm{$\QC^\ast_{\rho_{AB}}(O_A,O_B) = 0$ for arbitrary pairs of $O_A,O_B$} \notag \\
& \longrightarrow \textrm{$\rho_{AB}$ satisfies the PPT condition}.
\end{align}
\end{lemma}
From statement~\eqref{QC_ast_PPT_cond} and inequality~\eqref{main_inequality_lemma_2_QC_Fisher}, it is evident that the quantum Fisher information matrix also plays a role in quantum correlation measures.

\subsection{Proof of Lemma~\ref{lem:quantum_Fisher_matrix}}

%
%
%

Herein, consider the proof of Lemma~\ref{lem:quantum_correlation_1}. 
Consider the decomposition of $\rho$ as follows:
 \begin{align}
&\rho = \sum_m p_m  \ket{\phi_m}\bra{\phi_m},  \notag \\
&\ket{\phi_m}=\frac{1}{\sqrt{p_m}} \sqrt{\rho}\ket{\psi_m}, \quad p_m = \bra{\psi_m} \rho \ket{\psi_m},
\end{align} 
where $\ket{\psi_m}$ is chosen as the simultaneous eigenstates of $\mathcal{L}_{O_A}$ and $\mathcal{L}_{O_B}$ with the corresponding eigenvalues $\alpha_{1,m}$ and $\alpha_{2,m}$, respectively.
Then, an equation identical to~\eqref{Eq_phi_m_O_AO_B_phi_m_fin} is obtained:
 \begin{align}
  \label{Eq_phi_m_O_AO_B_phi_m_fin_fisher}
\bra{\phi_m}O_A\ket{\phi_m}\bra{\phi_m}O_B\ket{\phi_m}  = \alpha_{1,m}\alpha_{2,m}. 
 \end{align}

Next, consider the proof 
 \begin{align}
 \label{first_eq_Fisher_mat}
\sum_{m} p_m \bra{\phi_m}O_A\ket{\phi_m}\bra{\phi_m}O_B\ket{\phi_m}  = \frac{1}{2}\tr (\{ \rho, \mathcal{L}_{O_A}  \mathcal{L}_{O_B} \}),  
 \end{align}
 where $\{\cdot, \cdot\}$ is the anticommutator. 
By expanding the RHS in Eq.~\eqref{first_eq_Fisher_mat}, 
 \begin{align}
 \frac{1}{2}\tr (\{ \rho, \mathcal{L}_{O_A}  \mathcal{L}_{O_B} \}) 
&= \frac{1}{2}\sum_m \bra{\psi_m}  \{ \rho, \mathcal{L}_{O_A}  \mathcal{L}_{O_B} \} \ket{\psi_m}   \notag \\
&=\sum_m  \bra{\psi_m} \rho \ket{\psi_m} \alpha_{1,m}\alpha_{2,m},
 \end{align}
which reduces to the LHS in Eq.~\eqref{first_eq_Fisher_mat} from $p_m = \bra{\psi_m} \rho \ket{\psi_m}$ and Eq.~\eqref{Eq_phi_m_O_AO_B_phi_m_fin_fisher}

By contrast, using the spectral decomposition of $\rho=\sum_{s}\lambda_s \ket{\lambda_s} \bra{\lambda_s}$, 
 \begin{align}
   \label{first_eq_Fisher_mat_2}
&\frac{1}{2}\tr (\{ \rho, \mathcal{L}_{O_A}  \mathcal{L}_{O_B} \})  \notag \\
&= \sum_{s,s'} \frac{2\lambda_s \lambda_{s'}}{\lambda_s + \lambda_{s'}} \bra{\lambda_s} O_A \ket{\lambda_{s'}} \bra{\lambda_{s'}} O_B \ket{\lambda_{s}},
 \end{align}
where the form of $\mathcal{L}_O$ in Eq.~\eqref{def_L_O_lambda} is used. 
Further, by combining Eqs~\eqref{first_eq_Fisher_mat} and \eqref{first_eq_Fisher_mat_2}, 
 \begin{align}
 \label{first_eq_Fisher_mat_3}
&\sum_{m} p_m \bra{\phi_m}O_A\ket{\phi_m}\bra{\phi_m}O_B\ket{\phi_m}   \notag \\
&= \sum_{s,s'} \frac{2\lambda_s \lambda_{s'}}{\lambda_s + \lambda_{s'}} \bra{\lambda_s} O_A \ket{\lambda_{s'}} \bra{\lambda_{s'}} O_B \ket{\lambda_{s}}.  
 \end{align}

Finally, 
 \begin{align}
  \label{sec_eq_Fisher_mat}
& \sum_{m} p_m \bra{\phi_m}O_AO_B\ket{\phi_m}=\tr (\rho  O_A O_B) \notag \\
&= \sum_{s,s'} \frac{\lambda_s + \lambda_{s'}}{2} \bra{\lambda_s} O_A \ket{\lambda_{s'}} \bra{\lambda_{s'}} O_B \ket{\lambda_{s}},
 \end{align}
where $[O_A,O_B]=0$. 
Thus, by subtracting Eq.~\eqref{first_eq_Fisher_mat_3} from Eq.~\eqref{sec_eq_Fisher_mat}, 
\begin{align}
  \label{fin_eq_Fisher_mat}
& \sum_{m} p_m  \br{ \bra{\phi_m}O_AO_B\ket{\phi_m} - \bra{\phi_m}O_A\ket{\phi_m}\bra{\phi_m}O_B\ket{\phi_m}}  \notag \\
&= \sum_{s,s'} \frac{(\lambda_s - \lambda_{s'})^2}{2(\lambda_s + \lambda_{s'})} \bra{\lambda_s} O_A \ket{\lambda_{s'}} \bra{\lambda_{s'}} O_B \ket{\lambda_{s}} \notag \\
&=\frac{1}{4} \mathcal{F}_{\rho}(O_A,O_B) 
 \end{align}
is obtained.
Therefore, on applying the above equation to \eqref{def_Quantum_corr_ave}, inequality~\eqref{main_inequality_lemma_2_QC_Fisher} is proven. This completes the proof. $\square$

\subsection{Proof of Lemma~\ref{PPT_Fisher_Matrix}}

Consider the proof of the statement 
\begin{align}
\label{app_QC_ast_PPT_cond}
&\textrm{$\QC^\ast_{\rho_{AB}}(O_A,O_B) = 0$ for arbitrary pairs of $O_A,O_B$} \notag \\
& \longrightarrow \textrm{$\rho_{AB}$ satisfies the PPT condition}.
\end{align}
This statement can be easily evaluated via the following discussion.

First, if inequality~\eqref{ineq:prop:quantum_correlation_negativity} in Proposition~\ref{prop:quantum_correlation_negativity} can be proven by assuming inequality~\eqref{cond_M_Quantum_corr} for $\QC^\ast_\rho (O_A,O_B)$ instead of $\QC_\rho (O_A,O_B)$, the statement~\eqref{app_QC_ast_PPT_cond} is obtained. 
Second, in the proof of Proposition~\ref{prop:quantum_correlation_negativity}, inequality~\eqref{cond_M_Quantum_corr} is used only for deriving the upper bound~\eqref{inequality_quantum_correlation_used} for the proof of Lemma~\ref{two_qubits_cases_qc}.
From the second to the third lines in~\eqref{inequality_quantum_correlation_used}, $\QC_\rho (O_A,O_B)$ is used as an upper bound for
\begin{align}
\abs{\sum_s p_s \br{ \tr (\rho_{s,A} \Phi_{A} \Phi_{B}) }-  \tr (\rho_{s,A} \Phi_{A}) \tr (\rho_{s,B} \Phi_{B}) }; \notag 
\end{align}
however, $\QC^\ast_\rho (O_A,O_B)$ also serves as the upper bound for the above quantity.
Consequently, inequality~\eqref{ineq:prop:quantum_correlation_negativity} can be proven using the constraint on $\QC^\ast_\rho (O_A,O_B)$ alone. 
This completes the proof. $\square$

\clearpage
\section{Correction to the relation between quantum correlations and PPT entanglement}
\label{app:qc-ppt-correction}

This supplementary correction concerns Phys.\ Rev.\ X
\textbf{12}, 021022 (2022)~\cite{PhysRevX.12.021022}.
Equation and statement references below refer to the original part of
this manuscript unless they point within this correction.
The original text is retained; the statements identified below are
superseded by this correction. The mixed convex-roof
correlation $\mathrm{QC}_\rho$ does not imply the PPT condition. The
entanglement-clustering results are recovered using
$Q_\rho^{(1/2)}$ from Eq.~\eqref{QC_Wigner_Yanase_alpha}.

\subsection{Failure of the QC-to-PPT implication}

Equation~\eqref{two_qubit_cases_P_eta_purpose} is false: for
$|\eta\rangle=(|00\rangle+|11\rangle)/\sqrt{2}$ and
$\rho_A=\rho_B=(\hat{1}+\hat{\sigma}_y)/2$, its left-hand side is $1/2$,
not zero. More fundamentally, let $F$ be the swap on
$\mathbb{C}^9\otimes\mathbb{C}^9$ and set
\begin{equation}
 \rho_* = \frac{25\hat{1}-3F}{1998}.
 \label{eq:corr-counterexample}
\end{equation}
This state is positive and normalized. Since $F^{T_A}$ has one
nonzero eigenvalue, equal to $9$,
$\lambda_{\min}(\rho_*^{T_A})=-1/999$.
Nevertheless, $\mathrm{QC}_{\rho_*}(O_A,O_B)=0$ for every pair of
local operators, including non-Hermitian ones, as follows.

We identify the two local matrix spaces through the basis defining
$F$. Identity shifts do not change covariance, so take $O_A,O_B$
traceless and write $c=\operatorname{tr}(O_AO_B)$, with this trace
taken on the common nine-dimensional space. If $c=0$, the
covariance in $\rho_*$ itself vanishes. For $c\ne0$, there is a
traceless Hermitian $O$, with $\operatorname{tr}(O^2)=1$, such that
\begin{equation}
 \operatorname{tr}(O_AO)\operatorname{tr}(O_BO)=\theta c,
 \qquad \theta\ge\frac14.
 \label{eq:corr-direction}
\end{equation}
To see this, consider the real span of the Hermitian real and imaginary
parts of $O_A,O_B$, with inner product $\operatorname{tr}(XY)$.
Its dimension is at most four. On this space the real quadratic form
\[
 \operatorname{Im}\!\left[
 e^{-i\arg c}\operatorname{tr}(O_AO)\operatorname{tr}(O_BO)
 \right]
\]
has trace zero. A traceless real symmetric form has an orthonormal
basis with zero diagonal: choose a unit vector of zero quadratic form
between its extreme eigenvectors and repeat on the orthogonal
complement. In that basis the corresponding real parts sum to
$|c|$, so one basis vector gives Eq.~\eqref{eq:corr-direction}.

For this $O$, choose $t^2=74/\theta$ and define
\begin{equation}
 \rho_\pm=\frac{25\hat{1}-3F
       \pm t(O\otimes\hat{1}-\hat{1}\otimes O)}{1998}.
 \label{eq:corr-decomposition}
\end{equation}
These operators have trace one and average $\rho_*$.
If $o_i$ are the eigenvalues of $O$, their numerators have eigenvalues
$22$ on $|ii\rangle$ and
$25\pm\sqrt{9+t^2(o_i-o_j)^2}$ on each
$\operatorname{span}\{|ij\rangle,|ji\rangle\}$.
Since $(o_i-o_j)^2\le2$ and $t^2\le296$, all are positive.
Direct calculation gives
\begin{align}
 \operatorname{tr}(\rho_\pm O_AO_B)
 &= -\frac{3c}{1998},\notag\\
 \operatorname{tr}(\rho_\pm O_A)\operatorname{tr}(\rho_\pm O_B)
 &= -\frac{81t^2\theta c}{1998^2}
  =-\frac{3c}{1998}.
 \label{eq:corr-zero-covariance}
\end{align}
Thus both component covariances vanish. The pair-dependent
decomposition $\rho_*=(\rho_++\rho_-)/2$ is allowed by Eq.~\eqref{def_Quantum_corr},
proving the claim.

Consequently, Lemmas~\ref{lem:QC_Entanglement_condition_2},
\ref{lemma_Min_eigenvalue_T_A}, \ref{two_qubits_cases_qc}, and
\ref{PPT_Fisher_Matrix}, Proposition~\ref{prop:quantum_correlation_negativity}, and the general
$\mathrm{QC}$-based statement of Corollary~\ref{corol:quantum_correlation_negativity_2} are incorrect, and
Conjecture~\ref{conj_entanglement_exist} is false. For Lemma~\ref{PPT_Fisher_Matrix}, use
$\mathrm{QC}_\rho^*\le\mathrm{QC}_\rho$. Theorem~\ref{thm:quantum_correlation} and the
skew-information and quantum Fisher information bounds do not rely
on this erroneous implication.

\subsection{Corrected PPT bound}

Let $\rho$ be a state on $\Lambda$, and let $A,B$ be disjoint
subsystems. Recall Eq.~\eqref{QC_Wigner_Yanase_alpha} at $\alpha=1/2$:
\begin{equation}
 Q_\rho^{(1/2)}(O_A,O_B)
 =\operatorname{tr}(\rho O_AO_B)
  -\operatorname{tr}(\sqrt\rho\,O_A\sqrt\rho\,O_B).
 \label{eq:corr-square-root}
\end{equation}
Suppose that, for all local Hermitian observables,
\begin{equation}
 |Q_\rho^{(1/2)}(O_A,O_B)|
 \le\epsilon\|O_A\|\,\|O_B\|.
 \label{eq:corr-hypothesis}
\end{equation}
Then the following replace Lemma~\ref{lemma_Min_eigenvalue_T_A} and Proposition~\ref{prop:quantum_correlation_negativity}:
\begin{align}
 \delta&:=\max\{0,-\lambda_{\min}(\rho_{AB}^{T_A})\}
 \le\epsilon\min(\mathcal{D}_A,\mathcal{D}_B),
 \label{eq:corr-spectrum}\\
 E_R^{\mathrm{PPT}}(\rho_{AB})
 &\le\log[1+\mathcal{D}_{AB}\epsilon\min(\mathcal{D}_A,\mathcal{D}_B)].
 \label{eq:corr-relative-entropy}
\end{align}

\textit{Proof.}
In fixed product bases, form the normalized canonical purification
\[
 |\sqrt\rho\rangle\!\rangle
 =\sum_{i,j}(\sqrt\rho)_{ij}|i\rangle_\Lambda|j\rangle_{\Lambda'}.
\]
Let $\omega_{A'B}$ be its reduced state on the copied subsystem $A'$
and the physical subsystem $B$. Identify $A'$ with $A$ in the fixed
basis. Then $\omega_{A'B}\succeq0$ and
\begin{align}
 \operatorname{tr}[\omega_{A'B}(O_A^T\otimes O_B)]
 &=\operatorname{tr}(\sqrt\rho\,O_A\sqrt\rho\,O_B),\notag\\
 \operatorname{tr}[(\rho_{AB}^{T_A}-\omega_{A'B})
                 (O_A^T\otimes O_B)]
 &=Q_\rho^{(1/2)}(O_A,O_B).
 \label{eq:corr-purification}
\end{align}
The square root here is that of the full state $\rho$, not of its
marginal $\rho_{AB}$.

For any unit vector, use the Schmidt expansion
$|\eta\rangle=\sum_s\nu_s|s_A,s_B\rangle$, with
$\nu_s\ge0$ and $\sum_s\nu_s^2=1$, as in Eq.~\eqref{Schmidt_eta}.
Each off-diagonal matrix unit is $(X+iY)/2$ for Hermitian $X,Y$
of norm one. Bilinearity therefore extends
Eq.~\eqref{eq:corr-hypothesis} to the matrix units without increasing
$\epsilon$. Expanding $|\eta\rangle\langle\eta|$ yields
\begin{align}
 \bigl|\langle\eta|(\rho_{AB}^{T_A}-\omega_{A'B})|\eta\rangle\bigr|
 &\le\epsilon\Bigl(\sum_s\nu_s\Bigr)^2\notag\\
 &\le\epsilon\min(\mathcal{D}_A,\mathcal{D}_B).
 \label{eq:corr-schmidt}
\end{align}
Positivity of $\omega_{A'B}$ proves
Eq.~\eqref{eq:corr-spectrum}. The normalized state
\begin{equation}
 \sigma_{AB}=\frac{\rho_{AB}+\delta\hat{1}_{AB}}
                       {1+\mathcal{D}_{AB}\delta}
 \label{eq:corr-ppt-state}
\end{equation}
is PPT and obeys
$S(\rho_{AB}\|\sigma_{AB})\le\log(1+\mathcal{D}_{AB}\delta)$,
since $\rho_{AB}$ and $\rho_{AB}+\delta\hat{1}_{AB}$ commute
and the latter has larger eigenvalues. This proves
Eq.~\eqref{eq:corr-relative-entropy}.
Also, Eq.~\eqref{negativity_upper_bound_start} gives
$E_N(\rho_{AB})\le\log[1+2\mathcal{D}_{AB}\epsilon\min(\mathcal{D}_A,\mathcal{D}_B)]$,
replacing the general bound in Corollary~\ref{corol:quantum_correlation_negativity_2}.
\hfill$\square$

\subsection{Recovery of Corollary~\ref{corol:quantum_correlation_negativity} and Theorem~\ref{thm:quantum_correlation_negativity_1D}}

\textit{Corollary~\ref{corol:quantum_correlation_negativity}.}
Theorem~\ref{quantum_correlation_skey_type}, with the absolute-value estimate proved in
Eqs.~\eqref{QC_ineq_integrate_Lieb-Robinson_QC_WY}--\eqref{eq:orig-C24}, gives Eq.~\eqref{eq:corr-hypothesis} with
$\epsilon=C'_\beta\min(|\partial A|,|\partial B|)e^{-R/\xi'_\beta}$.
The constants are those of Eq.~\eqref{def_C'_beta_si_beta}. Inserting this $\epsilon$ into
Eq.~\eqref{eq:corr-relative-entropy} proves the original bound~\eqref{ineq_quantum_correlation_negativity},
using $C'_\beta\le C_\beta$, $\xi'_\beta\le\xi_\beta/2$, and
$\log(1+x)\le\sqrt{x}$ for $x\ge0$.

\textit{Theorem~\ref{thm:quantum_correlation_negativity_1D}.}
Keep the notation and construction of Appendix~\ref{Theorem_thm_quantum_correlation_PPT_rela_1D}.
Apply Theorem~\ref{quantum_correlation_skey_type} and Eq.~\eqref{eq:corr-spectrum} to the full
modified Gibbs state $\widetilde\rho_\beta$ in Eq.~\eqref{def:tilde_rho_beta}, before
tracing out the buffer. For $|A_2|=|B_2|=\ell$, this gives
\begin{align}
 \widetilde\delta
 &\le 2C'_\beta d_0^\ell e^{-R/\xi'_\beta}\notag\\
 &\le16C_\beta e^{-R/\xi_\beta+2\ell\log d_0}
 =\delta_{2,R}.
 \label{eq:corr-F21}
\end{align}
This replaces Eq.~\eqref{eq:orig-F21} without weakening its right-hand side.
The state $\widetilde\sigma_{AB}$ of Eq.~\eqref{def_tilde_sigma_AB_2} is PPT because
$\widetilde\Phi=\Phi_{A_1,A_2}\otimes\Phi_{B_1,B_2}$ is a local
product operator. In Eq.~\eqref{eq:orig-F14}, the spectral decomposition should
be of the partially transposed operator in Eq.~\eqref{eq:orig-F13}.

We also correct the missing factor $\mathcal{D}_{A_2B_2}=d_0^{2\ell}$ in
Eq.~\eqref{Ineq_for_the_sec_term_tilde_sigma}.
Using Eqs.~\eqref{ineq:lemma_quantum_belief},
\eqref{ineq:lemma_quantum_belief_2}, \eqref{def_tilde_sigma_AB_2}, and
\eqref{eq_negativity_belief_pro_8}, the unnormalized state satisfies
\begin{align}
 \|Z_{\widetilde\sigma}\widetilde\sigma_{AB}
       -\rho_{\beta,AB}\|_1
 &\le\delta_{1,\ell}
   +\widetilde\delta\widetilde Z
                  \|\widetilde\Phi\|^2\mathcal{D}_{A_2B_2}\notag\\
 &\le\delta_{1,\ell}
       +\delta_{2,R}d_0^{2\ell}e^{8gk\beta}
  =\bar\delta_{\ell,R}.
 \label{eq:corr-unnormalized}
\end{align}
Here $\bar\delta_{\ell,R}$ is the quantity already defined in
Eq.~\eqref{tilde_Z_tilde_Delta_Z_tilde_sigma}.
Taking the trace gives
$|Z_{\widetilde\sigma}-1|\le\bar\delta_{\ell,R}$, and hence
\begin{equation}
 \|\widetilde\sigma_{AB}-\rho_{\beta,AB}\|_1
 \le 2\bar\delta_{\ell,R}\le4\bar\delta_{\ell,R}.
 \label{eq:corr-normalized}
\end{equation}
Thus Eq.~\eqref{Ineq_for_the_term_tilde_sigma_fin_1_0} is recovered.
The same choice $\ell=\lceil R/[6\log(d_0)\xi_\beta]\rceil$ in
Eq.~\eqref{upp_delta_upp_delta}, followed by the smoothing step in
Eqs.~\eqref{eq:orig-F26}--\eqref{Ineq_for_the_term_tilde_sigma_fin_3},
recovers Eq.~\eqref{ineq_quantum_correlation_negativity_1D_main}
with its original constants.
The small-error regime required by the continuity bound can be assumed;
outside it the displayed estimate is weaker than
$E_R^{\mathrm{PPT}}\le\log\min(\mathcal{D}_A,\mathcal{D}_B)$.
Neither recovery uses Proposition~\ref{prop:quantum_correlation_negativity}
or Theorem~\ref{thm:quantum_correlation}.

\subsection{Vanishing square-root correlations do not imply separability}
\label{sec:corr-upb}

The PPT conclusion cannot be strengthened to separability. Consider
$P=\sum_{i=1}^5|a_i,b_i\rangle\langle a_i,b_i|$, the projector onto
the orthogonal Tiles unextendible product basis (UPB) on
$\mathbb{C}^3\otimes\mathbb{C}^3$~\cite{PhysRevLett.82.5385}.
Its normalized product vectors are
\[
\begin{gathered}
 |0\rangle\otimes\frac{|0\rangle-|1\rangle}{\sqrt2},\qquad
 \frac{|0\rangle-|1\rangle}{\sqrt2}\otimes|2\rangle,\\
 |2\rangle\otimes\frac{|1\rangle-|2\rangle}{\sqrt2},\qquad
 \frac{|1\rangle-|2\rangle}{\sqrt2}\otimes|0\rangle,\\
 \frac{|0\rangle+|1\rangle+|2\rangle}{\sqrt3}
 \otimes\frac{|0\rangle+|1\rangle+|2\rangle}{\sqrt3}.
\end{gathered}
\]
The complementary subspace contains no product vector. Thus
\begin{equation}
 \rho_{AB}=\frac{\hat1-P}{4},\qquad
 \sqrt{\rho_{AB}}=\frac{\hat1-P}{2}
 \label{eq:corr-upb-state}
\end{equation}
is entangled: a separable decomposition would require product
vectors in its support. All vectors above are real, so
$P^{T_A}=P$ and $\rho_{AB}^{T_A}=\rho_{AB}\succeq0$.

For arbitrary local operators, orthogonality of the product vectors
gives
\begin{align}
 \operatorname{tr}(P O_A P O_B)
 &=\sum_i\langle a_i|O_A|a_i\rangle
             \langle b_i|O_B|b_i\rangle\notag\\
 &=\operatorname{tr}(P O_AO_B).
 \label{eq:corr-upb-identity}
\end{align}
Indeed, the $i\ne j$ terms contain
$\langle a_j|a_i\rangle\langle b_i|b_j\rangle=0$.
Consequently,
\begin{align}
 Q_{\rho_{AB}}^{(1/2)}(O_A,O_B)
 &=\frac14\bigl[\operatorname{tr}(P O_AO_B)
                -\operatorname{tr}(P O_A P O_B)\bigr]\notag\\
 &=0.
 \label{eq:corr-upb-zero}
\end{align}
This holds also for non-Hermitian $O_A,O_B$. Hence vanishing
$Q^{(1/2)}$ for every local pair guarantees PPT, but not separability.

The counterexample is not restricted to rank-deficient states.
Replacing $\rho_{AB}$ by
$(\hat1-P+\eta P)/(4+5\eta)$ with $\eta>0$ gives a full-rank
PPT state. The same expansion replaces the prefactor $1/4$ in
Eq.~\eqref{eq:corr-upb-zero} by
$(1-\sqrt\eta)^2/(4+5\eta)$, so all $Q^{(1/2)}$ still vanish.
For sufficiently small $\eta>0$ these states remain entangled,
since the separable set is closed and the limit at $\eta=0$
is entangled.

\def\bibsection{\section*{References}}

\bibliography{Quantum_correlation}

\begin{thebibliography}{215}
\bibitem [{Note*3()}]{Note*3}%
  \BibitemOpen
  \bibinfo {note} {Choosing $R\gtrsim \protect \qopname \relax o{log}(n)$ in
  Ineq.~\protect \textup {\hbox {\mathsurround \z@ \protect \normalfont
  (\ignorespaces \ref {Product_high_temp}\unskip \@@italiccorr )}} yields
  $\delta _{\rho _{AB}} =1/{\protect \rm poly}(n)$ with $\delta _{\rho _{AB}} $
  defined in \protect \textup {\hbox {\mathsurround \z@ \protect \normalfont
  (\ignorespaces \ref {sep_closeness_ineq}\unskip \@@italiccorr )}}; hence, the
  continuity bound yields $E(\rho _{AB})\le 1/{\protect \rm poly}(n)$ as well.
  However, for specific entanglement measures, this discussion cannot be
  applied. For example, regarding the entanglement negativity, the distance
  $\delta _{\rho _{AB}}$ should be as small as $1/\protect \sqrt {\protect
  \mathcal {D}_A}$ to obtain a meaningful upper bound~\cite
  [Inequality~(16)]{PhysRevB.102.235110}), where $\protect \mathcal {D}_A$ may
  be as large as $e^{\protect \mathcal {O}(n)}$ if $|A|=\protect \mathcal
  {O}(n)$. Therefore, even at high temperatures, the possibility of certain
  entanglement measures satisfying the clustering property remains
  unclear.}\BibitemShut {Stop}%
\bibitem [{Note*4()}]{Note*4}%
  \BibitemOpen
  \bibinfo {note} {In Ref.~\cite {10.2307/1194711}, the quantum Fisher
  information is defined in a manner different to that in this study; it is
  defined as $\protect \mathcal {F}_{\rho _\beta } (K) /4$ according to this
  study’s notation.}\BibitemShut {Stop}%
\bibitem [{Note*9()}]{Note*9}%
  \BibitemOpen
  \bibinfo {note} {For example, consider the case in which $\rho _{AB}=|
  0^{\otimes n} \rangle \langle 0^{\otimes n} |$, with $AB=\Lambda $ and $\rho
  _{AB}'=(1-\epsilon )\rho _{AB}+\epsilon \rho _{\protect \rm r} $. Here, the
  state $\rho _{\protect \rm r}=| \protect \rm r \rangle \langle \protect \rm r
  |$ is a random pure state on $\Lambda $, which is orthogonal to $| 0^{\otimes
  n} \rangle $. We then obtain $E_N(\rho _{AB})=0$ and $E_N(\rho
  _{AB})=\protect \qopname \relax o{log}\left ( 1-\epsilon +\epsilon \protect
  \bigl \| \rho _{\protect \rm r}^{T_A}\protect \bigr \|_1 \right )$. Here,
  $\protect \bigl \| \rho _{\protect \rm r}^{T_A}\protect \bigr \|_1=e^{\Theta
  (n)}$~\cite {PhysRevA.81.052312}, and hence, provided $\epsilon =e^{-\protect
  \mathcal {O}(n^z)}$ ($0<z<1$), $| E_N(\rho _{AB}) - E_N(\rho '_{AB})| \propto
  n$ is obtained.}\BibitemShut {Stop}%
\end{thebibliography}
%
%
%




%
%
%
%
%
%
%
%
%
%

\end{document}